\documentclass[usenatbib]{mnras}
\usepackage{newtxtext,newtxmath}
\usepackage[T1]{fontenc}
\usepackage{ae,aecompl}
\usepackage{graphicx}	
\usepackage{amsmath}	
\usepackage{amssymb}	
\usepackage{subcaption}
\usepackage{color}
\usepackage{nccmath}
\usepackage{float}
\usepackage{placeins}
\usepackage{arydshln}
\usepackage{lscape}
\title[Clustering of radio sources in COSMOS]{The clustering and bias of radio-selected AGN and star-forming galaxies in the COSMOS field}

\author[Hale et al.]{C. L. Hale,$^{1}$\thanks{E-mail: catherine.hale@physics.ox.ac.uk}
M. J. Jarvis,$^{1,2}$ 
I. Delvecchio,$^{3}$
P. W. Hatfield,$^{1}$
M. Novak,$^{3}$
\newauthor V. Smol{\v c}i{\'c},$^{3}$
and G. Zamorani$^{4}$
\\ \\
$^{1}$Astrophysics, University of Oxford, Denys Wilkinson Building, Keble Road, Oxford, OX1 3RH, UK\\
$^{2}$Department for Physics, University of the Western Cape, Bellville 7535, South Africa\\
$^{3}$Department of Physics, Faculty of Science, University of Zagreb,  Bijeni\v{c}ka cesta 32, 10000  Zagreb, Croatia \\
$^{4}$INAF-Osservatorio Astronomico di Bologna, via Gobetti 93/3, I-40129 Bologna, Italy \\
}

\date{Accepted XXX. Received YYY; in original form ZZZ}

\pubyear{2017}

\begin{document}
\label{firstpage}
\pagerange{\pageref{firstpage}--\pageref{lastpage}}
\maketitle 

\begin{abstract}
Dark matter haloes in which galaxies reside are likely to have a significant impact on their evolution. We investigate the link between dark matter haloes and their constituent galaxies by measuring the angular two-point correlation function of radio sources, using recently released 3 GHz imaging over $\sim 2 \ \mathrm{deg}^2$ of the COSMOS field. We split the radio source population into Star Forming Galaxies (SFGs) and Active Galactic Nuclei (AGN), and further separate the AGN into radiatively efficient and inefficient accreters. 
Restricting our analysis to $z<1$, we find SFGs have a bias, $b = 1.5 ^{+0.1}_{-0.2}$, at a median redshift of $z=0.62$. On the other hand, AGN are significantly more strongly clustered with $b = 2.1\pm 0.2$ at a median redshift of 0.7. This supports the idea that AGN are hosted by more massive haloes than SFGs. We also find low-accretion rate AGN are more clustered ($b = 2.9 \pm 0.3$) than high-accretion rate AGN ($b = 1.8^{+0.4}_{-0.5}$) at the same redshift ($z \sim 0.7$), suggesting that low-accretion rate AGN reside in higher mass haloes. This supports previous evidence that the relatively hot gas that inhabits the most massive haloes is unable to be easily accreted by the central AGN, causing them to be inefficient. We also find evidence that low-accretion rate AGN appear to reside in halo masses of $M_{h} \sim 3-4 \times 10^{13}$\,$h^{-1}$\,M$_{\odot}$ at all redshifts. On the other hand, the efficient accreters reside in haloes of $M_{h} \sim 1-2 \times 10^{13}$\,$h^{-1}$\,M$_{\odot}$ at low redshift but can reside in relatively lower mass haloes at higher redshifts. This could be due to the increased prevalence of cold gas in lower mass haloes at $z \ge 1$ compared to $z<1$. 

\end{abstract}


\section{Introduction}
\label{sec:introduction}
For many years, the properties of Active Galactic Nuclei (AGN) were believed to be explained through a unified model \citep[][]{Urry}. This model involves an accreting black hole surrounded by a dusty torus, where in some instances the jets can form and produce bright radio emission. Differences in the observed AGN spectral properties were explained by variations of the orientation of the AGN with respect to the observer. In recent years evidence has suggested that two types of AGN may exist, with differences dependent on their accretion mode, and its efficiency \citep[see e.g.][]{Hardcastle2007,Mingo2014,HeckmanBest,Fernandes2015}. This classification distinguishes whether an AGN is radiatively efficient (radiative mode) or inefficient (jet mode). 

In the efficient mode (radiative), AGN accrete material from an accretion disk, aligned with the traditional orientation-based AGN unification \citep[][]{Urry}. These AGN have luminosities of $L \gtrsim 0.01L_{\rm Edd}$, where $L_{\rm Edd}$ is the Eddington luminosity. The inefficient mode (jet) however is thought to have an advection dominated accretion flow \citep[e.g.][]{Kording2006} with $L \lesssim 0.01 L_{\rm Edd}$; illustrations of these two modes can be found in Figure 3 of \cite{HeckmanBest}. For radio galaxies, the radiatively efficient radio-loud AGN are also known as High Excitation Radio Galaxies (HERGs), whereas the inefficient AGN are Low Excitation Radio Galaxies (LERGs). Due to the relationship between the accretion power and the ionisation properties, these two populations are likely to reside in different environments \citep[][]{Hardcastle2004,Janssen2012,Gendre2013}; evolve differently over time \citep[e.g.][]{Hardcastle2013,Best,Pracy2016} and have different properties such as stellar mass \citep{Herbert2011} and star-formation rates \citep[e.g.][]{Herbert2010,Smolcic2009,Hardcastle2013,Virdee2013}. 

However, it is important to measure these differences, as both HERGs and LERGs may be responsible for depositing energy into both their host galaxy's inter-stellar medium and the wider cluster-scale environment at both low \citep[e.g.][]{Fabian2002,Mcnamara2005,Mcnamara2007} and high redshifts \citep[e.g.][]{RawlingsJarvis2004,Hatch2014}. Such feedback from AGNs is now regularly invoked in various guises in both semi-analytic models \citep[SAMs; e.g. ][]{Bower2006,Croton2016} and hydrodynamical simulations \citep[e.g.][]{Beckmann2017,Mcalpine2017,Weinberger2017} of galaxy formation, to solve the overcooling problem and to quench or terminate star formation over cosmic time. Thus, determining the relationship between these forms of AGN activity and the star formation in galaxies is critical.
As we undertake deeper and wider area surveys, the increase in sample sizes for these two AGN populations, along with star-forming galaxies (SFGs) allows us to investigate fundamental properties of these sources in more detail. 

One key element of this picture is to understand the environment in which the different accretion-mode AGN reside, and to link this with the underlying dark matter distribution, allowing a more direct comparison with SAMs and simulations.
Galaxies are biased tracers of dark matter \citep[e.g.][]{Kaiser1984,CooraySheth,Mo2010}, and the mass of the haloes, their temperature as well as their structure can all affect their constituent galaxies and how they trace the dark matter \citep[e.g.][]{PressSchechter,LaceyCole1a,LaceyCole1b,ShethTormen}. One way to investigate the variety of environments and dark matter haloes of different galaxy types is through their clustering.

Measurement of the clustering of galaxies is frequently carried out by determining the two-point correlation function (TPCF). The TPCF provides a measurement of the clustering of galaxies at different scales, to do this it quantifies the excess number of galaxies compared to that of a random distribution of galaxies. This can be described both by the correlation over spatial scales (the spatial correlation function, $\xi(r)$), or angular scales (the angular correlation function, $\omega(\theta)$) using the projection of galaxies on the celestial sphere \citep{Peebles1980}.

Theory predicts power law distributions for the clustering of dark matter, and hence for galaxies over a broad range of scales \citep[e.g.][found $\xi(r) \propto r^{-1.8}$ and $\omega(\theta) \propto \theta^{-0.8}$]{Peebles1974}. This is reflected in later observational studies \citep[e.g.][]{Davis1983,Davis1988,RocheEales,Norberg2002}, which generally assume or measure a slope of $\sim -0.8$ in the angular TPCF over large angular scales. These large scales, where an approximate power law is observed, quantify the clustering of galaxies in different dark matter haloes (the `2-halo' term), and can be used to determine how biased (over-clustered) galaxies are with respect to the dark matter. Deep data with reasonable spatial resolution (often using optical and IR data) also show an excess of clustering at smaller scales due to the clustering of galaxies within a single dark matter halo \citep[the `1-halo term'; see e.g. ][]{CooraySheth,Yang2003,Zheng2007,Hatfield2016, Hatfield2017}.

Observations covering the whole of the electromagnetic spectrum have been used to investigate the relationship between AGN and star formation activity and their dark matter haloes. These studies find that AGN tend to be more highly clustered than the general galaxy population at similar redshifts, and inhabit similar haloes to the most massive galaxies \citep[][]{Mandelbaum2009,Gilli2009,Donoso2014}. Star forming galaxies tend to be less clustered than AGN, although the clustering appears to depend on the star formation rate (SFR); galaxies with high star-formation rates are more clustered than the low SFR counterparts \citep[e.g.][]{Lewis2002,Blain2004,Maddox2010,Amblard2011}. This can be interpreted in terms of the star-formation main sequence \citep[e.g.][]{Noeske2007,Whitaker2012,Johnston2015}, where the SFR is correlated with the stellar mass, and the stellar mass is more strongly correlated with the dark-matter halo mass. However, UV and optical tracers of star formation are susceptible to dust obscuration effects, where a true estimate of the SFR is difficult \citep[e.g.][]{Hao2011,Cucciati2012}.  Moreover, jet-mode AGN are extremely difficult to identify in optical imaging or spectroscopy as they essentially have similar properties to dead, red elliptical galaxies \citep{Best2005,Herbert2010}.

An efficient method of investigating the clustering of different accretion-mode AGN and the SFG population is to use radio surveys, where (at low frequency) synchrotron radiation from the jet emission from both high- and low-accretion rate AGN is visible \citep[e.g.][]{Best2005a,Pracy2016,Whittam2016} and the synchrotron emission from relativistic electrons accelerated in supernova remnants provides a reliable estimate of the SFR \citep{Yun2001,Bell2003,Jarvis2010,Davies2017}. Although there is a contribution at higher frequencies from Free-free emission \citep[see e.g.][]{Condon1992}.

Measurements of the clustering of radio sources have traditionally provided insights into how different types of AGN trace the underlying dark matter distribution \cite[e.g.][]{BlakeWall2,Lindsay2014a, NVSSClustering}. More recently however, the deeper radio surveys allow the clustering of less luminous AGN and SFGs to be studied in more detail \citep[e.g.][]{Lindsay2014,Magliocchetti}, and cross-correlation analyses with lensing of the cosmic microwave background provide an independent measurement of how radio sources trace the underlying dark matter distribution \citep[][]{PlanckLensing,Allison2015}.

Clustering is not only useful in assessing environmental influences on galaxy evolution, but also has cosmological importance. This large scale distribution relates to the matter power spectrum, which needs to be well understood for investigating the cosmological parameters that define our Universe. In recent years, there has been renewed interest in using radio sources as cosmological probes, mainly due to the fact that they can be observed to high redshift and do not suffer from dust obscuration, \citep[for a review see][]{JarvisSKAcosmo}, whilst there is also growing interest in using them for weak lensing experiments \citep[e.g.][]{BrownSKA,Harrison2016}.

In terms of the clustering, the large number of sources in deep radio surveys out to high redshifts makes them useful probes of the very large scales in the Universe, providing novel information on the amplitude of the power spectrum at $z \ge 1$ \citep[e.g.][]{Raccanelli2012,Camera2012}. The fact that radio surveys contain a range of populations with a large spread in the bias with respect to the dark matter, allows the multi-tracer technique \citep{Seljak2009} to be exploited for measuring e.g. the level of non-Gaussianity in the initial density fluctuations \citep[see e.g. ][]{Ferramacho2014,Raccanelli2015}. 

In this paper, we investigate the differences in the clustering of different radio source types in the COSMOS field using deep Karl G. Jansky Very Large Array (VLA)  3GHz data. We describe the methods and data used in this work in Section \ref{sec:methods}. The results of the clustering analysis are presented in Section \ref{sec:tpcf_results}, and discussed in Section \ref{sec:discussions}. Finally conclusions are provided in Section \ref{sec:conclusions}. We use $H_0=100h$ kms$^{-1}$Mpc$^{-1}$, $\Omega_M=0.3$, $\Omega_{\Lambda}=0.7$ and $\sigma_8=0.8$. For determining luminosities we assume $h=0.7$, for all clustering measurements we leave our values in terms of $h^{-1}$. 

\section{Methods}
\label{sec:methods}
\subsection{The Two-Point Correlation Function}
\label{sec:calctpcf}
The angular TPCF ($\omega(\theta)$) describes the probability of a pair of galaxies being observed, $\delta P$, over a certain solid angle element, $\delta \Omega$, for a given surface density, $\sigma$. 

\begin{ceqn}
\begin{align}
\label{eq:tpcf_def}
\delta P = \sigma [1+\omega(\theta)] \delta \Omega ,
\end{align}
\end{ceqn}
where $\omega(\theta)$ is often calculated using an estimator that counts pairs of galaxies within a given projected angular separation, $\theta$. The number of pairs is determined for both the real galaxies, $DD(\theta$), and randomly distributed galaxies, $RR(\theta$). A similar prescription to \cite{Lindsay2014} is followed here, where we use the estimator from \cite{LandySzalay}, i.e.

\begin{ceqn}
\begin{align}
\omega(\theta) = \frac{DD(\theta)-2DR(\theta)+RR(\theta)}{RR(\theta)} .
\label{eq:tpcf}  
\end{align}
\end{ceqn}

This takes into account the cross correlation between the data and random catalogue ($DR(\theta)$). The galaxy pairs $DD(\theta)$, $RR(\theta)$ and $DR(\theta)$ are normalised over all angles. 

The uncertainties using this estimator are naively Poisson. This, however, underestimates the errors: the counts in different bins are correlated and the size of the field observed is finite. In order to better estimate the errors, we use bootstrap resampling \citep[e.g.][]{LingBarrowFrenk}. \cite{Cress1996} and \cite{Gilli2009} both found that the bootstrap errors are about a factor of two larger than those calculated by the Poisson value. In this paper, the uncertainties on the correlation function are calculated using the 16$^{\textrm{th}}$ and 84$^{\textrm{th}}$ percentiles of the bootstrap samples.

\subsection{Data}
\label{sec:data}

The data used for this analysis is from \cite{Smolcic2017} where a complete description is provided, and we only provide a brief overview of these data here.  The radio data are from the VLA observations of the COSMOS field taken at 2-4~GHz, using a total time of 384 hours. The resulting map has a median r.m.s. of 2.3$\mu$Jy/beam and the catalogue is defined for sources which have a signal-to-noise ratio (SNR) > $5 \sigma$, where $\sigma$ is the local value of the r.m.s. noise at the position of the galaxy. We use a 5.5$\sigma$ cut on the final catalogue to minimize the number of spurious detections \citep[see][]{Smolcic2017}. This 5.5$\sigma$ catalogue contains a total of $8,928$ sources over $\sim 2$ deg$^2$. 

\cite{Smolcic2017sorting} crossmatched $\sim$86 per cent of these 5.5$\sigma$ galaxies with ancillary data (in the whole, unmasked field), and $\sim$79 per cent of the whole (5.5$\sigma$) sample were classified into either being a SFG or AGN using the wealth of multi-wavelength ancillary data. The AGN are then classified as radiatively efficient (or High to moderate Luminosity AGN; HLAGN) or radiatively inefficient (or Moderate to low Luminosity AGN; MLAGN). HLAGN and MLAGN are designed to be roughly equivalent to High (HERGs) and Low Excitation Radio Galaxies (LERGs), respectively \citep[see e.g.][]{Hardcastle2007,HeckmanBest}. 

HLAGN were defined as the collective sample of AGN identified from at least one of the following diagnostics: X-ray luminosity \citep[L$_X$>1$0^{42}$~erg~s$^{-1}$; e.g. ][]{Szokoly2004}, mid-IR colours \citep[][]{Donley2012} or SED-fitting decomposition \citep{Delvecchio2017}. The latter criteria are mainly sensitive to radiative emission likely arising from accretion onto the central SMBH rather than from star formation. For the remainder of the sample, \cite{Smolcic2017sorting} used the rest frame (dust-corrected) [NUV-r] colours to separate between quiescent galaxies or those dominated by star formation. Sources with a UV excess (NUV-r<3.5) were classified as SFGs; red, quiescent sources (NUV-r>3.5) and those with a 1.4 GHz radio luminosity excess relative to the galaxy's SFR\footnote{The SFR estimates were derived from the total infrared luminosity ($L_{\textrm{IR}}$, over the rest-frame 8-1000$\mu$m) of each best-fit galaxy template obtained from SED-fitting. Each $L_{\textrm{IR}}$ estimate was converted to a SFR via a Kennicutt \citep[][]{Kennicutt1998} conversion factor, and scaled to a Chabrier initial mass function \citep[][]{Chabrier2003}. A threshold in log$_{10}(L_{\rm 1.4 GHz}/SFR_{IR})$ was used to define an excess, see \cite{Smolcic2017sorting}.} were classified as quiescent MLAGN and radio-excess MLAGN, respectively. Since more than 60 per cent of the radio-excess MLAGN sources also fulfilled SFG requirements, we excluded these from our SFG subsample. Our total AGN population consists of the combination of HLAGN, quiescent MLAGN and radio-excess MLAGN populations. For our AGN subsamples we consider only the HLAGN and quiescent MLAGN.

\begin{figure*}
\begin{center}
\begin{minipage}[b]{.5\textwidth}
\centering
\includegraphics[height=7.cm]{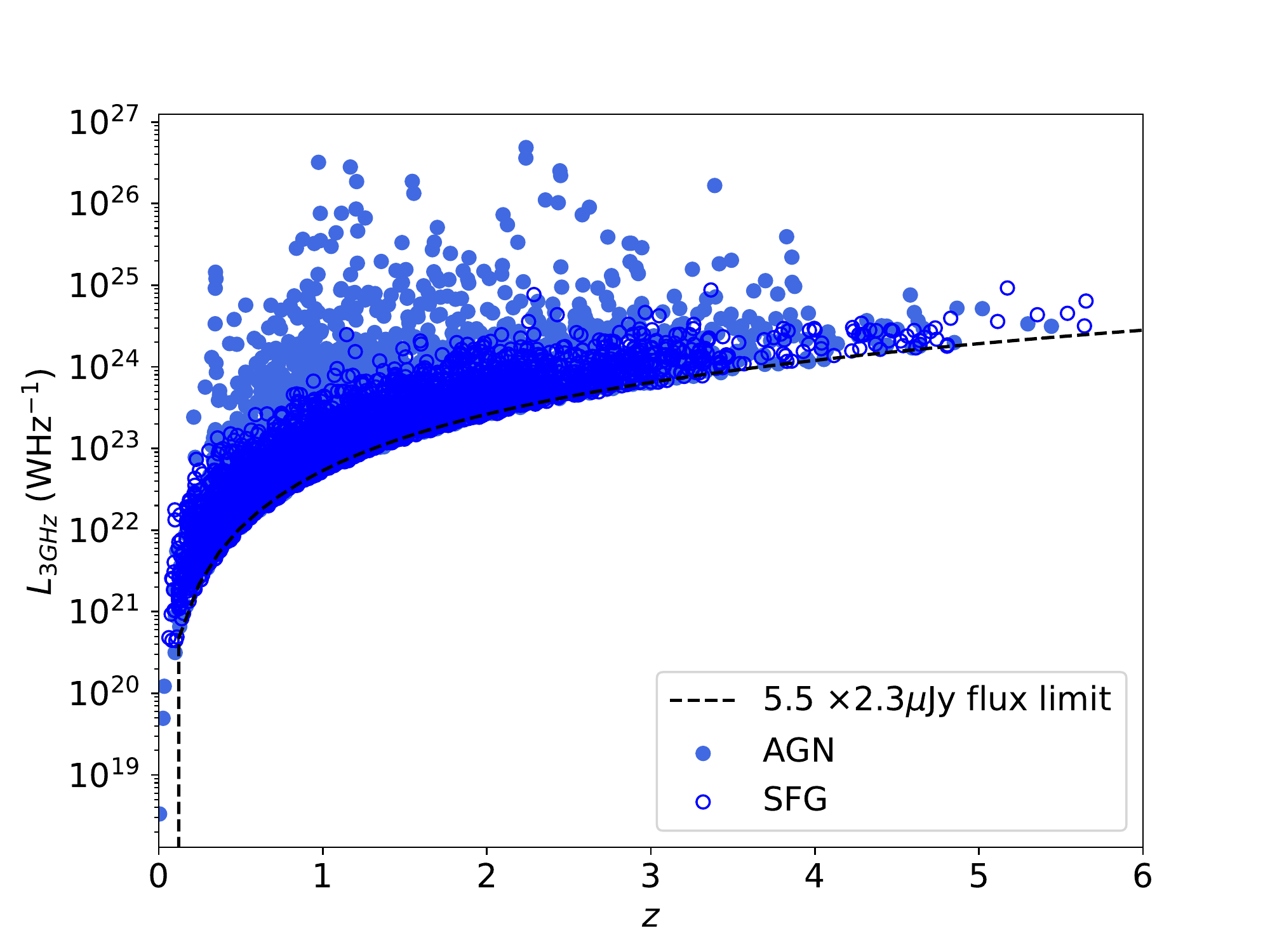}
\subcaption{}
\end{minipage}%
\begin{minipage}[b]{.5\textwidth}
\centering
\includegraphics[height=7.cm]{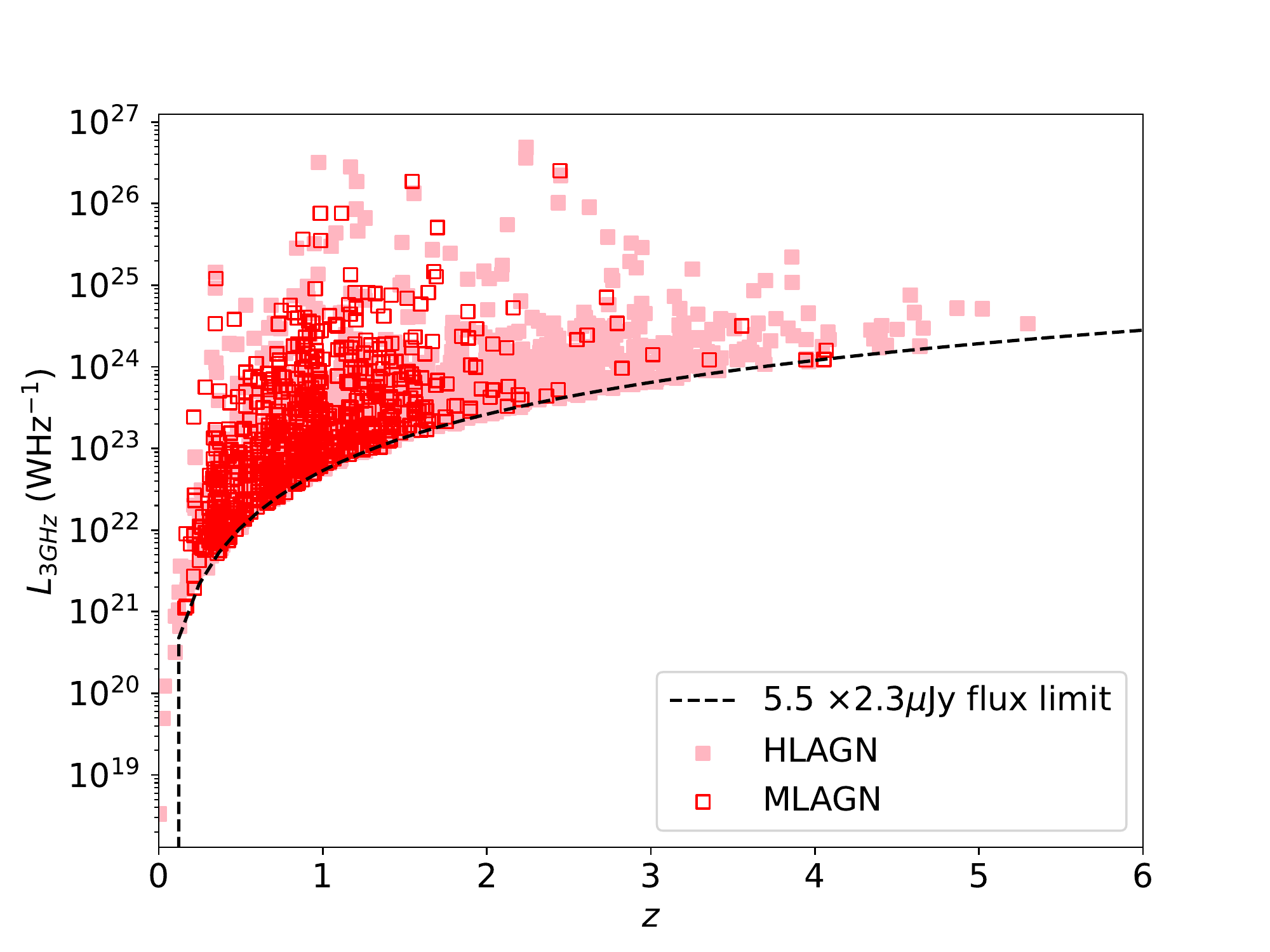}
\subcaption{}
\end{minipage}%
\end{center}
\caption{The luminosity distribution of (a) AGN (light blue filled circles) and SFGs (dark blue open circles) and (b) HLAGN (light red filled squares) and MLAGN (open red squares) with redshift. Also shown (black dashed line) is the luminosity limit for the flux limit of 5.5 $\times$ 2.3 $\mu$Jy for a source with a spectral index, $\alpha =0.7$.}
\label{fig:luminositydist}
\end{figure*}

In Sections ~\ref{sec:tpcf_results} and \ref{sec:discussions} we also measure the clustering of AGN and SFGs split by luminosity. The luminosity distribution with redshift for these sources can be seen in Figure \ref{fig:luminositydist}(a). For completeness we also show the luminosity distribution for HLAGN and MLAGN in Figure \ref{fig:luminositydist}(b). 
\subsubsection{Masking}
\label{sec:mask}
In their classification of sources, \cite{Smolcic2017sorting} used the photometric redshifts from \cite{Laigle2016} to assign redshifts to the radio sources. As such, a mask needed to be used in order to accurately account for the possibility of a radio source being unclassified due to its proximity to a bright star for example. We also reduce the size of the field slightly to mitigate the larger noise variation at the very edges of the field, this results in a slightly reduced size for the parent catalogue from 8,982 to 8,887. 

\begin{figure}
\begin{center}
\includegraphics[height=6cm]{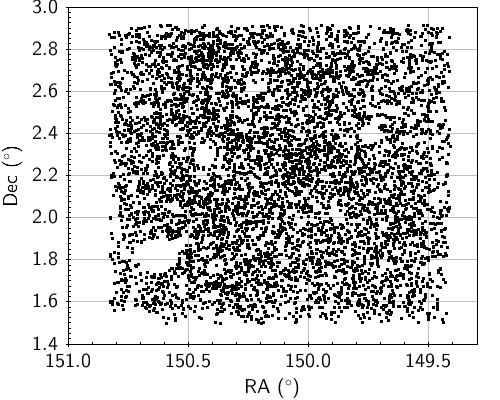}
\end{center}
\caption{The locations of the radio sources used in this work after the mask from \protect \cite{Laigle2016}, defined using the optical and near-infrared data, has been applied.}
\label{fig:locations}
\end{figure}

The resulting sky distribution of the radio sources used in this work is shown in Figure~\ref{fig:locations}. From the trimmed region we have a total of 8,887 sources, and in our final, masked catalogue 3,704 SFGs without a radio-excess and 2,937 robustly classified AGN. Splitting by luminosity, 1,654 SFGs and 747 AGN have $L_{\rm 3GHz} < 10^{23}$ WHz$^{-1}$. Full details of the numbers in each subsample are given in Table~\ref{table:clustering}.

\subsection{Random Catalogue}
\label{sec:random} 
As the radio map has non-uniform noise, this will affect the detection of sources as a function of their position. Therefore, when generating a random catalogue we can not solely rely on generating random positions, we also need to determine whether a source would be detectable, given the r.m.s. at that position. 

To obtain the random catalogue, we therefore used the Square Kilometre Array (SKA) Design Study Simulated Skies simulations \citep[$S^3$; ][hereafter $S^3$]{Wilman2008,Wilman2010}. $S^3$ are semi-empirical simulations that have been widely used to help in the design of the SKA. Thus, to create the random catalogue, a random position (RA, Dec) within the field was chosen, along with a random flux density (using the 1.4 GHz fluxes in $S^3$ scaled to a 3 GHz flux using a spectral index $\alpha = 0.7$, where $S_\nu \propto \nu^{-\alpha}$) and redshift using a source from the $S^3$ catalogue\footnote{For the full radio sample, some sources do not have either photometric or spectroscopic redshifts. For these we use the redshift for the random sources, generated from the $S^3$ simulation, to determine $N(z)$, and hence calculate $r_0$.}. To remain in the random catalogue the flux of the source plus a randomly assigned noise value was required to be above the local value of 5.5$\sigma$. The noise was determined from sampling from a gaussian distribution with a $\sigma$ of the r.m.s. at that location. This determined whether a source would be observable at its given location, taking into account any noise peaks or troughs. This process was repeated until the number of random sources was 20 times the number of real radio sources to ensure that the errors were dominated by the data, not the random catalogue.

When the random catalogues were generated for galaxies of different types an extra condition was needed. This is because some less luminous populations will be more heavily weighted around the flux limit of the survey, and thus be more affected by Malmquist and Eddington biases. To overcome this issue we made use of the $S^3$ catalogue, which contains information regarding whether the radio source is a SFG or AGN. When generating the random samples (see Section~\ref{sec:clustype}), only $S^3$ galaxies corresponding to the same type were permitted in the random catalogue. The classification in $S^3$ does not classify the radio sources as HLAGN and MLAGN, so the $S^3$ AGN catalogue was used for these sub-populations.

\begin{figure*}
\begin{center}
\begin{minipage}[b]{.3\textwidth}
\centering
\includegraphics[height=5.2cm]{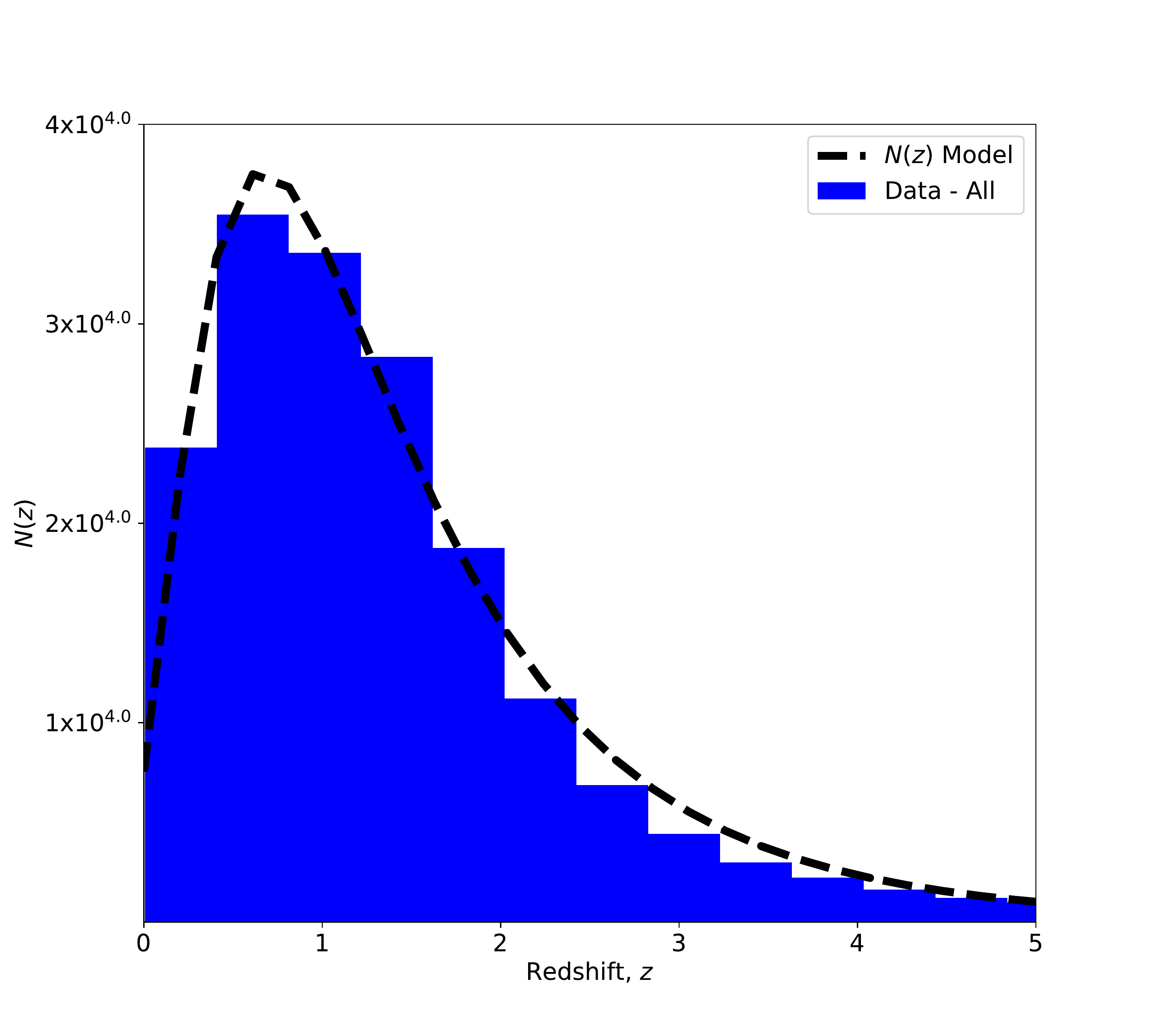}
\subcaption{(a)}
\end{minipage}%
\begin{minipage}[b]{.3\textwidth}
\centering
\includegraphics[height=5.2cm]{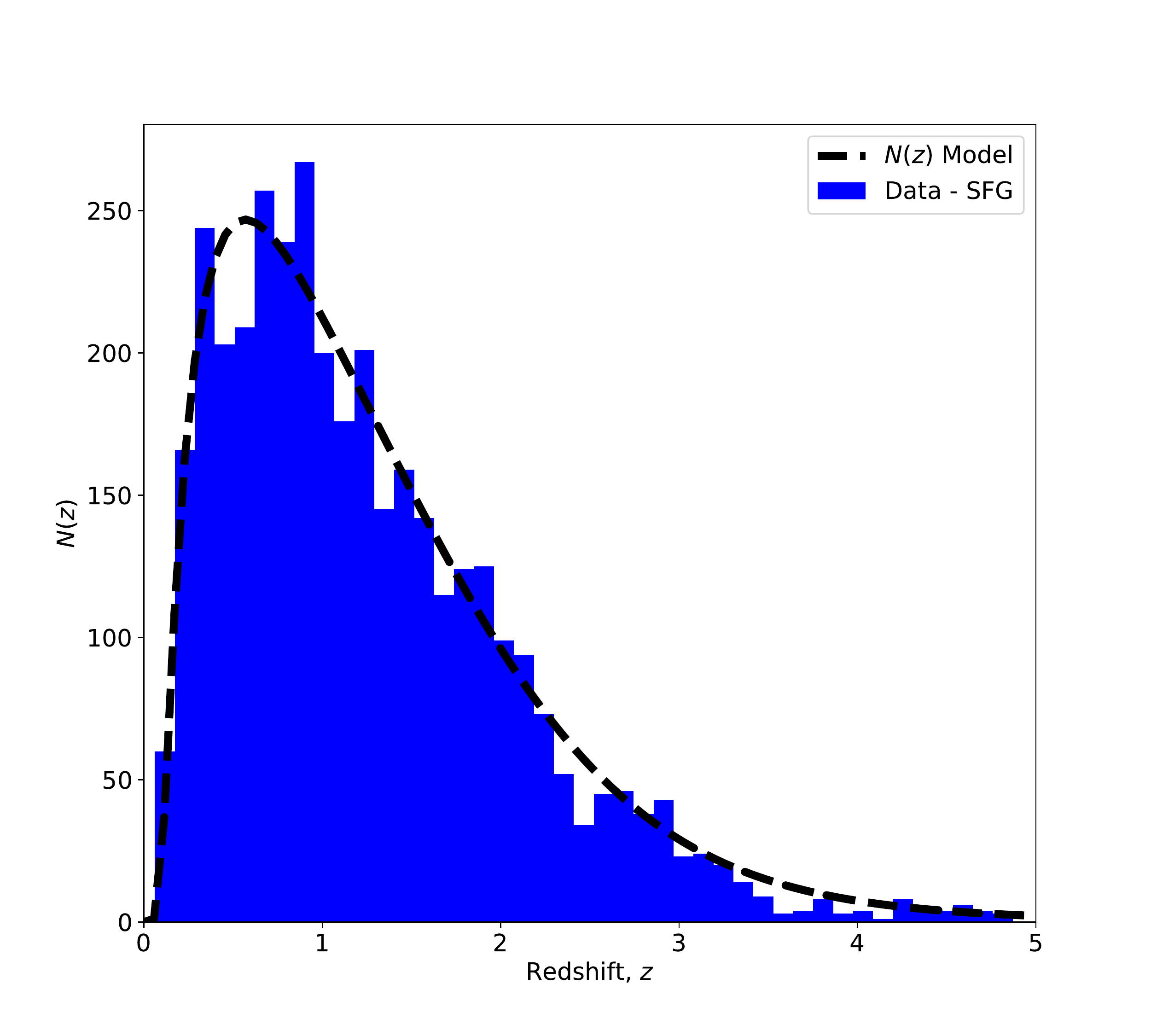}
\subcaption{(b)}
\end{minipage}%
\begin{minipage}[b]{.3\textwidth}
\centering
\includegraphics[height=5.2cm]{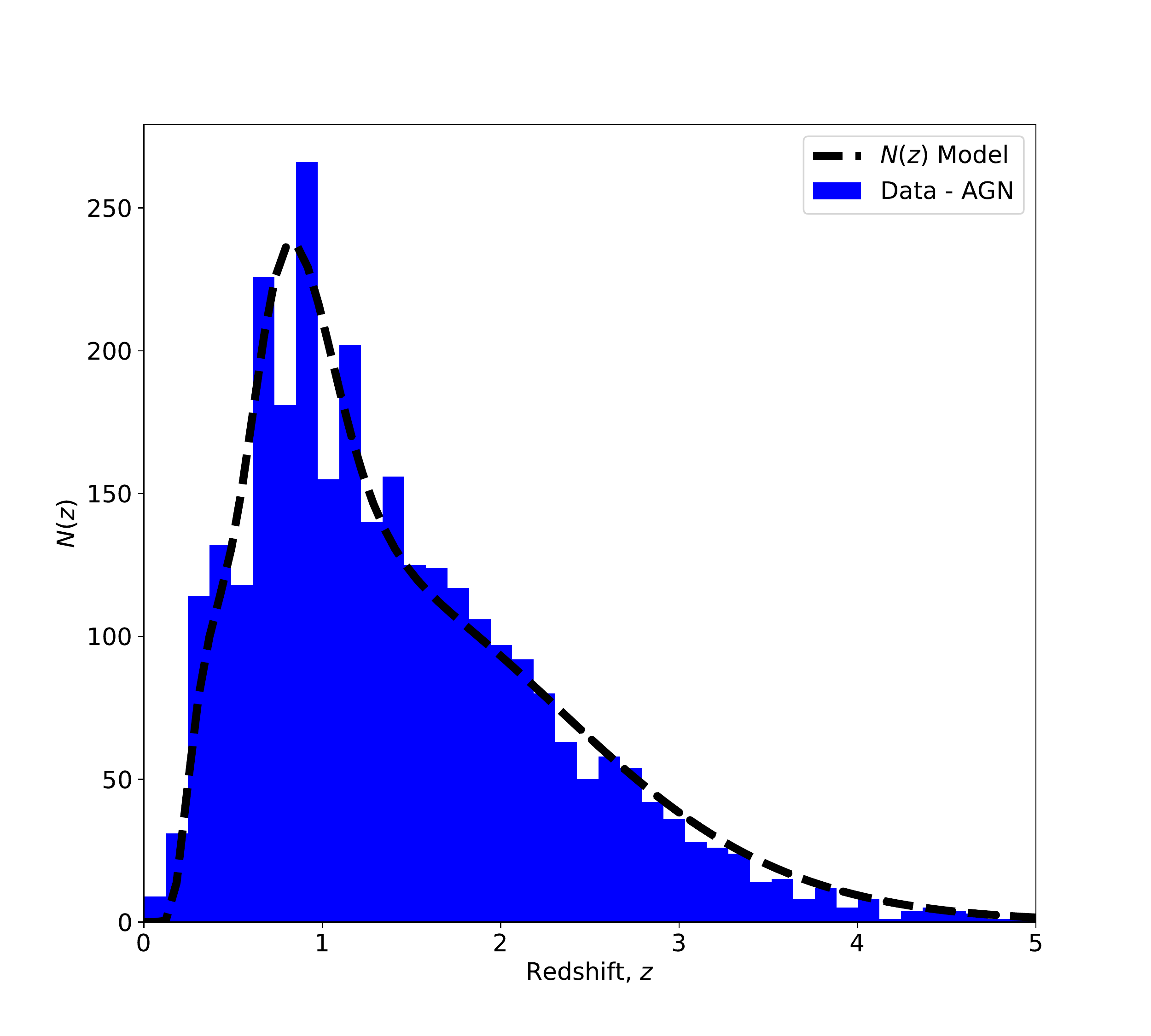}
\subcaption{(c)}
\end{minipage}%
\newline \newline
\begin{minipage}[b]{.3\textwidth}
\centering
\includegraphics[height=5.2cm]{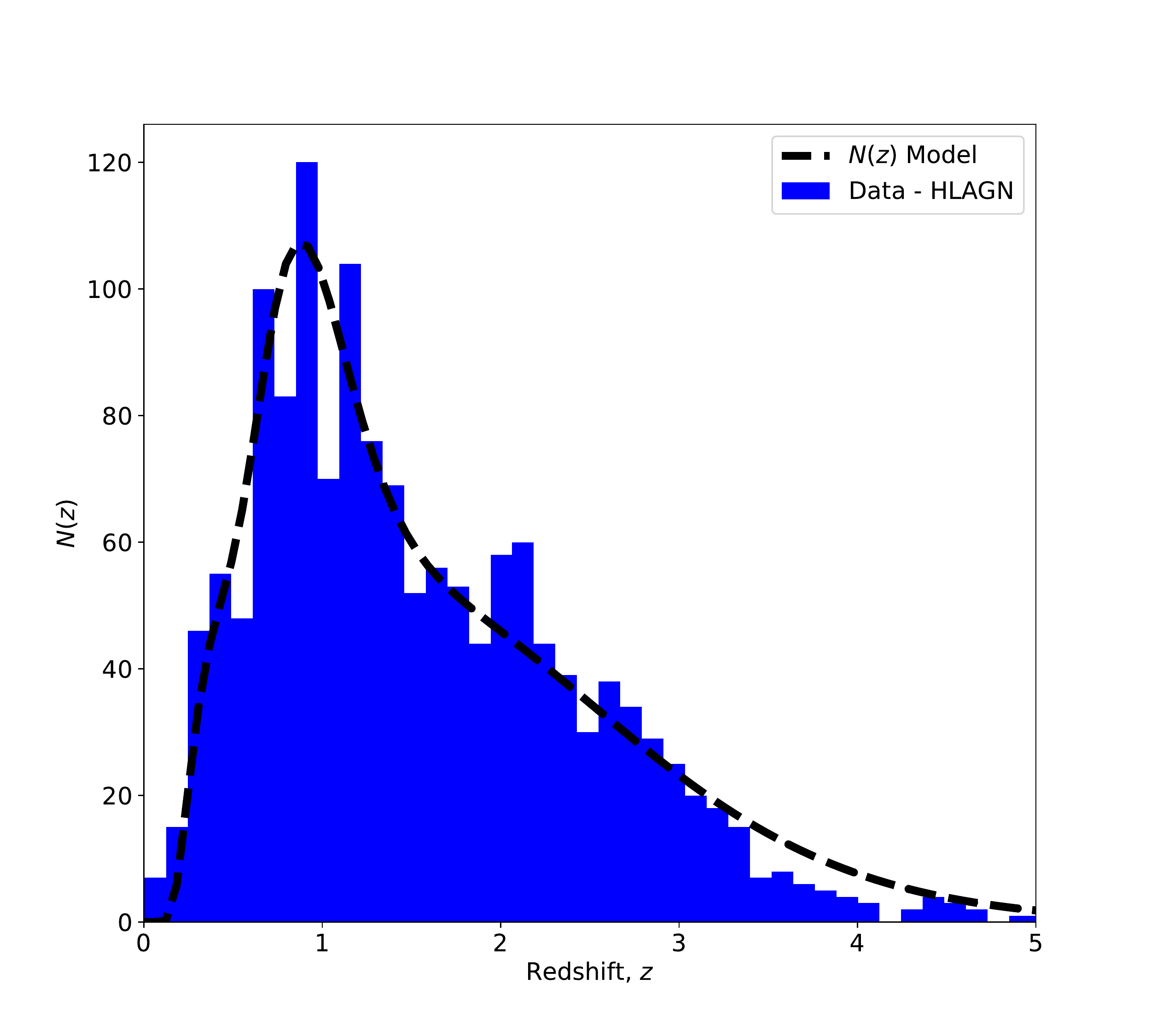}
\subcaption{(d)}
\end{minipage}%
\begin{minipage}[b]{.3\textwidth}
\centering
\includegraphics[height=5.2cm]{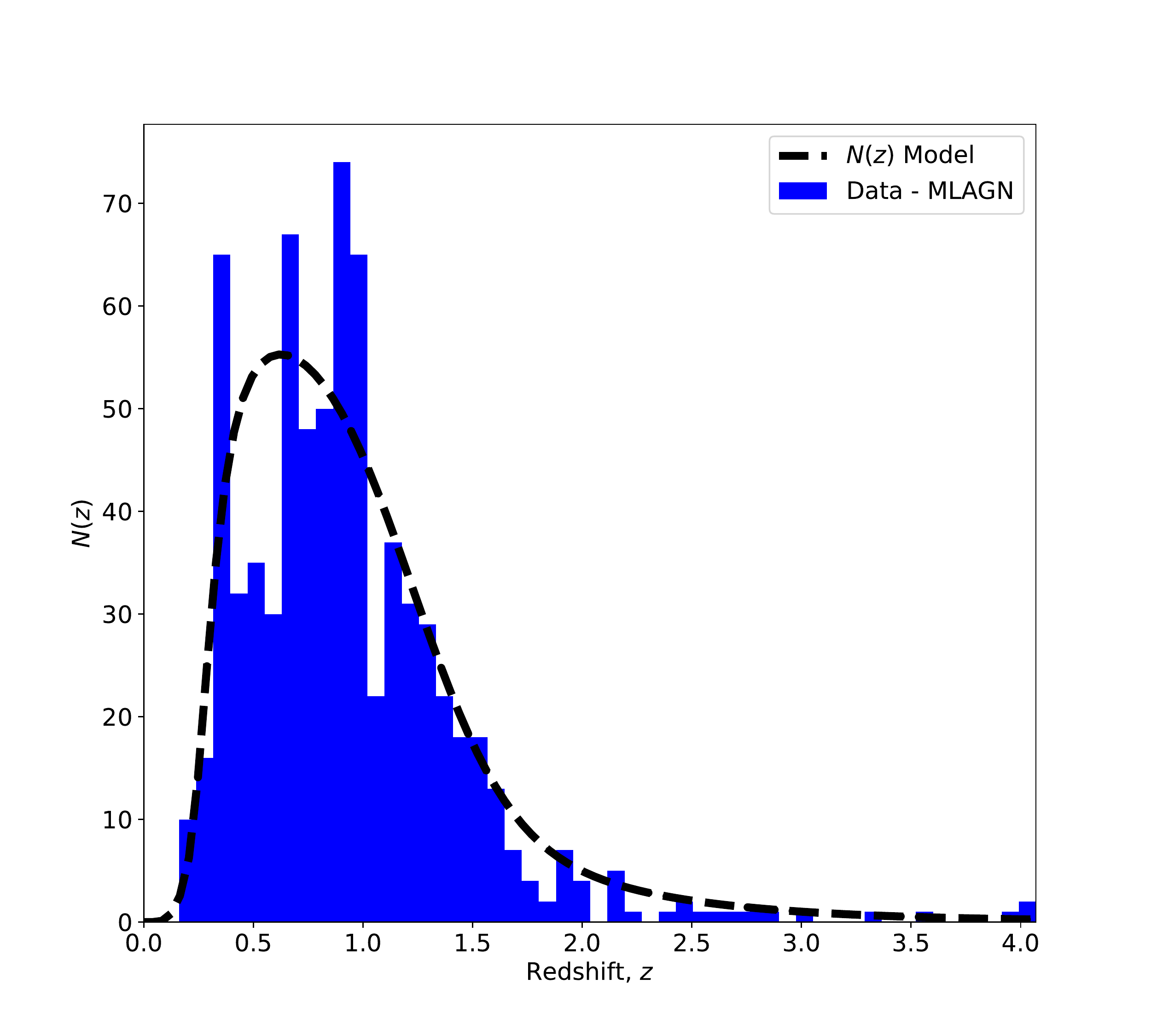}
\subcaption{(e)}
\end{minipage}%
\begin{minipage}[b]{.3\textwidth}
\centering
\includegraphics[height=5.2cm]{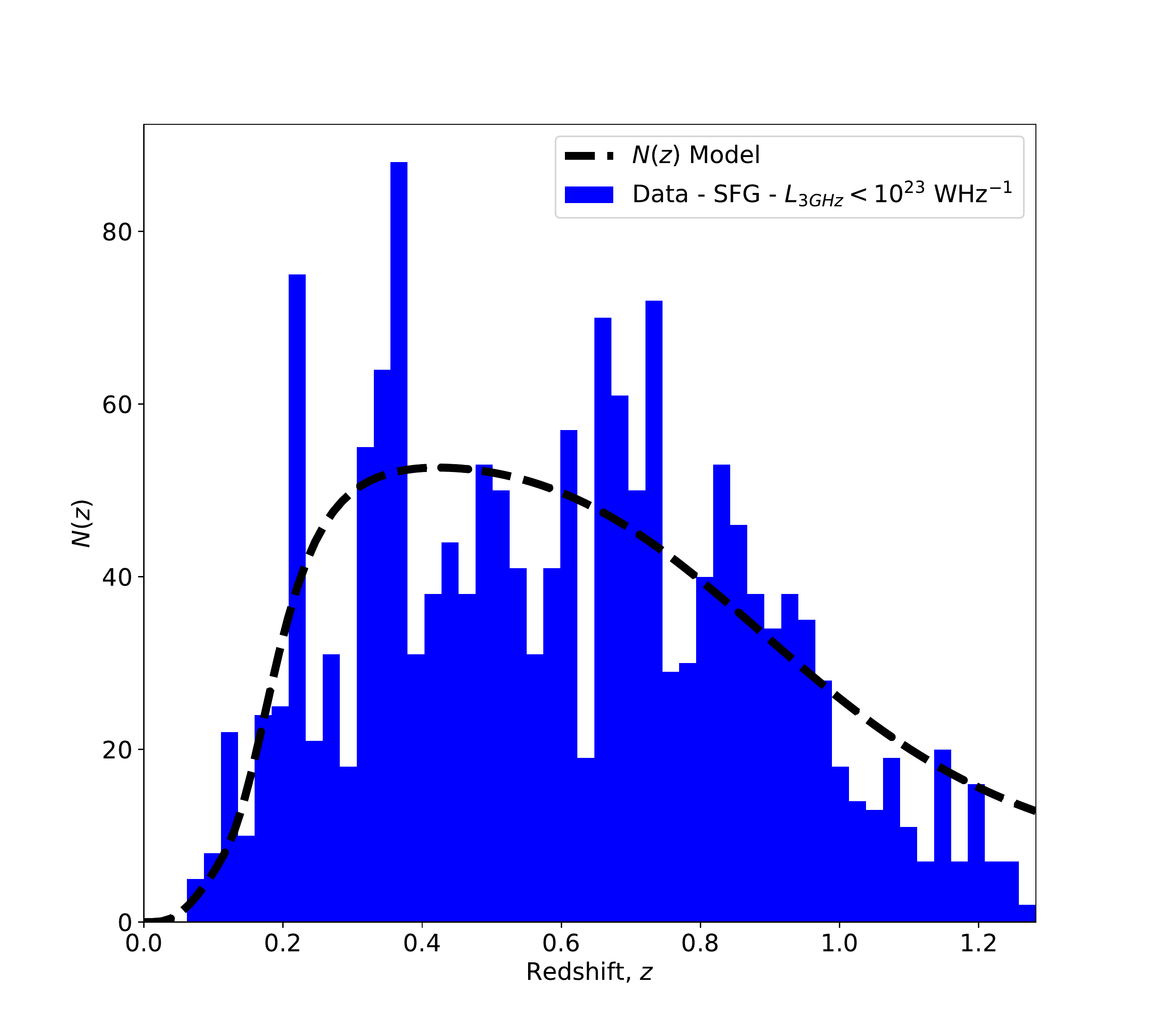}
\subcaption{(f)}
\end{minipage}%
\newline \newline 
\begin{minipage}[b]{.3\textwidth}
\centering
\includegraphics[height=5.2cm]{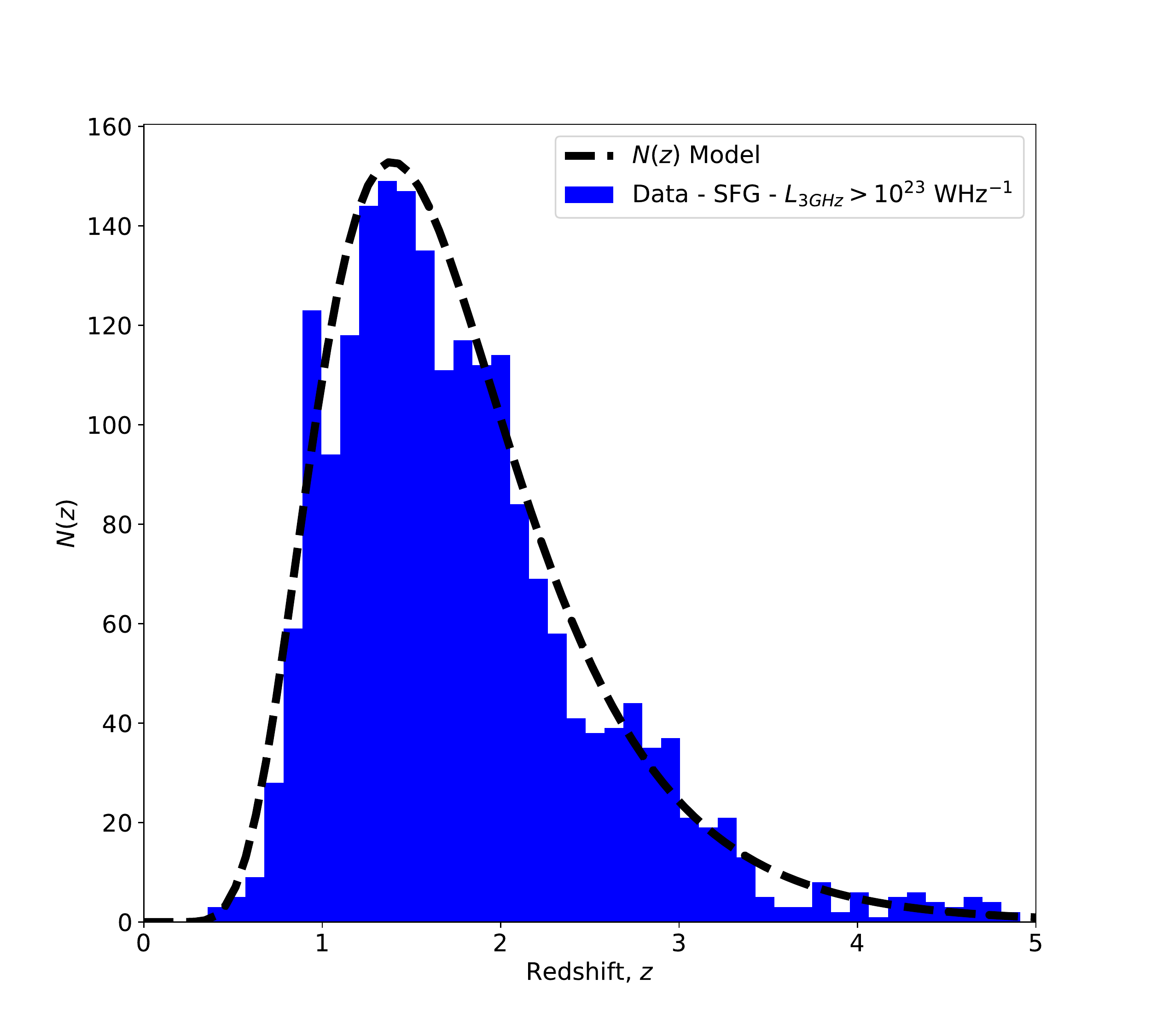}
\subcaption{(g)}
\end{minipage}%
\begin{minipage}[b]{.3\textwidth}
\centering
\includegraphics[height=5.2cm]{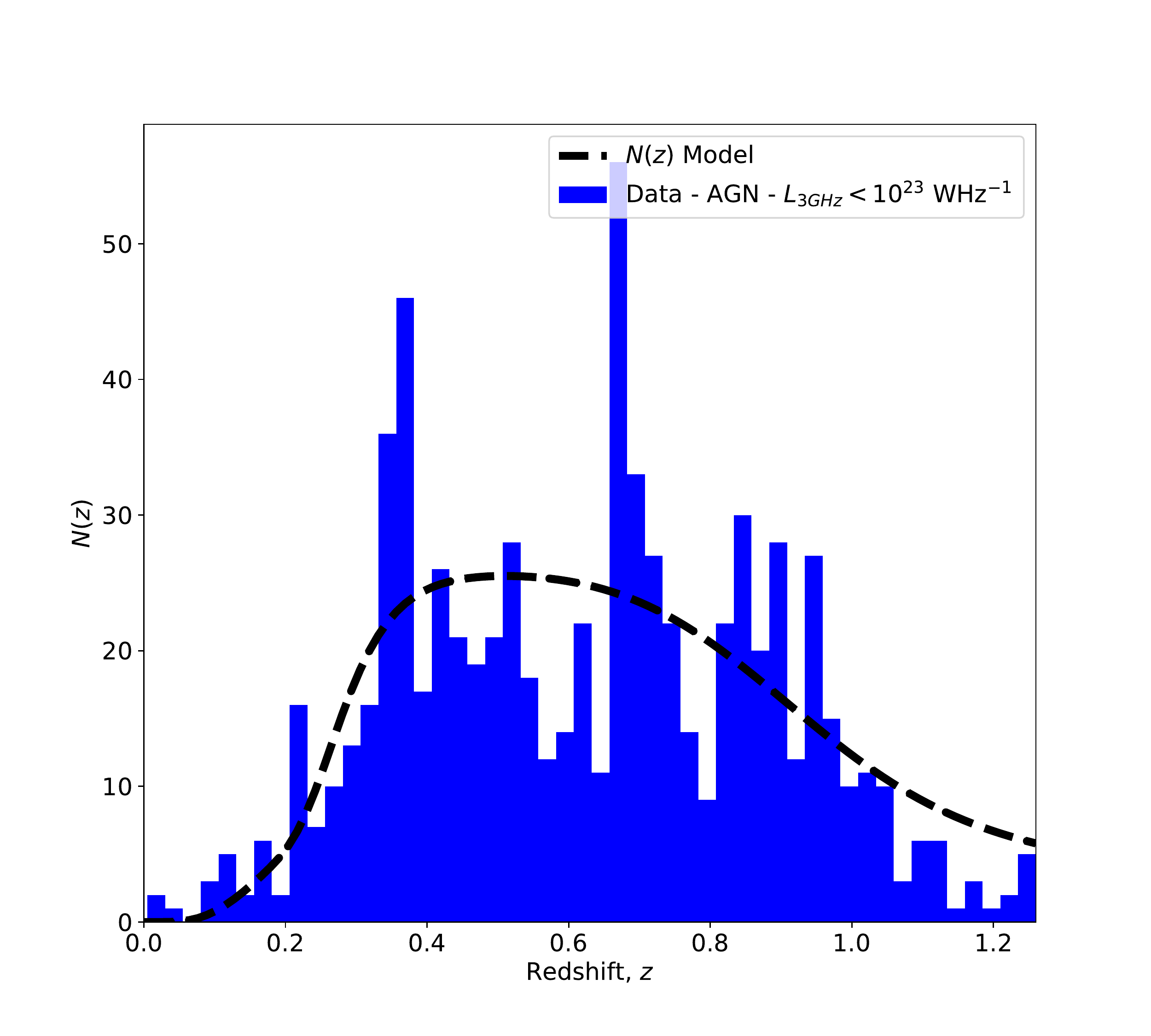}
\subcaption{(h)}
\end{minipage}%
\begin{minipage}[b]{.3\textwidth}
\centering
\includegraphics[height=5.2cm]{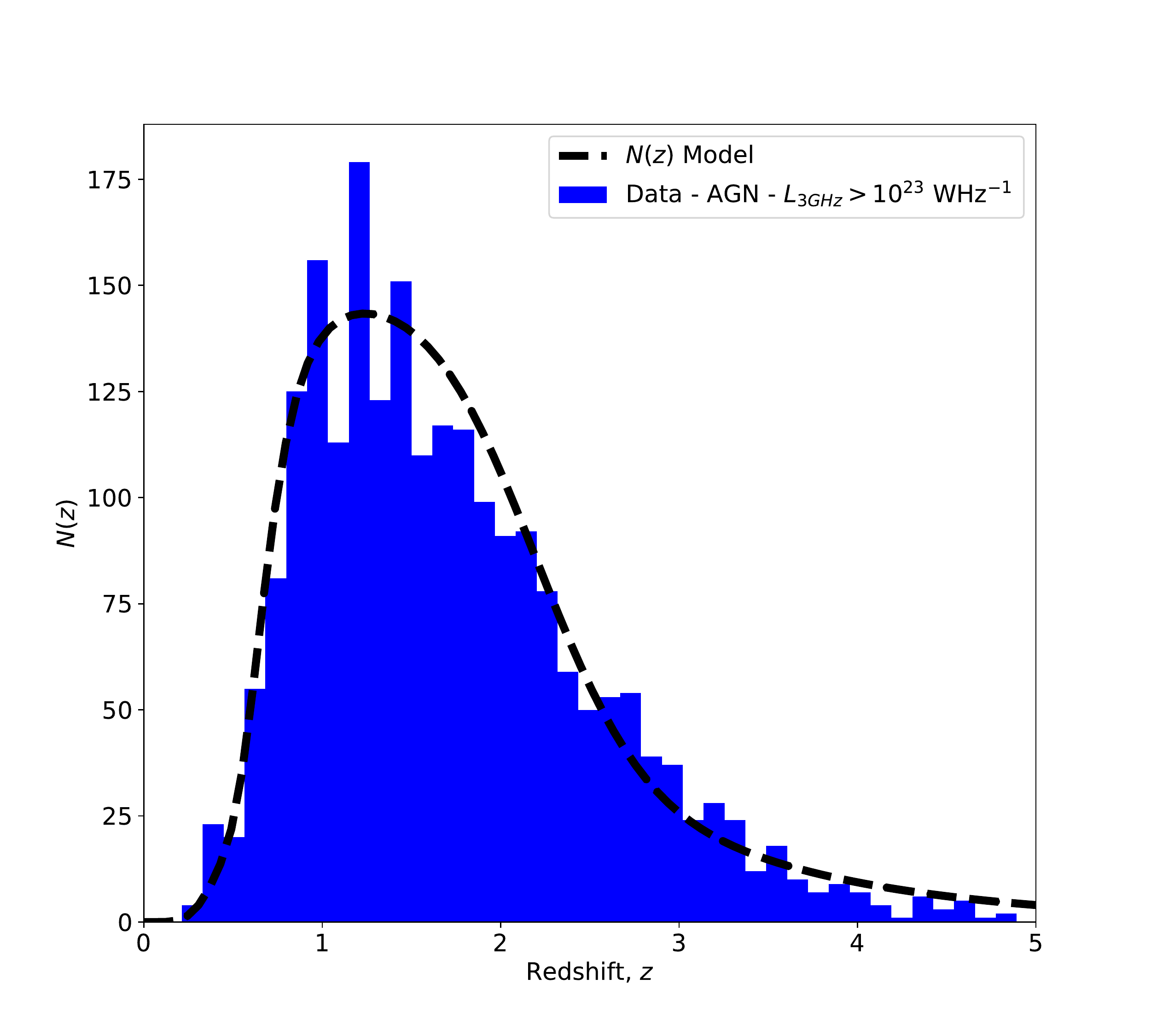}
\subcaption{(i)}
\end{minipage}%
\newline
\end{center}
\caption{The observed (histogram) and modelled (black line) redshift distribution ($N(z)$) of radio sources used in the clustering analysis for (a) All galaxies, (b) SFGs, (c) AGN, (d) HLAGN, (e) MLAGN, (f) low luminosity SFGs ($L_{\rm 3GHz} < 10^{23}$ WHz$^{-1}$), (g) high luminosity SFGs ($L_{\rm 3GHz} > 10^{23}$ WHz$^{-1}$), (h) low luminosity AGN ($L_{\rm 3GHz} < 10^{23}$ WHz$^{-1}$) and (i) high luminosity AGN ($L_{\rm 3GHz} > 10^{23}$ WHz$^{-1}$). For (a) the histogram corresponds to the redshift distribution of the associated random catalogue. In all other cases it is the histogram of the real, observed redshifts. $N(z)$ is defined here as the number of sources in the given redshift bin. }
\label{fig:redshiftdistributionall}
\end{figure*}

\subsection{Fitting and the Integral Constraint}
\label{sec:integral_constraint}

The angular TPCF, $\omega(\theta)$, quantifies the projected clustering of galaxies on the celestial sphere, whereas $\xi(r)$ quantifies the real space clustering of sources. If redshifts of sources were known accurately, $\xi(r)$ could be calculated for the galaxies directly, using their angular positions and redshifts. However, the synchrotron emission from radio sources is featureless, and redshifts can only be measured by cross-matching the radio data to ancillary data, usually at optical and near-infrared wavelengths. Moreover, as we probe fainter galaxies, photometric redshifts are used, as the time required to obtain spectroscopic redshifts for large samples of faint galaxies can be prohibitively long \citep[but see][]{Smith2016}. Photometric redshifts may not provide enough accuracy to directly calculate the spatial correlation function \citep[but see][]{Allevato2016}. Furthermore, not all sources detected in the radio will be detected in data at other wavelengths. The combination of these issues leads us to calculate the angular correlation function, $\omega(\theta)$\footnote{From hereon any mention of the TPCF refers to the angular TPCF, $\omega({\theta})$.}.

To fit the TPCF we follow similar analyses on deep radio survey fields \citep[e.g.][]{Lindsay2014, Magliocchetti} and assume a power law for the large scale correlation function. However, we note that \cite{BlakeWall2} highlight the addition of a second power law with a steeper negative slope with $\omega(\theta) \propto \theta^{-3.4}$ at small angles. This arises from multi-component sources that have not been identified as one source. We use a single power law distribution for the TPCF as the sources have already been matched with their optical IDs. This means that any multi-component sources should have already been associated as such. Furthermore, for this flux-density limit and area, the number of double sources is expected to be very low \citep[$N<10$; ][]{Wilman2008}. 

As the field has a finite size, the TPCF is underestimated at relatively large angular scales. This is due to a diminishing number of galaxy pairs at scales similar to that of the survey area. To account for this finite aperture size effect we incorporate an integral constraint when fitting $\omega(\theta)$ \citep[see][]{GrothPeebles}. The integral constraint, $\sigma^2$ is defined in Equation \ref{eq:integral_constraint}, following \cite{GrothPeebles}, where the model ($\omega_{model}$) is the fit to the data. From this we find the true TPCF ($\omega_{true}$) which is the TPCF that would be measured if we did not have a finite field size,

\begin{ceqn}
\begin{align}
\label{eq:integral_constraint}
\omega_{model}(\theta) = \omega_{true}(\theta) - \sigma^2 .
\end{align}
\end{ceqn}
The integral constraint can be approximated analytically, following \cite{RocheEales}:
\begin{ceqn}
\begin{align}
\label{eq:integral_constraint_approx}
\sigma^2 = \frac{\sum RR(\theta) \omega_{true}(\theta)}{\sum RR(\theta)}.
\end{align}
\end{ceqn}
As discussed earlier, a single power law distribution was assumed to be the underlying distribution: 
\begin{ceqn}
\begin{align}
\label{eq:w_pl}
\omega_{true}(\theta) = A \theta^{1-\gamma},
\end{align}
\end{ceqn}
where $A$ describes the amplitude of the TPCF, and 1-$\gamma$ the slope. The values are then found by minimizing $\chi^2$, defined by:
\begin{ceqn}
\begin{align}
\label{eq:chi}
\chi^2 =  \sum \frac{ \left[\omega_{obs} - (\omega_{true}(\theta) - \sigma^2)\right]^2}{\Delta \omega ^2} ,
\end{align}
\end{ceqn}
where $\Delta \omega$ is the bootstrap resampled error, as discussed in Section \ref{sec:calctpcf}.

We fit $\omega(\theta)$ over the range $\theta \sim 10^{-3} - 0.5 ^{\circ}$, in equal logarithmically-spaced bins.

\subsection{Spatial Correlation Function, $\xi(r)$, and Clustering Length, $r_0$}
\label{sec:r0}

We use Limber inversion \citep[see e.g.][]{Limber1953,Peebles1980,Overzier2003} to relate the projected angular clustering to the spatial clustering. The clustering length is related to the spatial correlation function through:
$\xi(r) = \left(\frac{r_0}{r}\right)^{\gamma}$, and $r_0$ is determined using:
\begin{ceqn}
\begin{align}
A={r_0}^{\gamma}H_{\gamma}\bigg(\frac{H_0}{c}\bigg) \frac{{\int_0}^{\infty} N^2(z)(1+z)^{\gamma - (3+\epsilon)}\chi^{1-\gamma}(z)E(z)dz}{\left[{\int_0}^{\infty} N(z)dz\right]^2},
\label{eq:limber_inv}
\end{align}
\end{ceqn}
where
\begin{ceqn}
\begin{align}
H_{\gamma} =\frac{\Gamma(\frac{1}{2})\Gamma(\frac{\gamma-1}{2})}{\Gamma(\frac{\gamma}{2})},
\label{eq:h-gamma}
\end{align}
\end{ceqn}
$N(z)$ is the redshift distribution of the sources and $\Gamma(x)$ is the gamma function and the comoving line-of-sight distance, $\chi(z)$ is given  by \citep{Hogg}:
\begin{ceqn}
\begin{align}
\chi(z) =\frac{c}{H_0} {\int_0}^z \frac{dz'}{E(z')} ,
\end{align}
\end{ceqn}
where 
\begin{ceqn}
\begin{align}
E(z)=[\Omega_{m,0}(1+z)^3 + \Omega_{k,0}(1+z)^2+\Omega_{\Lambda, 0}]^{\frac{1}{2}}.
\end{align}
\end{ceqn} 

In this work, values for $r_0$ as well as its associated uncertainties are found through sampling from the Probability Density Functions (P.D.F.s) in $A$, with $\epsilon = \gamma -3$ (comoving clustering). The median $r_0$ values are recorded in Table \ref{table:clustering}, with the associated uncertainties calculated from the 16$^{\textrm{th}}$ and 84$^{\textrm{th}}$ percentiles of the samples.

\subsection{Bias}

The bias relates how galaxies cluster spatially compared to the underlying dark matter, defined in Equation \ref{eq:bias} \citep[see e.g.][]{PeacockSmith2000}. As dark matter is assumed to be governed only by gravity, it is well understood through theory and simulations, therefore by tracing galaxies we can investigate how clustered these are compared to the underlying dark matter. 

\begin{ceqn}
\begin{align}
b^2(z)=\frac{\xi_{gal}(r,z)}{\xi_{DM}(r,z)}
\label{eq:bias}
\end{align}
\end{ceqn}
We determine the bias using:

\begin{ceqn}
\begin{align}
b(z)=\left( \frac{r_0(z)}{8} \right)^{\gamma/2} \frac{J_2^{1/2}}{\sigma_8 D(z)/D(0)}.
\end{align}
\end{ceqn}
\citep[e.g.][]{Lindsay2014a}. Here $D(z)$ is the growth factor at a given redshift \citep[][]{CPT1992}, calculated using the formula in \cite{Hamilton2001} and  $J_2 = 72/[(3-\gamma)(4-\gamma)(6-\gamma)2^{\gamma}]$. 

For this work, the bias is always evaluated at the median redshift of the sample. Again we use the median, 16$^{\textrm{th}}$ and 84$^{\textrm{th}}$ percentiles from the distribution of $r_0$ (see above) to quantify the bias and its uncertainty.

\subsection{Redshift distributions}
\label{sec:redshifts}
To find $r_0$, the redshift distribution for the radio sources, $N(z)$, is required. When the clustering of all sources (no redshift or galaxy type cuts) was investigated the redshift distribution of the random sample, generated using $S^3$ was used, as we do not have redshifts for all sources. This should mimic the underlying redshift distribution of the data, as many studies have shown that the source counts and redshift distribution from $S^3$ describe the overall properties of the radio source population very well \citep[see e.g.][]{mcalpine2013}, although we note there is some evidence $S^3$ may under-predict the source counts at the faintest fluxes \citep{Smolcic2017}. When the clustering of the different sub-samples of radio sources are investigated separately, we use the actual redshift distribution of these sources\footnote{We note that as a check we found very little difference in our inferred values for $r_0$ and $b(z)$ using the actual and $S^3$ redshift distributions when we considered the clustering of the sub-populations over the whole redshift range.}.
In all cases the redshift distributions were modelled by a smooth functional form, shown Figure~\ref{fig:redshiftdistributionall}. For the low and high redshift samples we used the same redshift distribution, but restricted to the relevant redshift range.

\section{Results}
\label{sec:tpcf_results}

In this section we present the analysis of the clustering of all sources in the field. We then investigate the clustering as a function of radio source type and redshift.
When considering the different source populations, the low density of sources leads to large uncertainties. We therefore use a fixed slope for the power law of the TPCF, of the form $\omega(\theta) \propto \theta^{-0.8}$. 

\begin{figure*}
\begin{center}
\begin{minipage}[b]{.5\textwidth}
\centering
\includegraphics[height=7.cm]{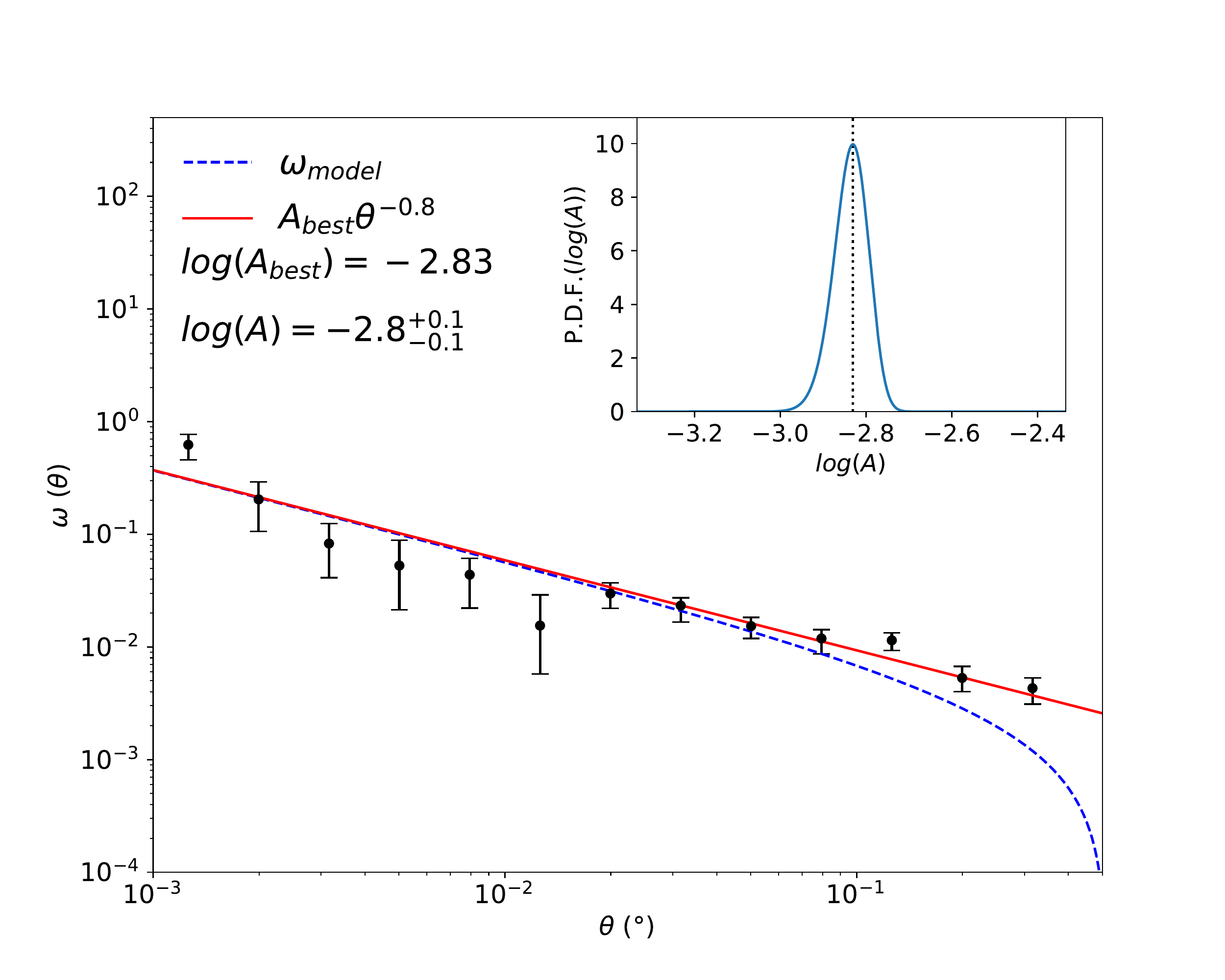}
\end{minipage}%
\begin{minipage}[b]{.5\textwidth}
\centering
\includegraphics[height=7.cm]{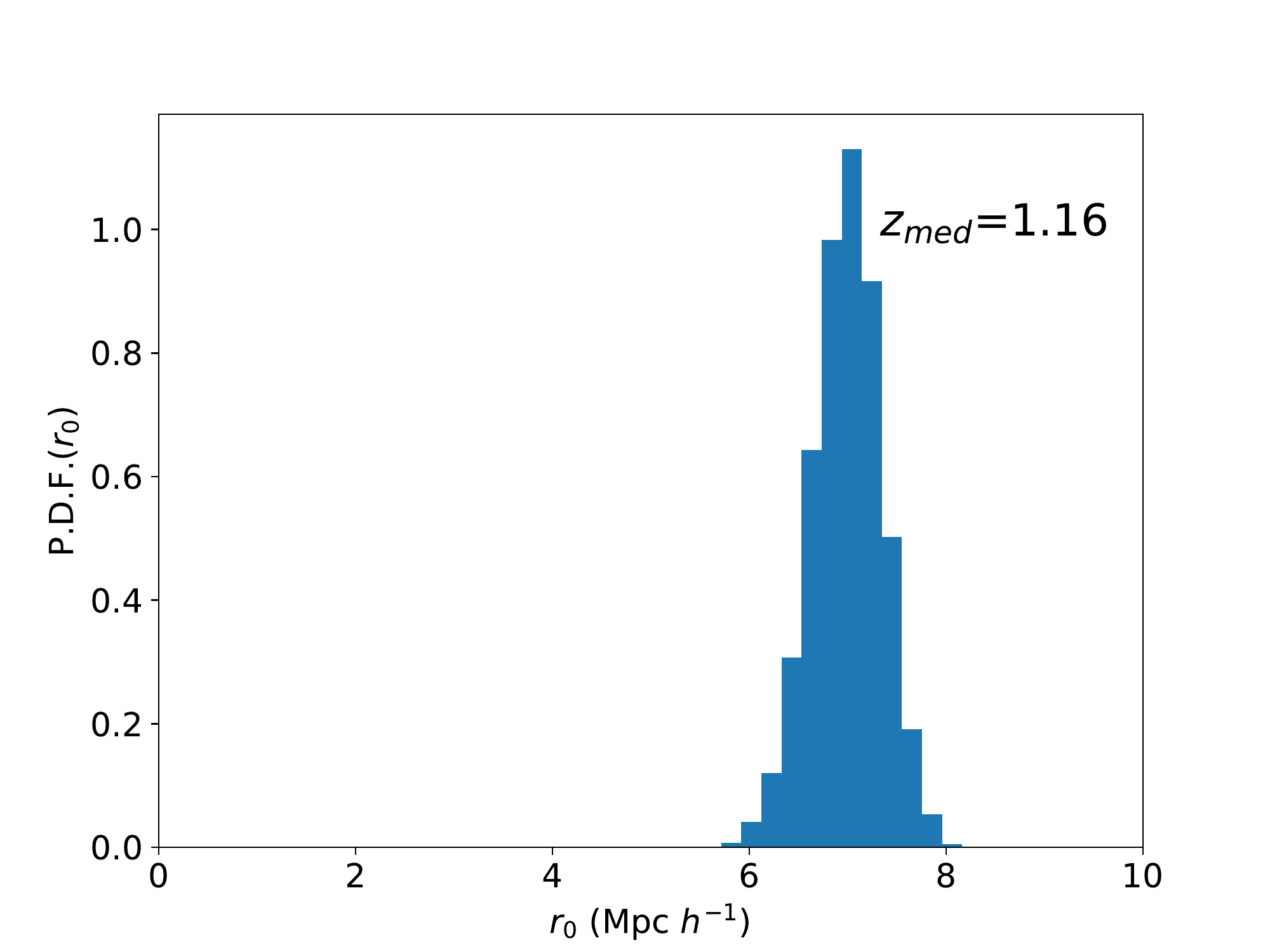}
\end{minipage}%
\end{center}
\caption{TPCF and fit for all COSMOS radio sources. The left panel shows the TPCF, the red solid line represents the best fit power law to $\omega(\theta)$ and the dashed blue line is the same as the red line but with the integral constraint, $\sigma^2$ subtracted from it. The inset in the top right corner shows the probability density function (P.D.F.) for the value of $A$ to fit the TPCF, with the dashed line showing the best fit value. The right panel shows the corresponding P.D.F. for $r_0$.}
\label{fig:clusteringall}
\end{figure*}

\begin{figure*}
\begin{center}
\begin{minipage}[b]{.5\textwidth}
\centering
\includegraphics[height=7.cm]{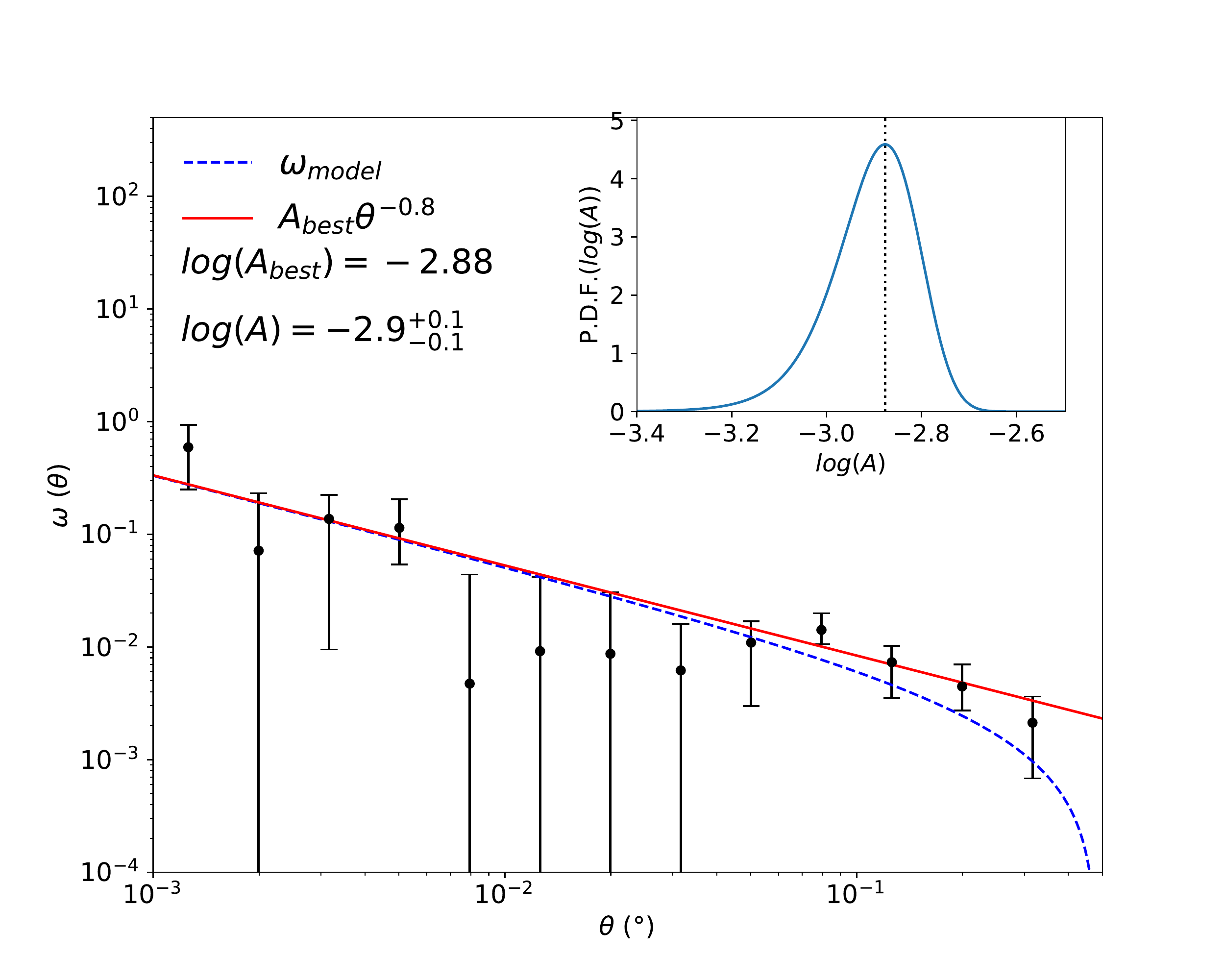}
\end{minipage}%
\begin{minipage}[b]{.5\textwidth}
\centering
\includegraphics[height=7.cm]{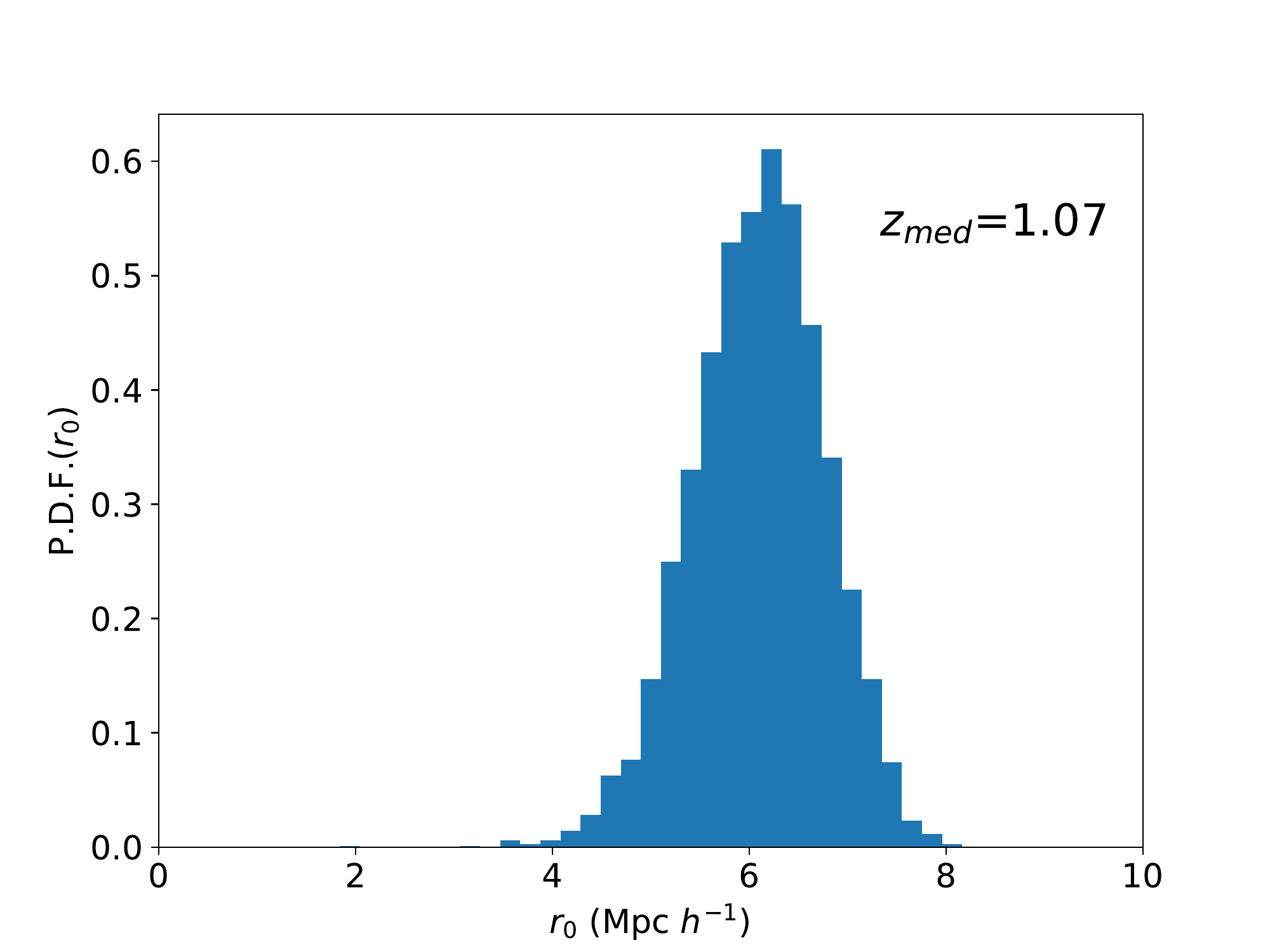}
\end{minipage}%
\newline
\begin{minipage}[b]{.5\textwidth}
\centering
\includegraphics[height=7.cm]{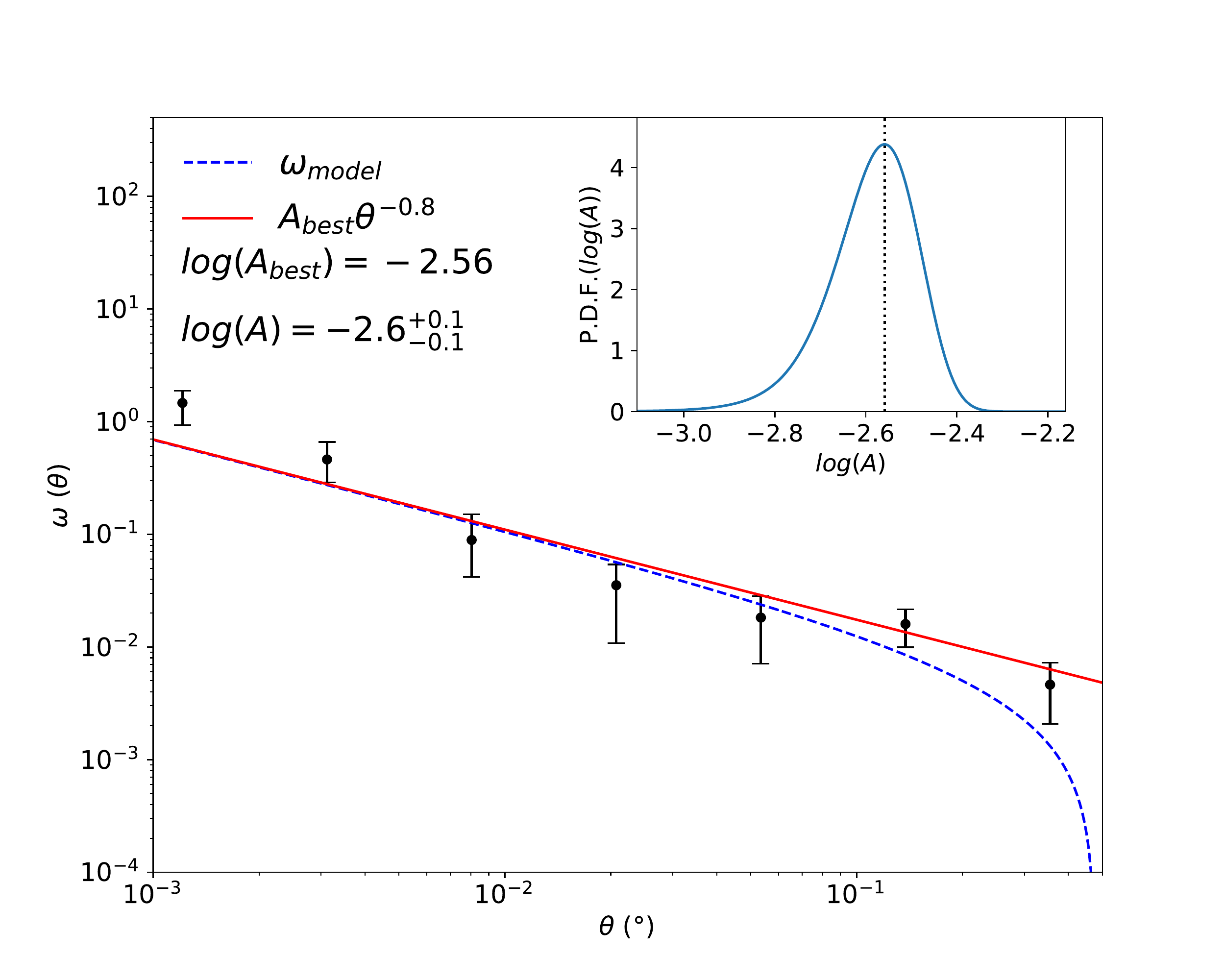}
\end{minipage}%
\begin{minipage}[b]{.5\textwidth}
\centering
\includegraphics[height=7.cm]{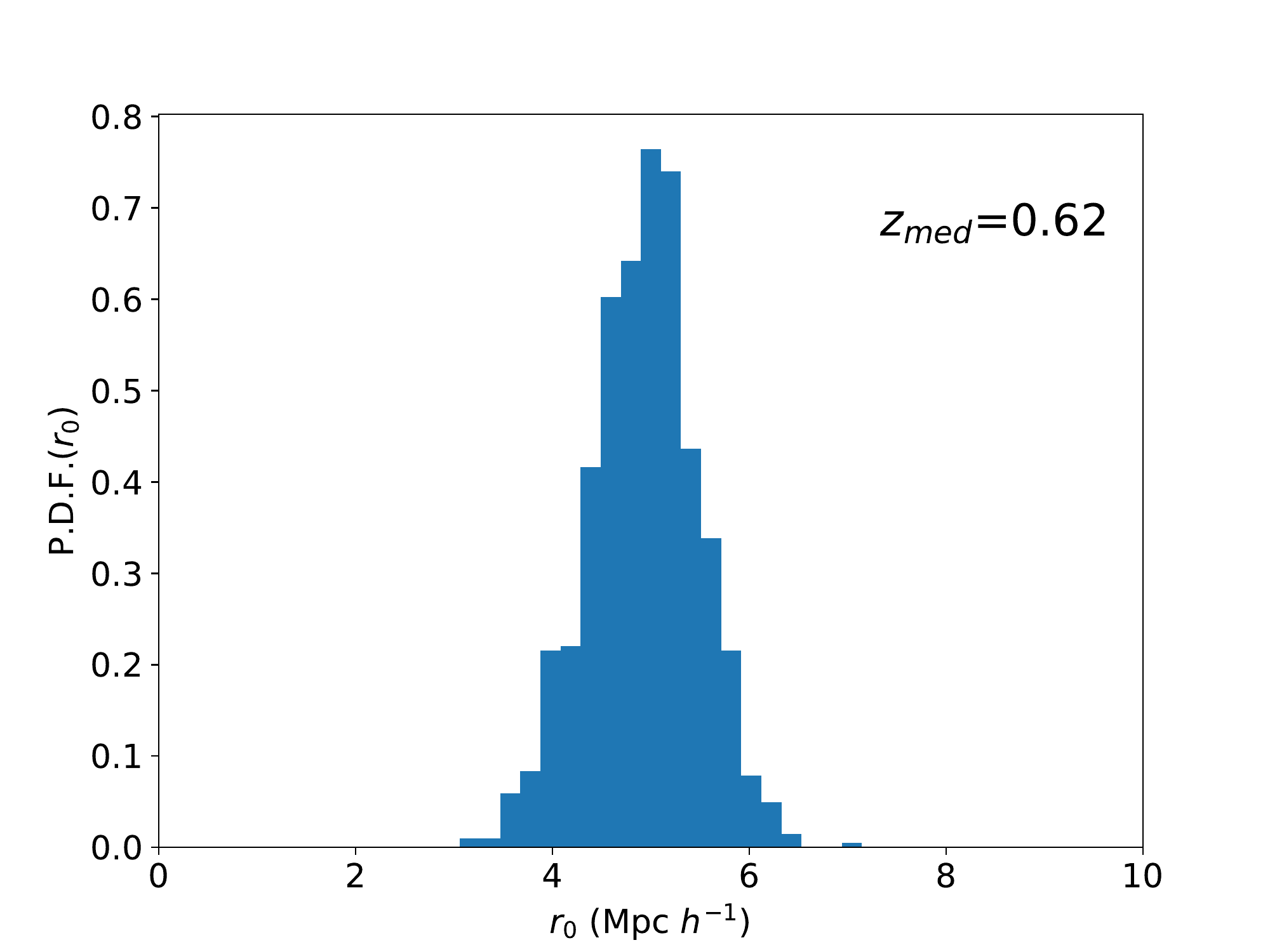}
\end{minipage}%
\end{center}
\caption{TPCF and fit for All SFGs (top panel), SFGs with $z<1$ (bottom panel). The left panel shows the TPCF, the red solid line represents the power law best fit to $\omega(\theta)$ and the dashed blue line is the same as the red line but with the integral constraint, $\sigma^2$ subtracted from it. The inset in the top right corner shows the probability density function for the value of $A$ to fit the TPCF, with the dashed line showing the best fit value. The right panel shows the corresponding P.D.F. for $r_0$. }
\label{fig:clusteringSFG}
\end{figure*}

Our results for the TPCF and the best fits can be seen in the left hand panels of Figures~\ref{fig:clusteringall}--\ref{fig:clusteringMLAGN}, and the corresponding right hand panels show the associated P.D.F. for $r_0$.
The relevant parameters for the clustering analyses are presented in Table \ref{table:clustering}. This includes the number of galaxies used in each analysis and their median redshift and luminosity. For each investigation the best-fit value of $A$ as well as the associated median and uncertainties are presented\footnote{We allowed log$_{10}(A$) to vary in the range [-6,0].} and the values of $r_0$ and $b(z)$ are also given.

%%%%%%% SFG FIGURES %%%%%%%%%

%%%%%%%%%%%%%%%%%%%%%%%%%%%%%

%%%%%%%%%%% Table %%%%%%%%%%%%%%%

\begin{table*} 
\small
\begin{centering}
\begin{tabular}{l c c c c c c c c c c c c c c} 

Galaxy  &redshift& N  &  Median redshift&Median Luminosity&$	\textrm{log}_{10}(A)$&$	\textrm{log}_{10}(A)$ &Best $r_0$&Median $r_0$&$b(z_{med})$   \\  
Type&&&($z_{med}$)&WHz$^{-1}$&Best&Median&  (Mpc $h^{-1}$)&  (Mpc $h^{-1}$)& \\  \hline \hline
All & All& 8887 & 1.16 & - &-2.83&$ -2.8 ^{+ 0.1}_{- 0.1 }$&$ 7.0 $&$ 7.0 ^{+ 0.3 }_{- 0.4 }$ &$ 2.7 ^{+ 0.1 }_{- 0.1 }$\\ \\  \hline 
SFG & All& 3704 & 1.07 &1.28$\times 10^{23}$&-2.88&$ -2.9 ^{+ 0.1}_{- 0.1 }$&$ 6.1 $&$ 6.1 ^{+ 0.6 }_{- 0.7 }$ &$ 2.3 ^{+ 0.2 }_{- 0.2 }$\\ \\  \hline 
AGN & All& 2937 & 1.24 &3.22$\times 10^{23}$&-2.57&$ -2.6 ^{+ 0.1}_{- 0.1 }$&$ 9.6 $&$ 9.6 ^{+ 0.7 }_{- 0.7 }$ &$ 3.6 ^{+ 0.2 }_{- 0.2 }$\\ \\  \hline 
HLAGN & All& 1456 & 1.35 &3.48$\times 10^{23}$&-2.59&$ -2.6 ^{+ 0.1}_{- 0.1 }$&$ 9.8 $&$ 9.8 ^{+ 1.3 }_{- 1.4 }$ &$ 3.9 ^{+ 0.5 }_{- 0.5 }$\\ \\  \hline 
MLAGN & All& 726 & 0.88 &1.28$\times 10^{23}$&-2.33&$ -2.4 ^{+ 0.1}_{- 0.2 }$&$ 9.5 $&$ 9.5 ^{+ 1.5 }_{- 1.8 }$ &$ 3.1 ^{+ 0.4 }_{- 0.5 }$\\ \\  \hline 
SFG & All& 1654 & 0.60 &3.55$\times 10^{22}$&-2.64&$ -2.7 ^{+ 0.1}_{- 0.1 }$&$ 4.9 $&$ 4.9 ^{+ 0.6 }_{- 0.7 }$ &$ 1.5 ^{+ 0.2 }_{- 0.2 }$\\ 
\tiny ($L_{\rm 3GHz} < 10^{23}$ WHz$^{-1}$) &&&&&&& \\ \hline 
SFG & All& 2050 & 1.64 &3.28$\times 10^{23}$&-3.22&$ -3.7 ^{+ 0.6}_{- 1.5 }$&$ 3.8 $&$ 4.0 ^{+ 1.3 }_{- 1.5 }$ &$ 1.9 ^{+ 0.6 }_{- 0.7 }$\\
\tiny ($L_{\rm3GHz} > 10^{23}$ WHz$^{-1}$) &&&&&&& \\ \hline 
AGN & All& 747 & 0.66 &4.35$\times 10^{22}$&-2.62&$ -3.1 ^{+ 0.5}_{- 1.8 }$&$ 4.9 $&$ 5.0 ^{+ 1.2 }_{- 1.5 }$ &$ 1.6 ^{+ 0.3 }_{- 0.4 }$\\   
\tiny ($L_{\rm3GHz} < 10^{23}$ WHz$^{-1}$) &&&&&&& \\ \hline 
AGN & All& 2190 & 1.56 &5.14$\times 10^{23}$&-2.54&$ -2.5 ^{+ 0.1}_{- 0.1 }$&$ 9.9 $&$ 9.9 ^{+ 0.9 }_{- 0.9 }$ &$ 4.2 ^{+ 0.3 }_{- 0.3 }$\\   
\tiny ($L_{\rm3GHz} > 10^{23}$ WHz$^{-1}$) &&&&&&& \\  \hdashline  \\
SFG & z<1& 1756 & 0.62 &3.81$\times 10^{22}$ &-2.56&$ -2.6 ^{+ 0.1}_{- 0.1 }$&$ 5.0 $&$ 5.0 ^{+ 0.5 }_{- 0.6 }$ &$ 1.5 ^{+ 0.1 }_{- 0.2 }$\\ \\  \hline 
AGN & z<1& 1126 & 0.70 &6.85$\times 10^{22}$ &-2.25&$ -2.3 ^{+ 0.1}_{- 0.1 }$&$ 6.9 $&$ 6.9 ^{+ 0.6 }_{- 0.7 }$ &$ 2.1 ^{+ 0.2 }_{- 0.2 }$\\ \\  \hline 
HLAGN & z<1& 488 & 0.69 &6.38$\times 10^{22}$ &-2.40&$ -2.9 ^{+ 0.5}_{- 1.9 }$&$ 5.7 $&$ 5.8 ^{+ 1.4 }_{- 1.8 }$ &$ 1.8 ^{+ 0.4 }_{- 0.5 }$\\ \\  \hline 
MLAGN & z<1& 485 & 0.70 &6.82$\times 10^{22}$ &-2.00&$ -2.0 ^{+ 0.1}_{- 0.1 }$&$ 9.7 $&$ 9.7 ^{+ 1.2 }_{- 1.3 }$ &$ 2.9 ^{+ 0.3 }_{- 0.3 }$\\ \\  \hline 
SFG & z<1& 1520 & 0.57 &3.17$\times 10^{22}$ &-2.56&$ -2.6 ^{+ 0.1}_{- 0.2 }$&$ 4.8 $&$ 4.8 ^{+ 0.6 }_{- 0.7 }$ &$ 1.4 ^{+ 0.2 }_{- 0.2 }$\\ 
\tiny ($L_{\rm3GHz} < 10^{23}$ WHz$^{-1}$) &&&&&&& \\ \hline 
SFG & z<1& 236 & 0.91 &1.42$\times 10^{23}$ &-2.13&$ -3.1 ^{+ 1.0}_{- 1.9 }$&$ 5.9 $&$ 5.9 ^{+ 1.8 }_{- 2.1 }$ &$ 2.0 ^{+ 0.5 }_{- 0.7 }$\\
\tiny ($L_{\rm3GHz} > 10^{23}$ WHz$^{-1}$) &&&&&&& \\ \hline 
AGN & z<1& 694 & 0.61 &4.02$\times 10^{22}$ &-2.39&$ -2.5 ^{+ 0.2}_{- 0.6 }$&$ 5.9 $&$ 5.9 ^{+ 1.0 }_{- 1.2 }$ &$ 1.8 ^{+ 0.3 }_{- 0.3 }$ \\
\tiny ($L_{\rm3GHz} < 10^{23}$ WHz$^{-1}$) &&&&&&& \\ \hline 
AGN & z<1& 432 & 0.84 &2.61$\times 10^{23}$ &-1.90&$ -1.9 ^{+ 0.1}_{- 0.1 }$&$ 9.0 $&$ 8.9 ^{+ 0.9 }_{- 1.2 }$ &$ 2.9 ^{+ 0.3 }_{- 0.3 }$\\
\tiny ($L_{\rm3GHz} > 10^{23}$ WHz$^{-1}$) &&&&&&& \\ \hdashline \\
AGN & z $\ge$ 1& 1811 & 1.77 &5.63$\times 10^{23}$&-2.58&$ -2.6 ^{+ 0.1}_{- 0.1 }$&$ 7.3 $&$ 7.3 ^{+ 0.9 }_{- 0.9 }$ &$ 3.5 ^{+ 0.4 }_{- 0.4 }$\\ \\  \hline 
HLAGN & z $\ge$ 1& 968 & 1.85 &5.60$\times 10^{23}$&-2.68&$ -3.1 ^{+ 0.4}_{- 1.7 }$&$ 6.5 $&$ 6.5 ^{+ 1.5 }_{- 1.9 }$ &$ 3.2 ^{+ 0.7 }_{- 0.9 }$\\ \\  \hline 
MLAGN & z $\ge$ 1& 241 & 1.31 &3.30$\times 10^{23}$&-2.02&$ -2.5 ^{+ 0.5}_{- 2.1 }$&$ 11.3 $&$ 11.3 ^{+ 2.5 }_{- 3.0 }$ &$ 4.3 ^{+ 0.9 }_{- 1.1 }$\\ \\  \hline 
AGN & z $\ge$ 1& 1758 & 1.79 &5.80$\times 10^{23}$&-2.57&$ -2.6 ^{+ 0.1}_{- 0.1 }$&$ 8.4 $&$ 8.4 ^{+ 0.8 }_{- 0.9 }$ &$ 4.0 ^{+ 0.3 }_{- 0.4 }$\\ 
\tiny ($L_{\rm3GHz} > 10^{23}$ WHz$^{-1}$) &&&&&&& \\  \hline \\

\end{tabular}
\caption{Parameters describing the fits to the TPCF analyses: galaxy type; redshift range; number of sources (N); median redshift ($z_{med}$); median luminosity; best fit $A$; median $A$; best fit $r_0$, median $r_0$; best fit bias ($b$) and median $b$. The values of the bias have been evaluated at the median redshift of the sample being investigated. }
\label{table:clustering}
\end{centering}
\end{table*}

%%%%%%%%%%%%%%%%%%%%%%%%%%%

\subsection{Clustering of All Radio Sources}
\label{sec:clusall}

The TPCF for all radio sources in the COSMOS field is shown in Figure \ref{fig:clusteringall}. We find an amplitude of the TPCF of  $\log_{10}(A)=-2.8 \pm 0.1$. Using the redshift distribution shown in Figure~\ref{fig:redshiftdistributionall}(a), this corresponds to $r_0= 7.0 ^{+ 0.3 }_{- 0.4 }$\,Mpc\,$h^{-1}$ ($b=2.7 \pm 0.1$, $z_{med} = 1.16$). This measurement of the clustering is difficult to interpret given the mix of SFGs and AGN in the sample, but is given for completeness and is useful in making comparisons with previous clustering measurements in the literature.

\subsection{Clustering by Type}
\label{sec:clustype}

We know from previous work that at these faint fluxes, the radio source counts are dominated by SFGs, although there is still a non-negligible contribution from AGN, and possibly importantly from those AGN traditionally classified as radio-quiet \citep[e.g.][]{JarvisRawlings2004,White2015, White2017}. By using the excellent ancillary data coverage in the COSMOS field we are able to determine the clustering properties of both SFGs and AGN. We are also able to consider how the accretion efficiency of AGN affects their clustering by considering HLAGN and MLAGN separately.

We also investigate how the clustering of these sources varies with redshift, allowing us to determine how the bias and clustering length evolve. This also provides a natural way to compare the clustering of different source populations at the same cosmic epoch. Finally, we also investigate how the clustering of the two populations (SFGs and AGN) depends on their intrinsic luminosities.

\subsubsection{Clustering of SFGs}
\label{sec:clusSFG}

The TPCF for all SFGs is shown in Figure~\ref{fig:clusteringSFG} (top panel). 
When SFGs across all redshifts are considered, we find $\log_{10}A = -2.9\pm 0.1$, which using the redshift distribution shown in Figure~\ref{fig:redshiftdistributionall}(b) gives $r_0 = 6.1 ^{+ 0.6 }_{- 0.7 }$\,Mpc\,$h^{-1}$ ($b(z)=2.3 \pm 0.2$, $z_{med} \sim 1.1$). However, this relatively high value for the bias of SFGs is strongly influenced by the long tail of sources to relatively high redshift, when compared to SFGs selected at other wavelengths \citep[e.g.][]{Gilli2007, Magliocchetti2013}. We therefore re-evaluate the clustering of SFGs by restricting the redshift range to $z<1$. This mitigates any possible mis-identification of SFGs and AGN (mis-identifications are more likely at the higher redshifts as they are predominantly fainter across all wavebands) whilst also providing a benchmark at a moderate redshift that enables comparison with work on the clustering of SFGs at other wavelengths.

The TPCF for SFGs at $z<1$ is shown in the bottom panel of Figure~\ref{fig:clusteringSFG}. We find a lower value for the clustering length, with  $r_0 = 5.0 ^{+ 0.5 }_{- 0.6 }$\,Mpc\,$h^{-1}$ ($b = 1.5 ^{+ 0.1 }_{- 0.2 }$, $z_{med} = 0.62$). These clustering measurements are similar to those found using surveys covering similar redshifts at other wavelengths \citep{Gilli2007, Starikova2012, Magliocchetti2013} and in the radio \citep{Magliocchetti} and suggests that SFGs are not strongly biased tracers of the dark matter density field at these moderate redshifts. 

We are also able to split the SFG sample by their intrinsic luminosity at 3\,GHz, to determine whether the level of clustering is dependent on the SFR. 
We divide the SFG into two sub-samples at $L_{\rm 3GHz} = 10^{23}$~WHz$^{-1}$, which corresponds to a SFR $\sim 100$\,M$_{\odot}$yr$^{-1}$, using the relation from \cite{Bell2003}. When considering the high- and low-luminosity SFGs, very different ranges in redshift are probed by the two sub-populations, with the low luminosity sources lying predominantly at lower redshifts ($z_{med} = 0.60$) compared to the higher luminosity SFGs ($z_{med} = 1.64$).We restrict both sub-samples to $z<1$ to make more comparable clustering measurements. At $z<1$, the low luminosity SFGs have $r_0 = 4.8^{+ 0.6 }_{- 0.7 }$\,Mpc\,$h^{-1}$ ($b =1.4 \pm 0.2$), whereas the high luminosity SFGs have $r_0 = 5.9^{+ 1.8 }_{- 2.1 }$\,Mpc\,$h^{-1}$ ($b = 2.0^{+ 0.5 }_{- 0.7 }$). The high $L_{\rm 3GHz}$ SFGs are again at the higher median redshift of $z=0.91$, compared to the median redshift of the low luminosity SFGs of $z=0.57$.  However, regardless of the difference in median redshift, this suggests that the clustering of the radio-selected SFGs in COSMOS sample is not strongly dependent on the SFR. 

When fitting the clustering of SFGs at high redshift we found the fitting relatively ill-constrained and so defer the investigation to future deep and wide surveys \citep[see e.g.][]{Jarvis2017}, which will better be able to investigate the bias of SFGs at high $z$.

\subsubsection{Clustering of AGN}
\label{sec:clusAGN}

\begin{figure*}
\begin{center}
\begin{minipage}[b]{.5\textwidth}
\centering
\includegraphics[height=7.cm]{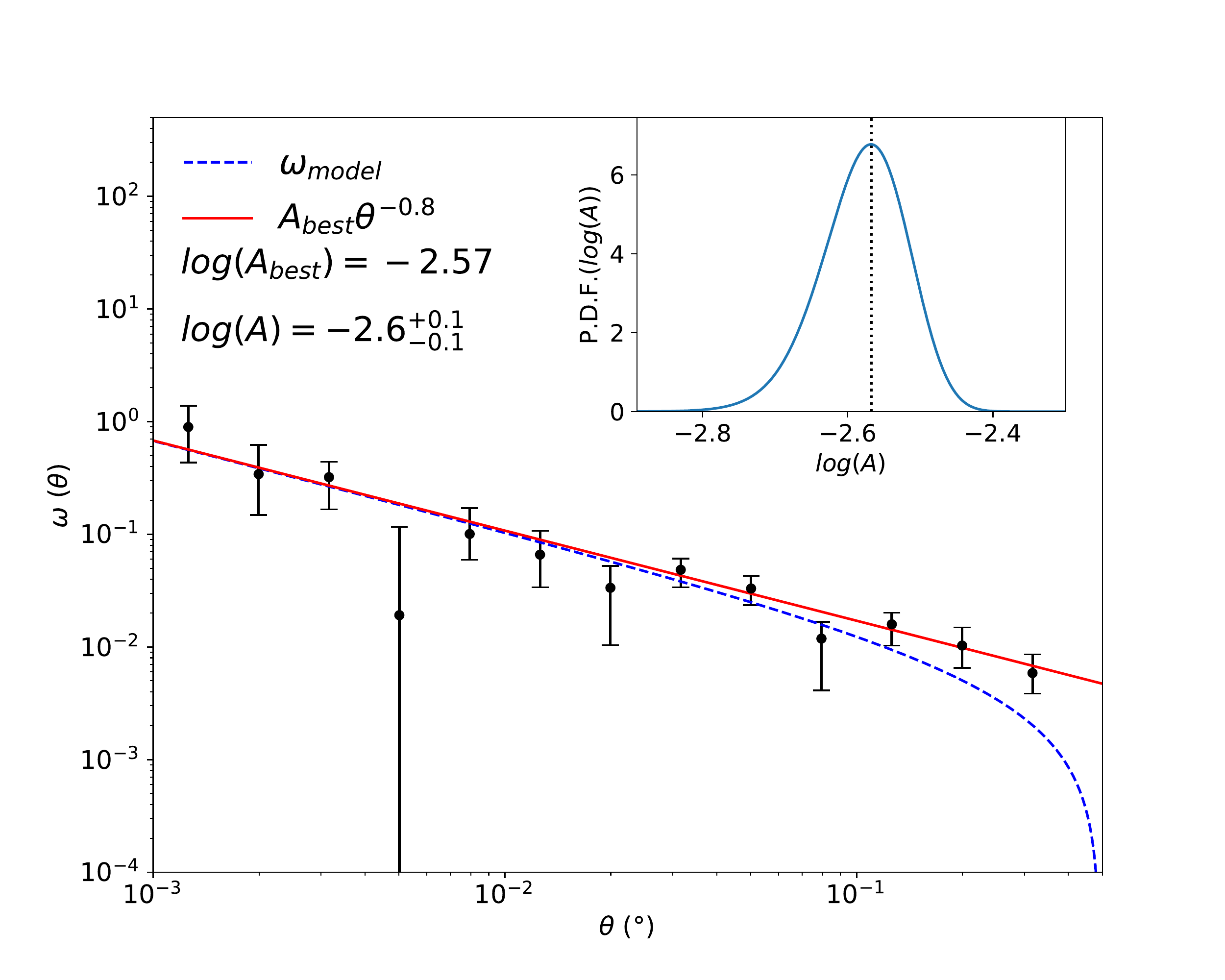}
\end{minipage}%
\begin{minipage}[b]{.5\textwidth}
\centering
\includegraphics[height=7.cm]{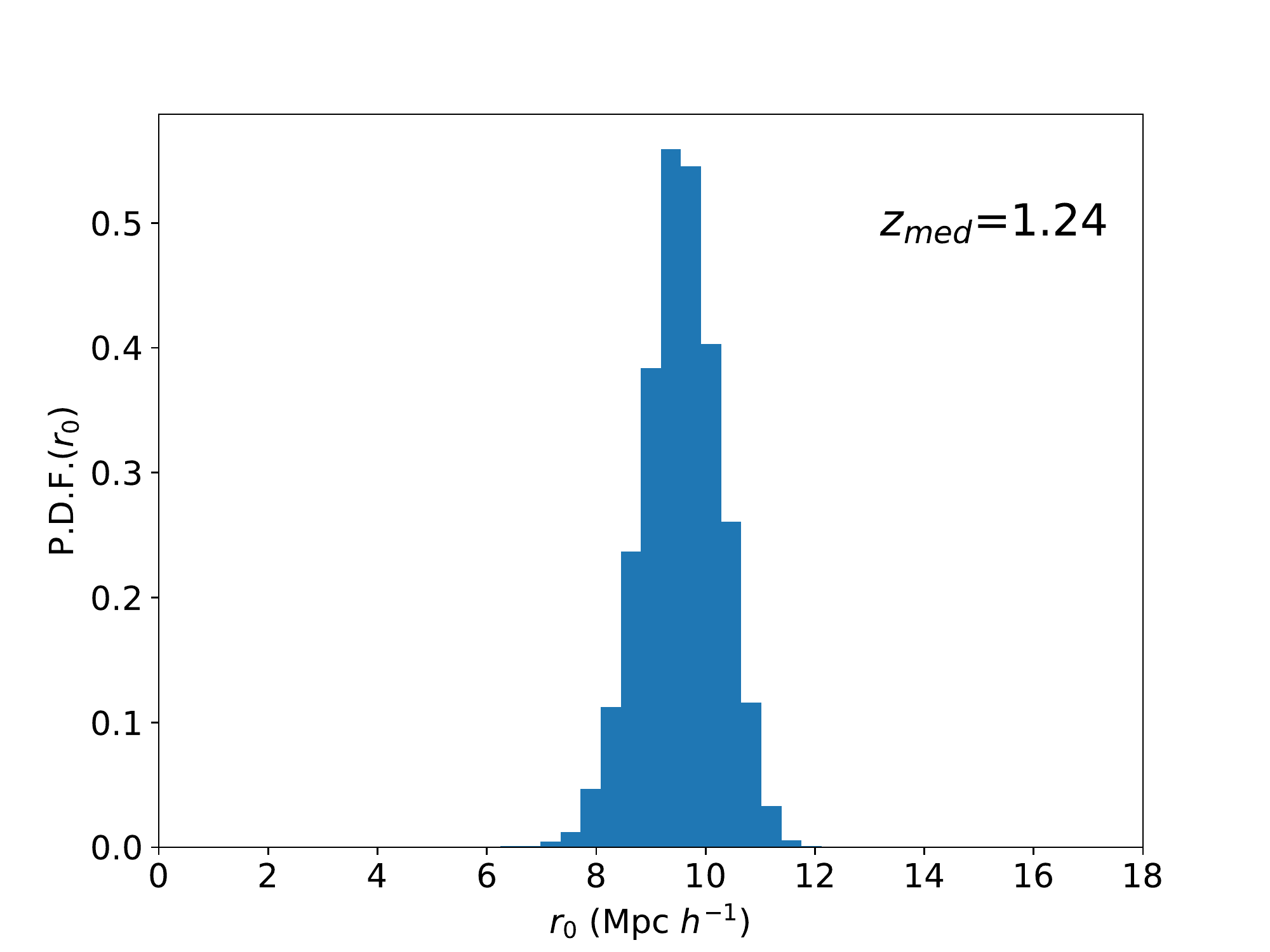}
\end{minipage}%
\newline
\begin{minipage}[b]{.5\textwidth}
\centering
\includegraphics[height=7.cm]{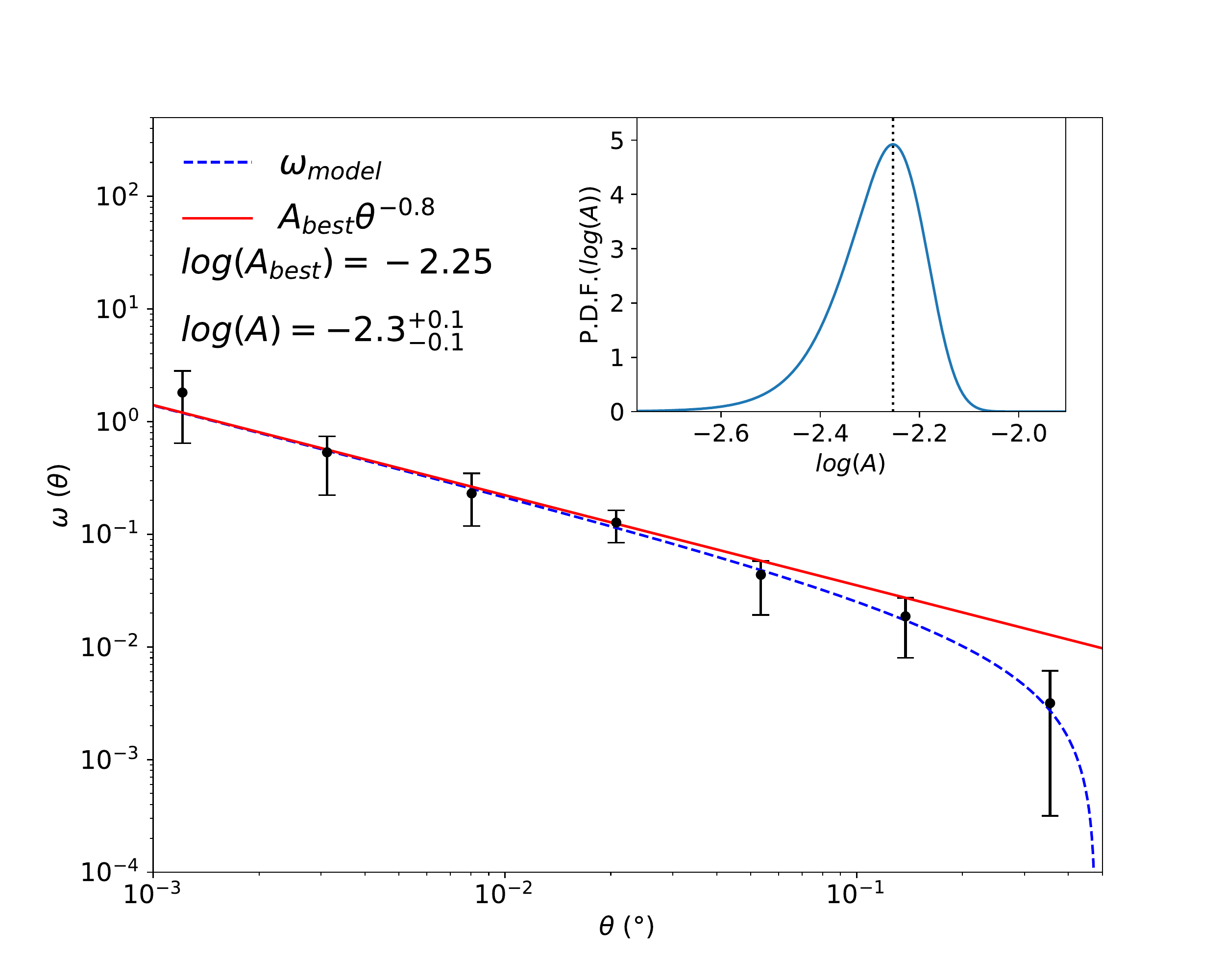}
\end{minipage}%
\begin{minipage}[b]{.5\textwidth}
\centering
\includegraphics[height=7.cm]{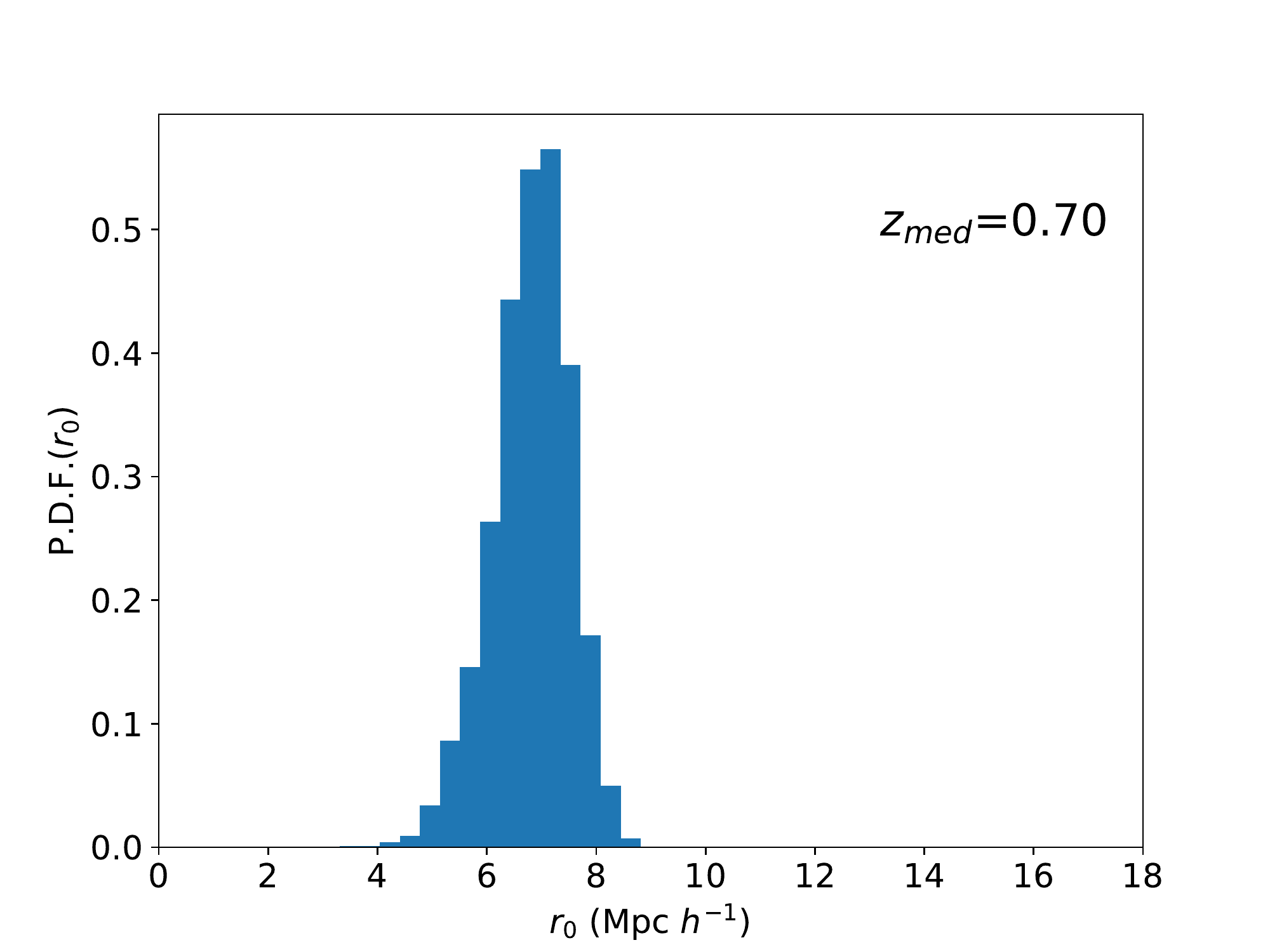}
\end{minipage}%
\newline
\begin{minipage}[b]{.5\textwidth}
\centering
\includegraphics[height=7.cm]{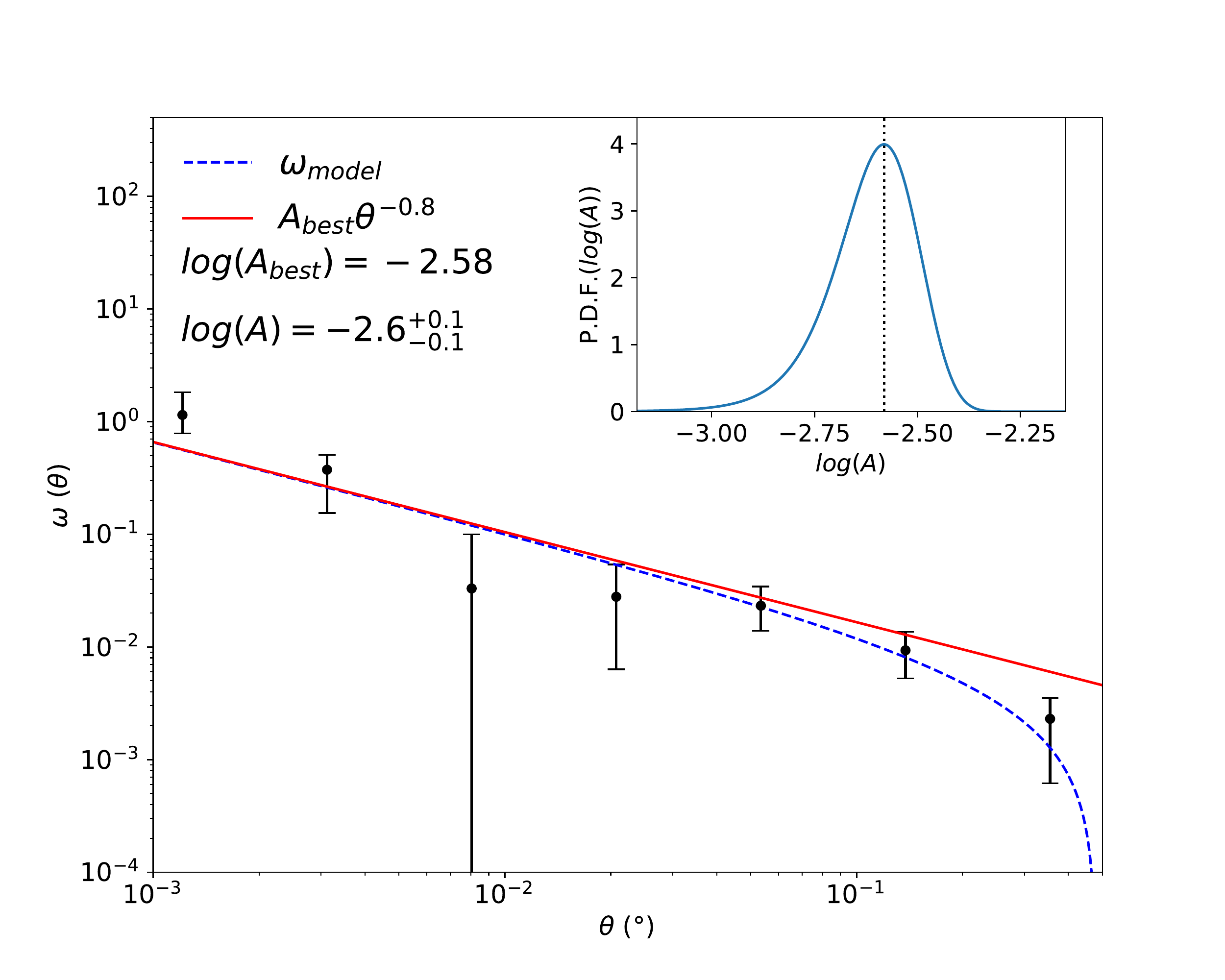}
\end{minipage}%
\begin{minipage}[b]{.5\textwidth}
\centering
\includegraphics[height=7.cm]{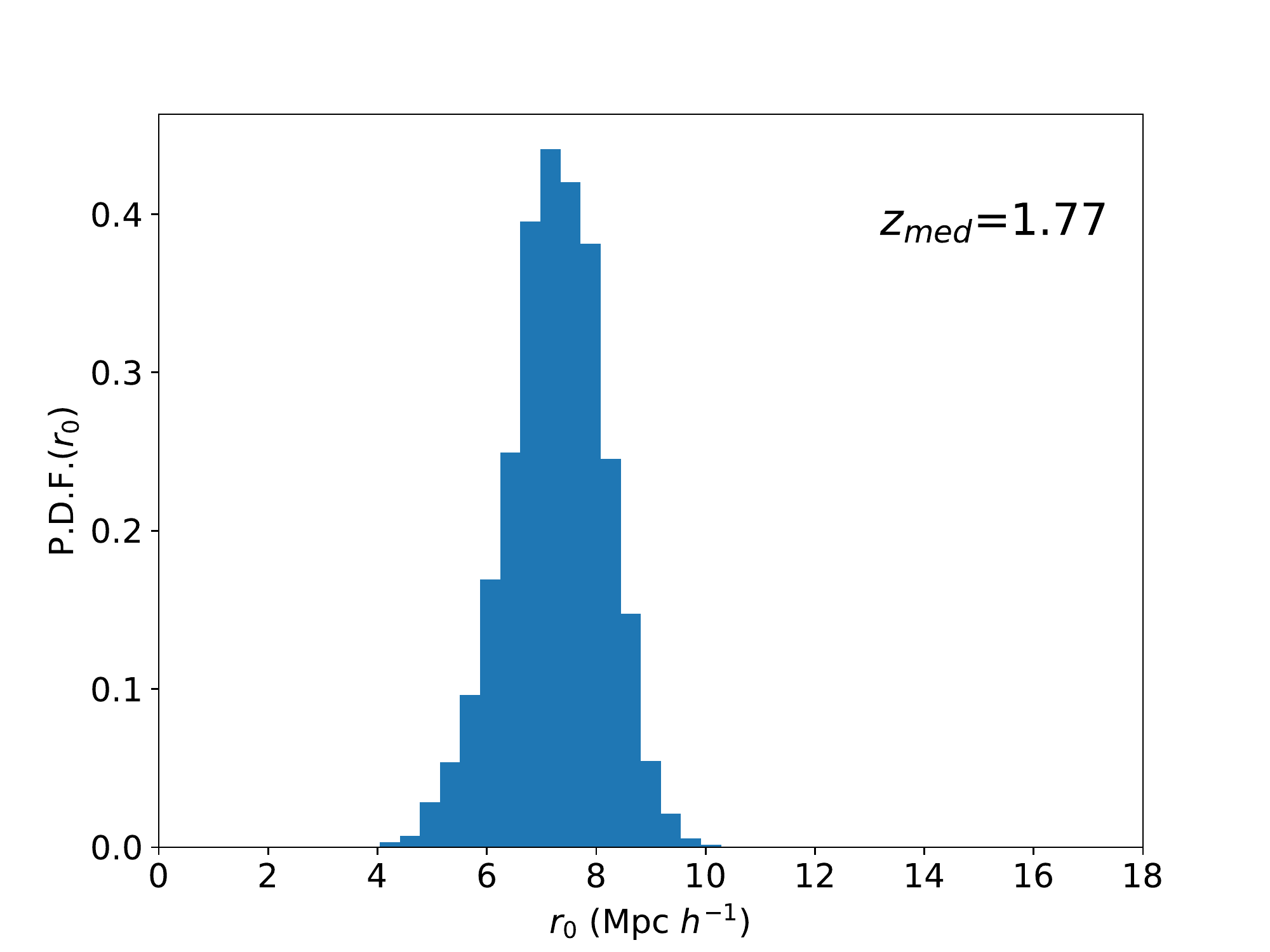}
\end{minipage}%
\end{center}
\caption{TPCF and fits for All AGN (top panel), AGN with $z<1$ (middle panel) and AGN with $z \ge 1$ (bottom panel). The left panel shows the TPCF, the red solid line represents the power law best fit to $\omega(\theta)$ and the dashed blue line is the same as the red line but with the integral constraint, $\sigma^2$ subtracted from it. The inset in the top right corner shows the probability density function for the value of $A$ to fit the TPCF, with the dashed line showing the best fit value. The right panel shows the corresponding P.D.F. for $r_0$. }
\label{fig:clusteringAGN}
\end{figure*}

For all AGN (Figure~\ref{fig:clusteringAGN}, top panel), we find $\log_{10}A = -2.6\pm 0.1$, corresponding to $r_0 = 9.6 \pm 0.7$\,Mpc\,$h^{-1}$ ($b = 3.6 \pm 0.2$, $z_{med} \sim 1.2$). 
Our clustering measurement for all AGN is derived from a sample with a similar redshift distribution to that of \cite{Magliocchetti}, who used a shallower radio survey over the COSMOS field, and our value of $r_0$ is comparable to the one derived in their work, $r_0 = 7.84 ^{+ 1.75}_{-2.31}$\,Mpc\,$h^{-1}$ at $z_{med} \sim 1.2$.

Restricting our comparisons to $z<1$, our clustering results can be seen in the middle panel of Figure \ref{fig:clusteringAGN}, for these AGN we find $\log_{10}A = -2.3\pm 0.1$ corresponding to $r_0 = 6.9 ^{+ 0.6 }_{- 0.7 }$\,Mpc\,$h^{-1}$ ($b = 2.1 \pm 0.2$, $z_{med}\sim 0.7$). This is similar to X-ray selected AGN from the work of \cite{Gilli2005,Gilli2009}, suggesting that the radio selected sample and the X-ray selected AGN trace similar halo mass distributions.
We also measure the high redshift ($z \ge 1$) clustering, (bottom panel of Figure \ref{fig:clusteringAGN}), and find a clustering amplitude of $\log_{10}A = -2.6\pm 0.1$, which corresponds to a similar clustering length of $r_0 = 7.3 \pm 0.9$\,Mpc\,$h^{-1}$ ($b = 3.5 \pm 0.4$, $z_{med} \sim 1.8$).

Given the depth of the radio survey, we are also able to split the AGN by their 3\,GHz luminosity (at $10^{23}$ WHz$^{-1}$) to investigate whether higher luminosity AGN preferentially lie in more highly biased environments, as one would expect if radio luminosity scaled with e.g. stellar mass. A split into high- and low-luminosity AGN has been used to investigate the different evolutionary pathways of such sources, with a wealth evidence showing that high-luminosity AGN evolve more rapidly than their low-luminosity counterparts \citep[e.g.][]{ClewleyJarvis2004,Sadler2007,Prescott2016}, therefore one may expect their clustering to also evolve differently.
 
Restricting the analysis of the high- and low-luminosity AGN to $z<1$, to make the comparison at similar cosmic epochs, we find for the high-luminosity AGN, $r_0 = 8.9 ^{+ 0.9 }_{- 1.2 } $\,Mpc\,$h^{-1}$ ($b = 2.9 \pm 0.3$, $z_{med} \sim 0.8$) and for the low-luminosity AGN, $r_0 = 5.9^{+ 1.0 }_{- 1.2 } $\,Mpc\,$h^{-1}$ ($b = 1.8 \pm 0.3$, $z_{med}=0.61$). Therefore, the high-luminosity AGN are significantly more clustered than their low-luminosity counterparts. 
Using the $z \ge 1$ high-luminosity sub-sample we are also able to determine the clustering length of these sources at much earlier epochs, and find $r_{0} = 8.4^{+0.8}_{-0.9}$\,Mpc\,$h^{-1}$ ($b = 4.0 ^{+ 0.3}_{- 0.4 } $, $z_{med} \sim 1.8$), which provides evidence for strong evolution in the bias of these sources at high redshift. Unfortunately, we have too few low-luminosity AGN at $z \ge 1$ to measure their clustering properties.

\begin{figure*}
\begin{center}
\begin{minipage}[b]{.5\textwidth}
\centering
\includegraphics[height=7.cm]{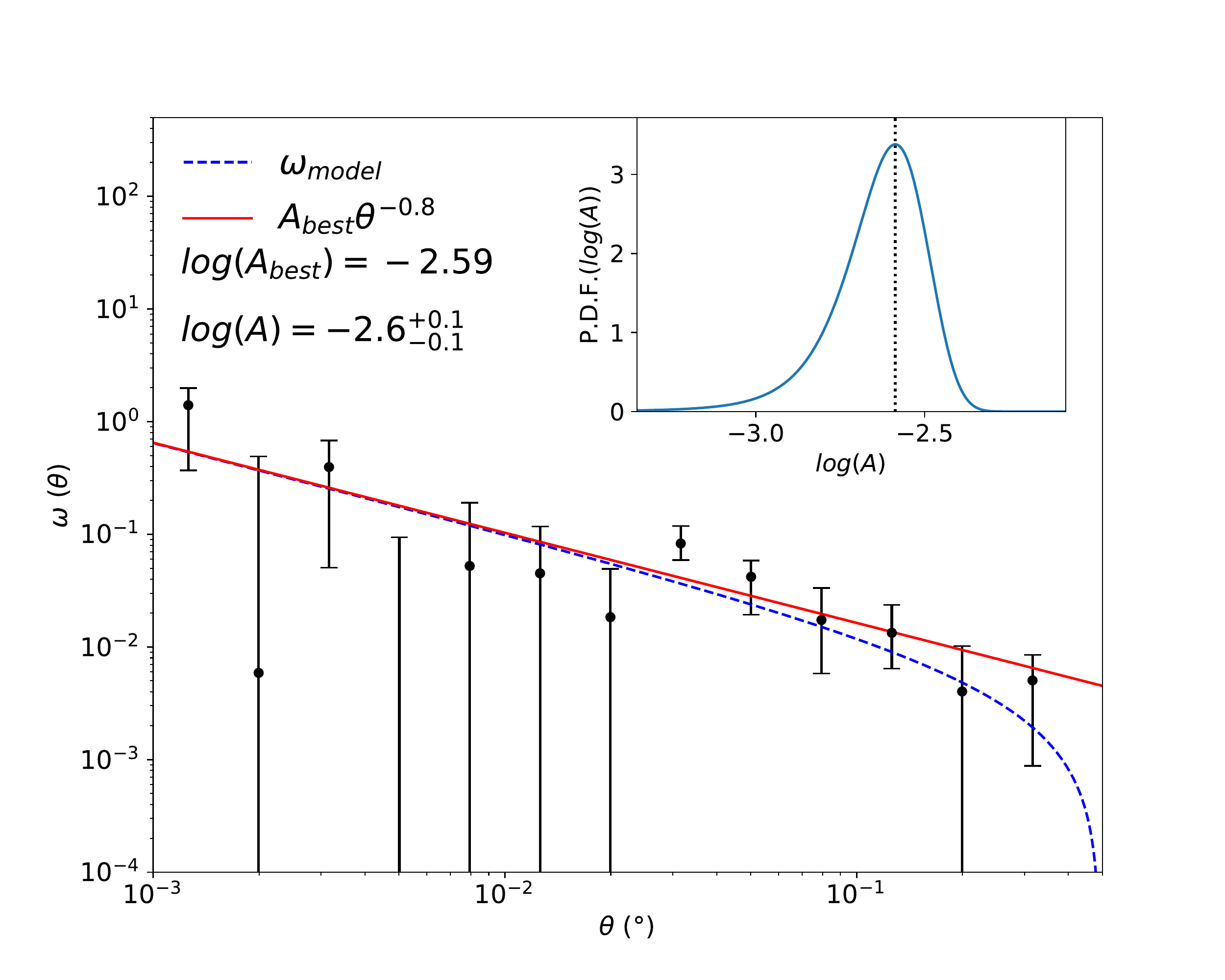}
\end{minipage}%
\begin{minipage}[b]{.5\textwidth}
\centering
\includegraphics[height=7.cm]{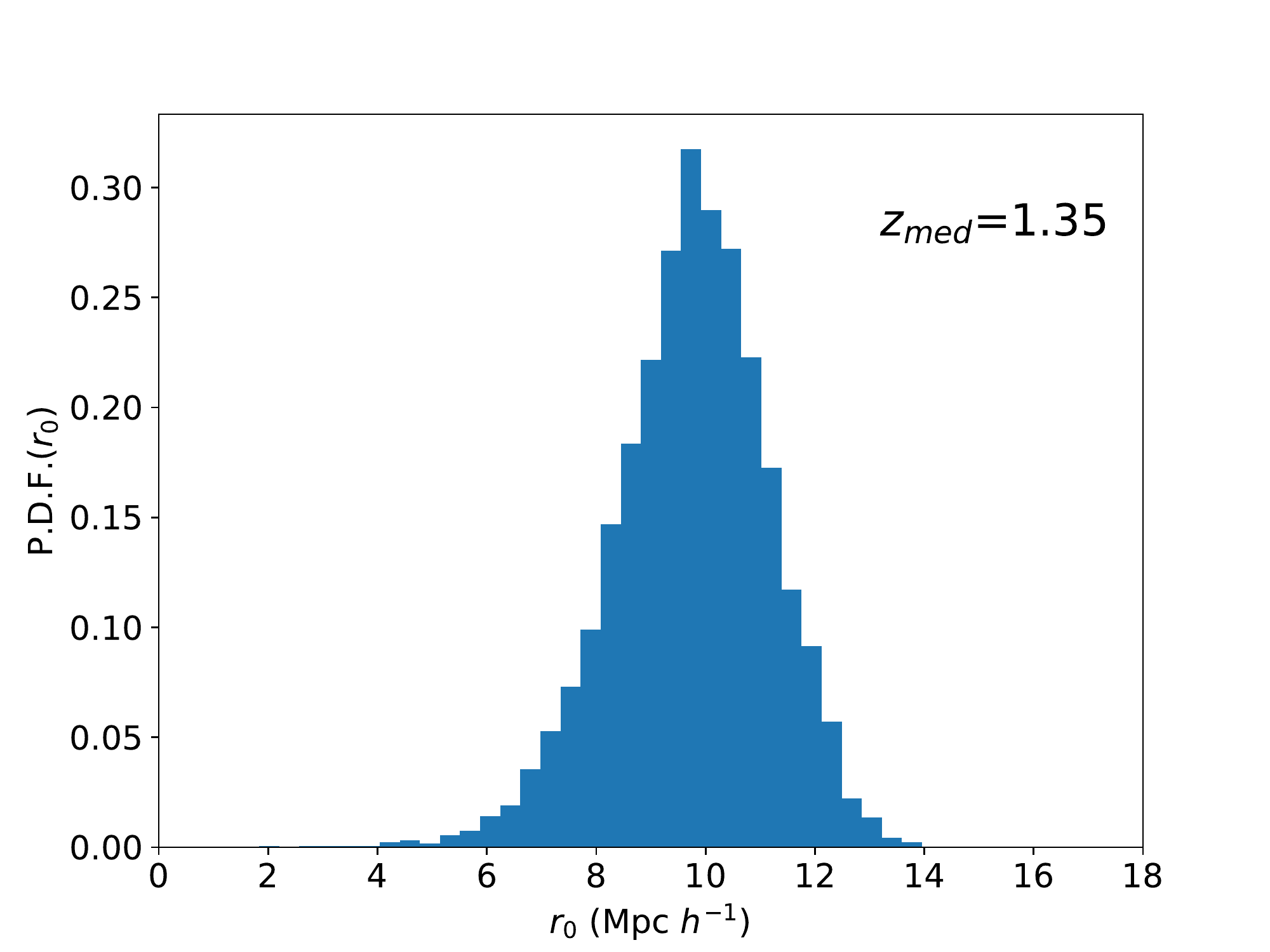}
\end{minipage}%
\newline
\begin{minipage}[b]{.5\textwidth}
\centering
\includegraphics[height=7.cm]{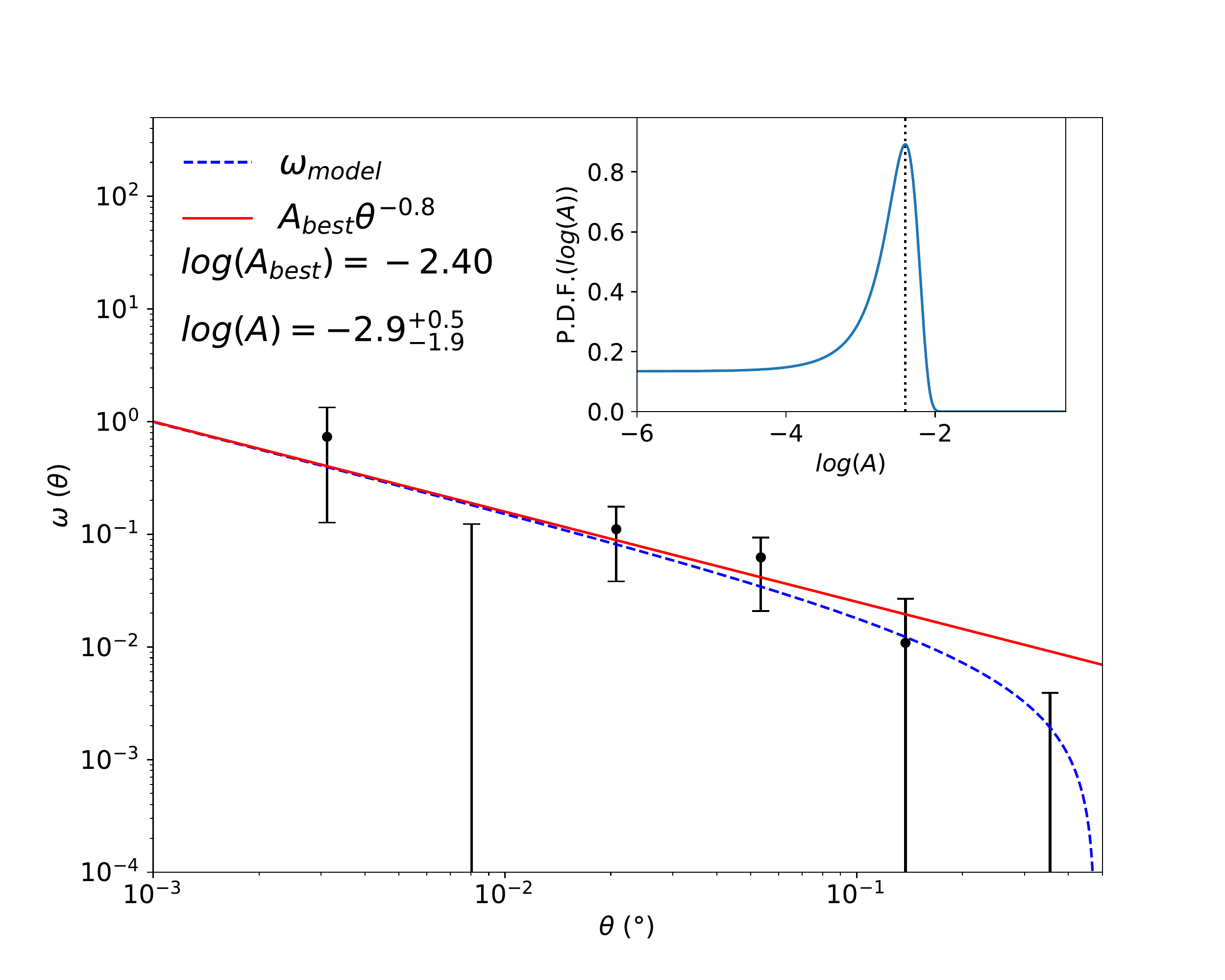}
\end{minipage}%
\begin{minipage}[b]{.5\textwidth}
\centering
\includegraphics[height=7.cm]{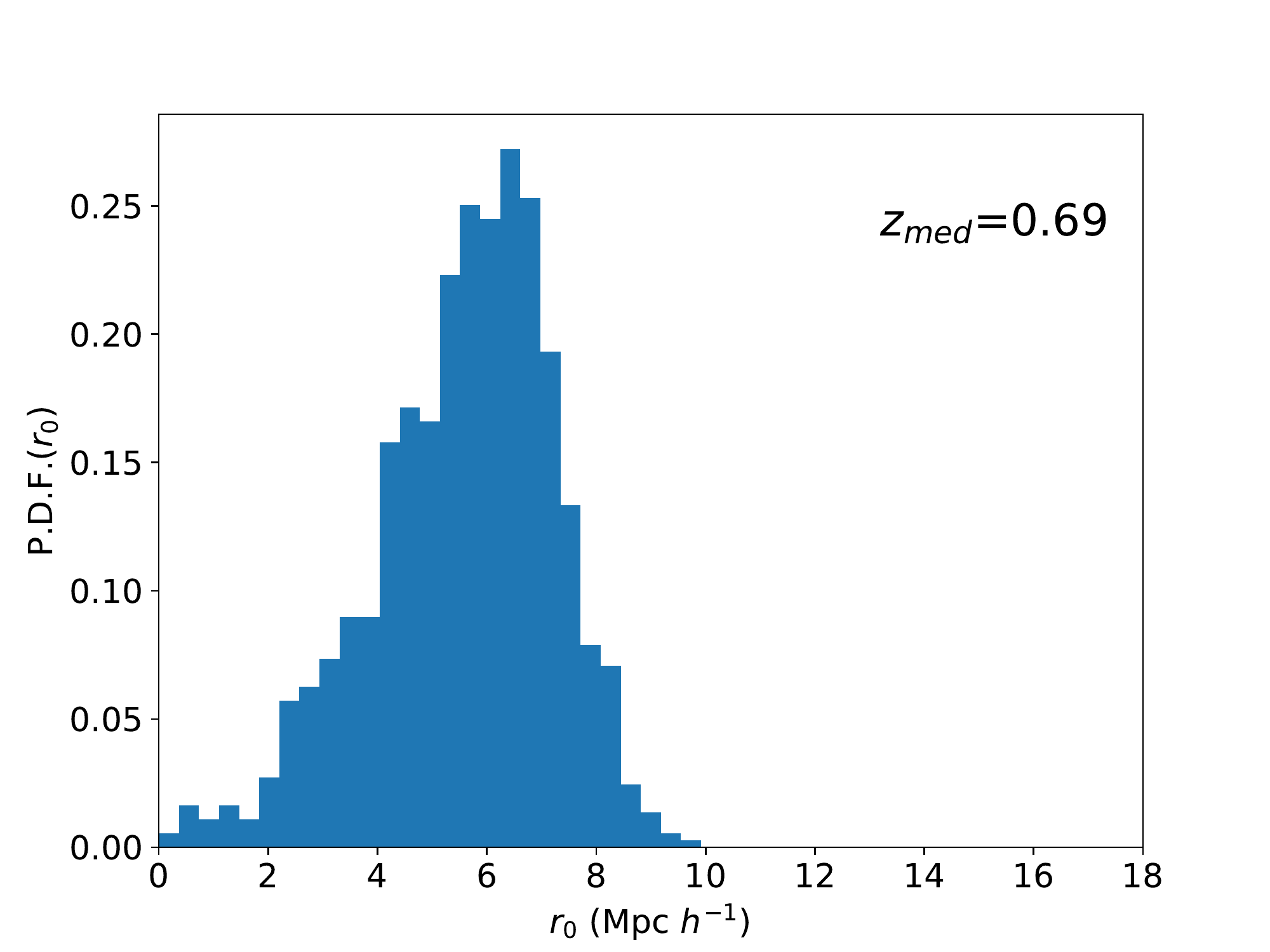}
\end{minipage}%
\newline
\begin{minipage}[b]{.5\textwidth}
\centering
\includegraphics[height=7.cm]{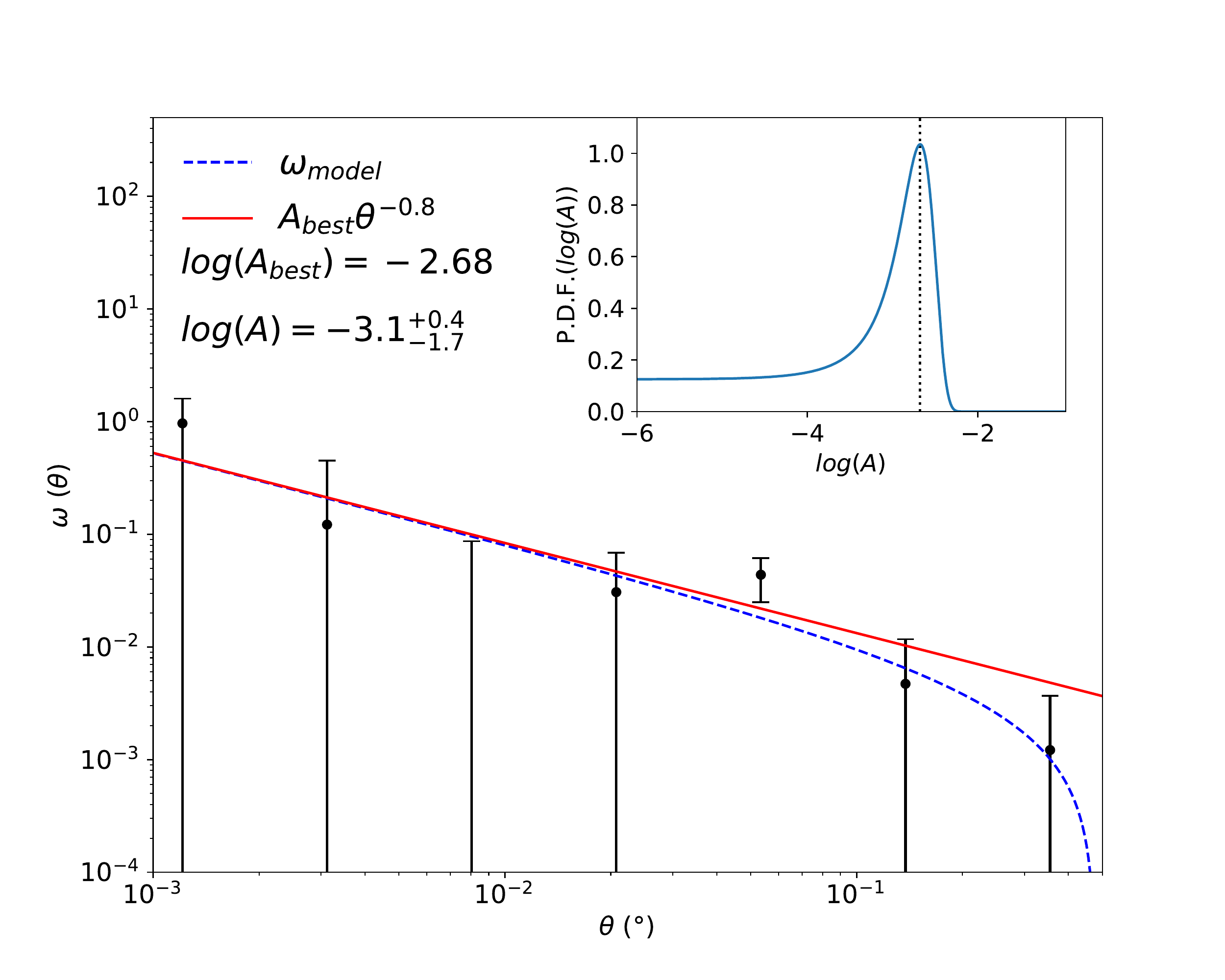}
\end{minipage}%
\begin{minipage}[b]{.5\textwidth}
\centering
\includegraphics[height=7.cm]{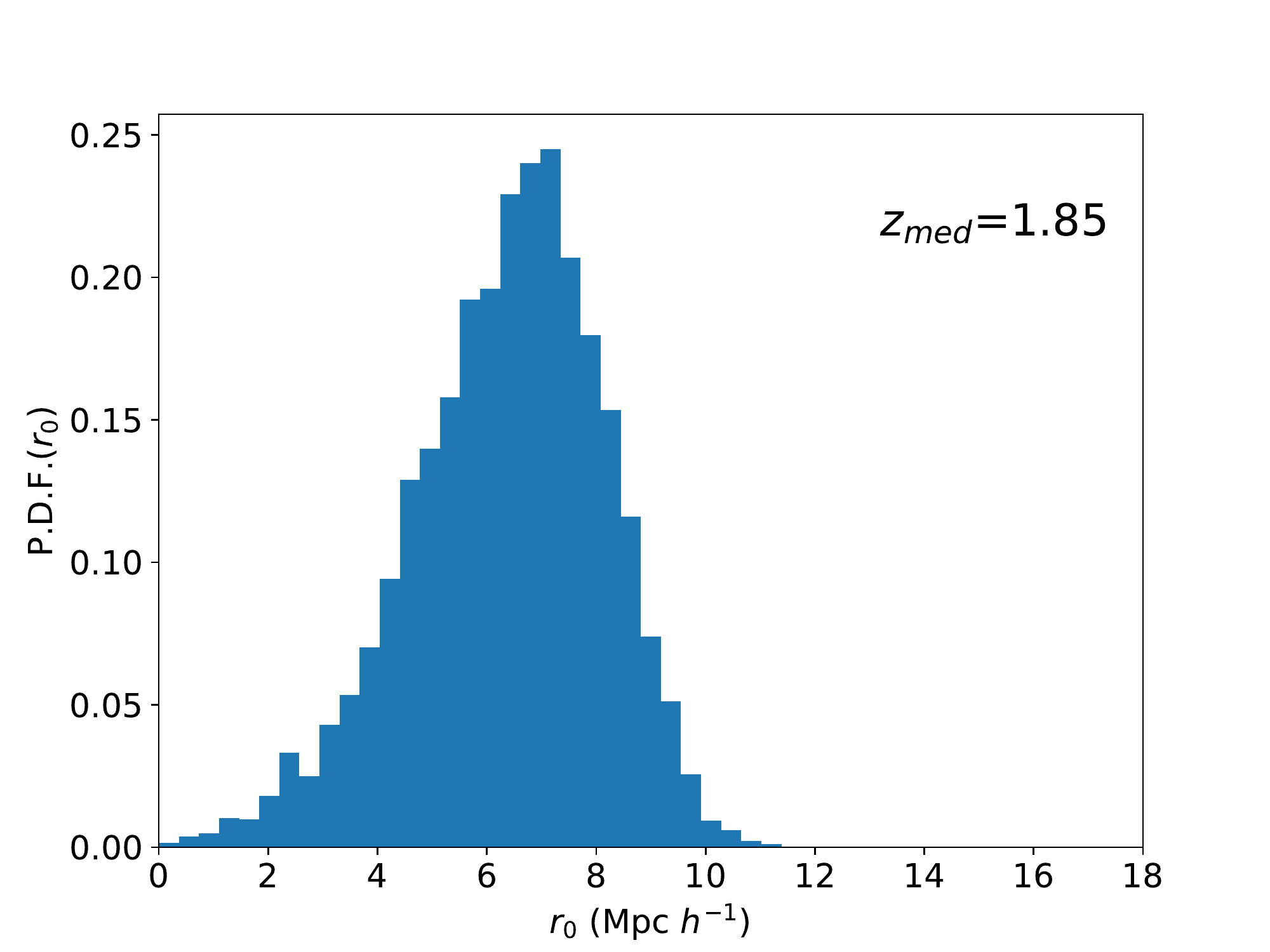}
\end{minipage}%
\end{center}
\caption{TPCF and fits for All HLAGN (top panel), HLAGN with $z<1$ (middle panel) and HLAGN with $z \ge 1$ (bottom panel). The left panel shows the TPCF, the red solid line represents the power law best fit to $\omega(\theta)$ and the dashed blue line is the same as the red line but with the integral constraint, $\sigma^2$ subtracted from it. The inset in the top right corner shows the probability density function for the value of $A$ to fit the TPCF, with the dashed line showing the best fit value. The right panel shows the corresponding P.D.F. for $r_0$. }
\label{fig:clusteringHLAGN}
\end{figure*}

\begin{figure*}
\begin{center}
\begin{minipage}[b]{.5\textwidth}
\centering
\includegraphics[height=7.cm]{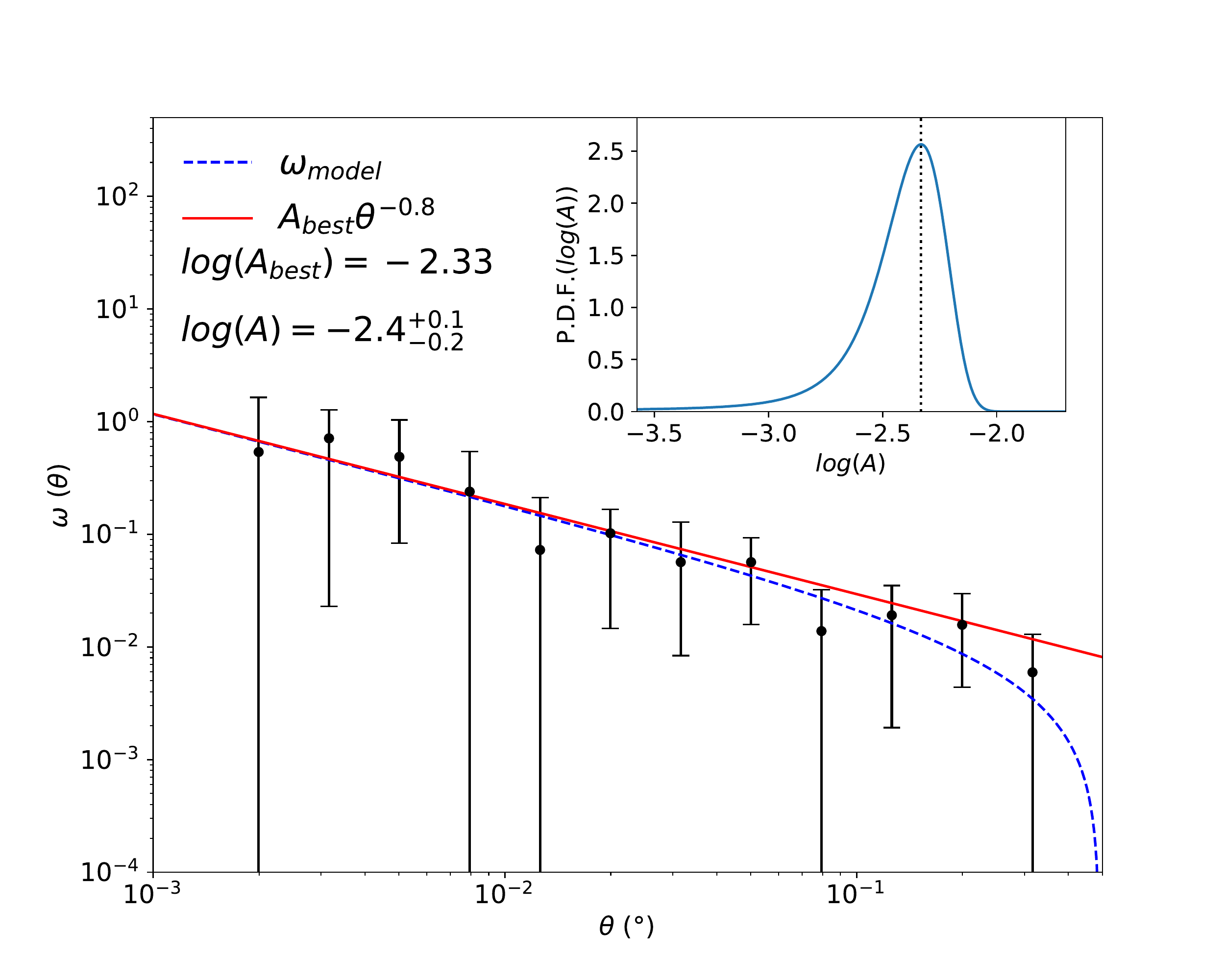}
\end{minipage}%
\begin{minipage}[b]{.5\textwidth}
\centering
\includegraphics[height=7.cm]{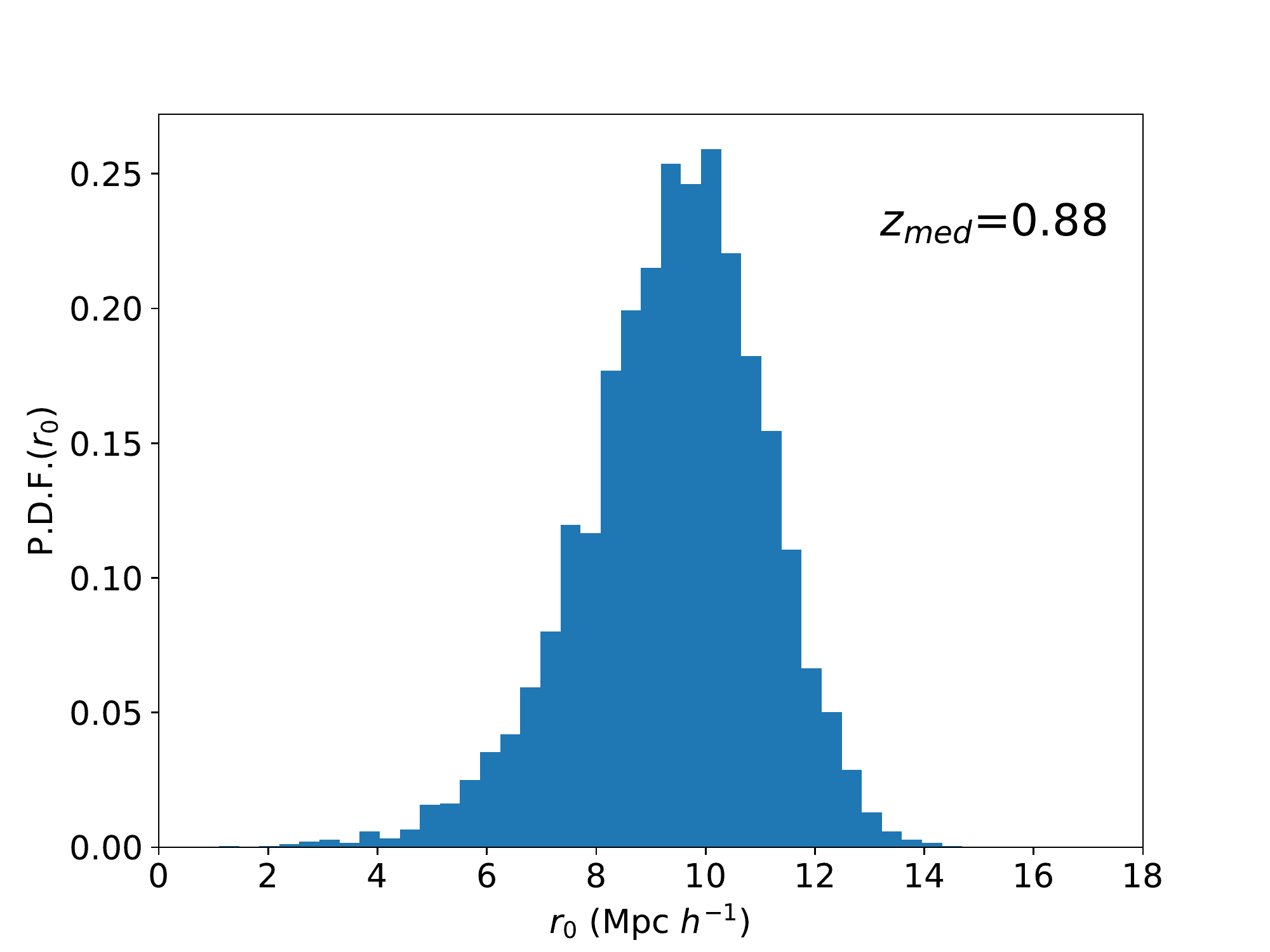}
\end{minipage}%
\newline
\begin{minipage}[b]{.5\textwidth}
\centering
\includegraphics[height=7.cm]{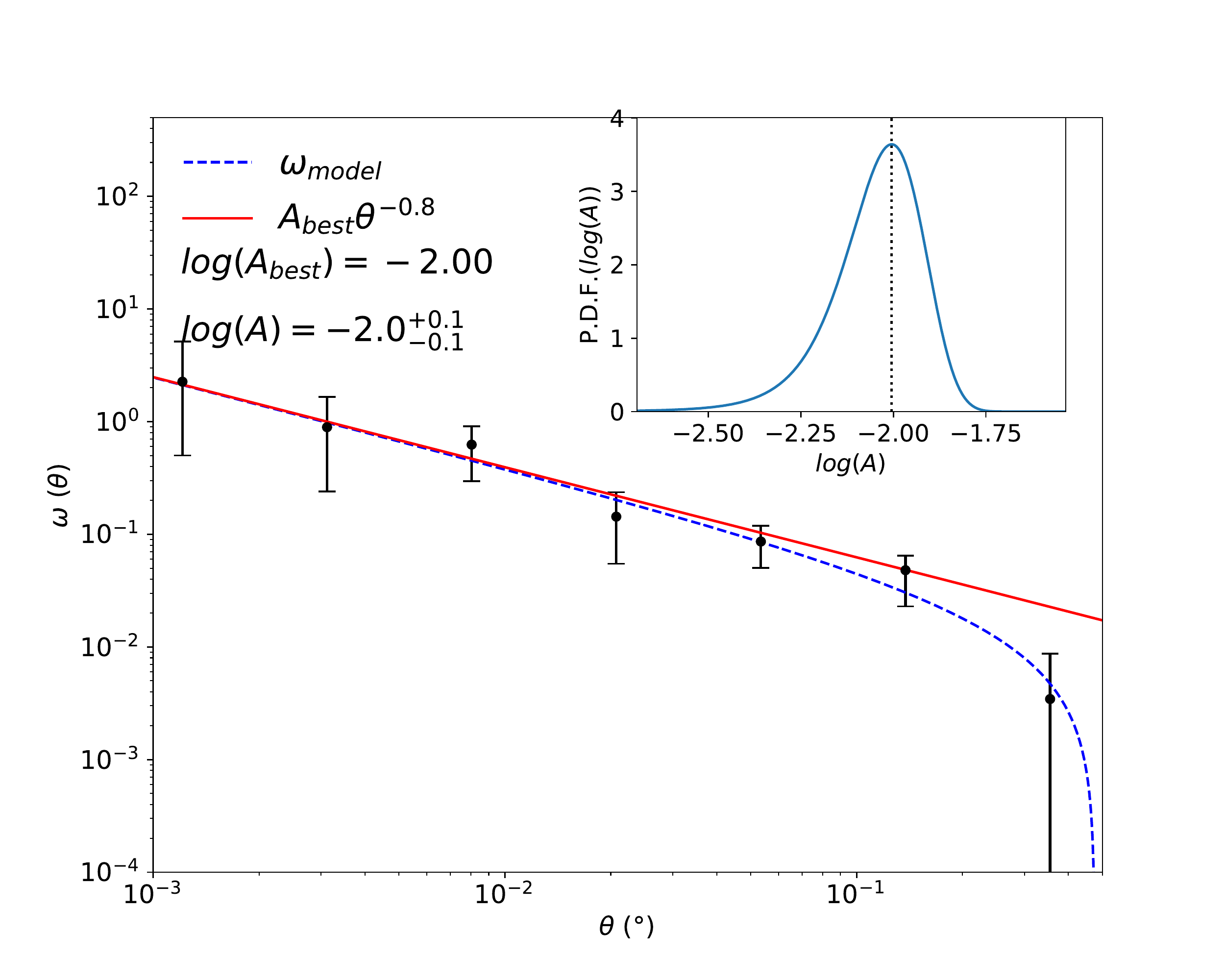}
\end{minipage}%
\begin{minipage}[b]{.5\textwidth}
\centering
\includegraphics[height=7.cm]{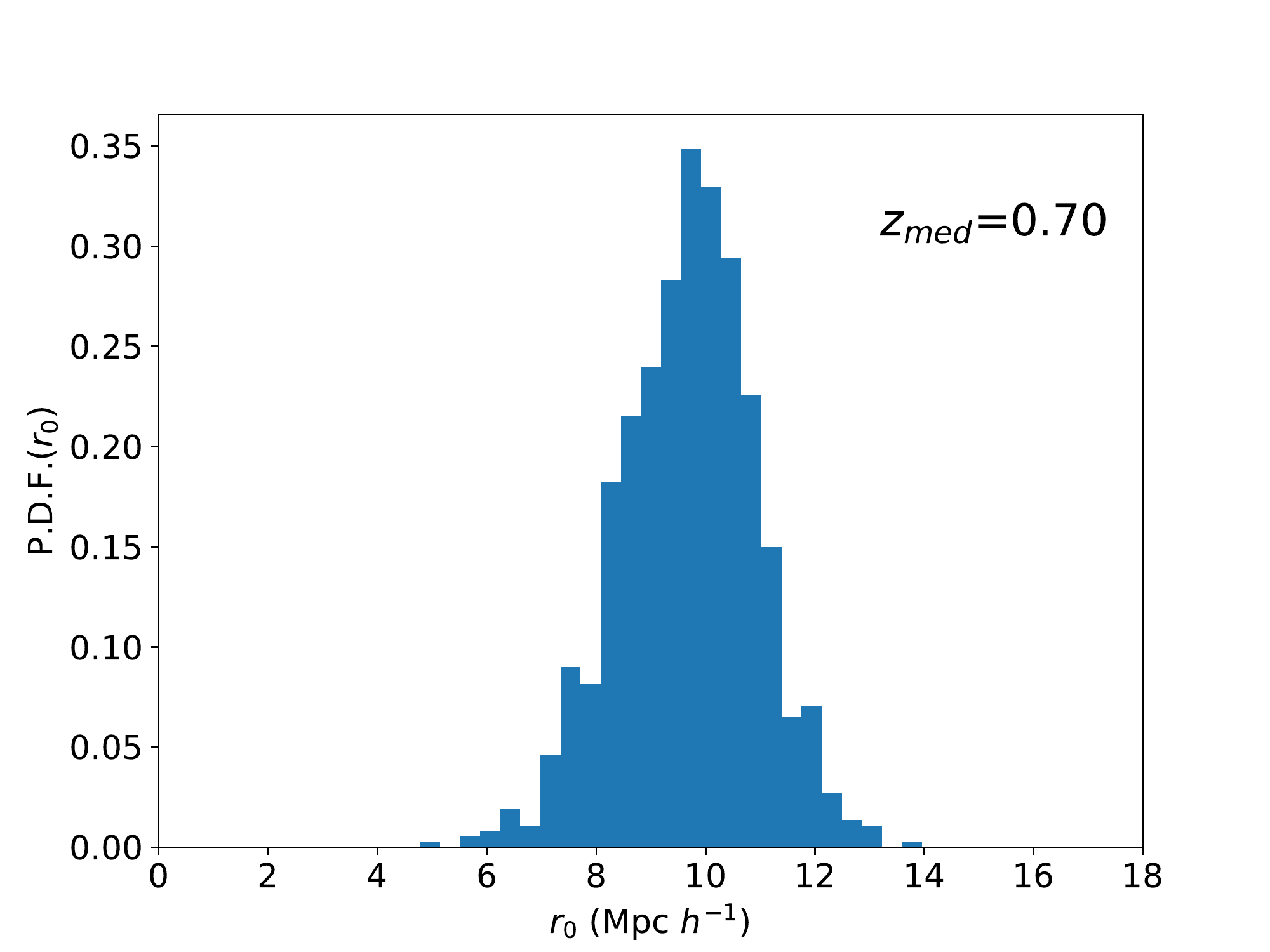}
\end{minipage}%
\newline
\begin{minipage}[b]{.5\textwidth}
\centering
\includegraphics[height=7.cm]{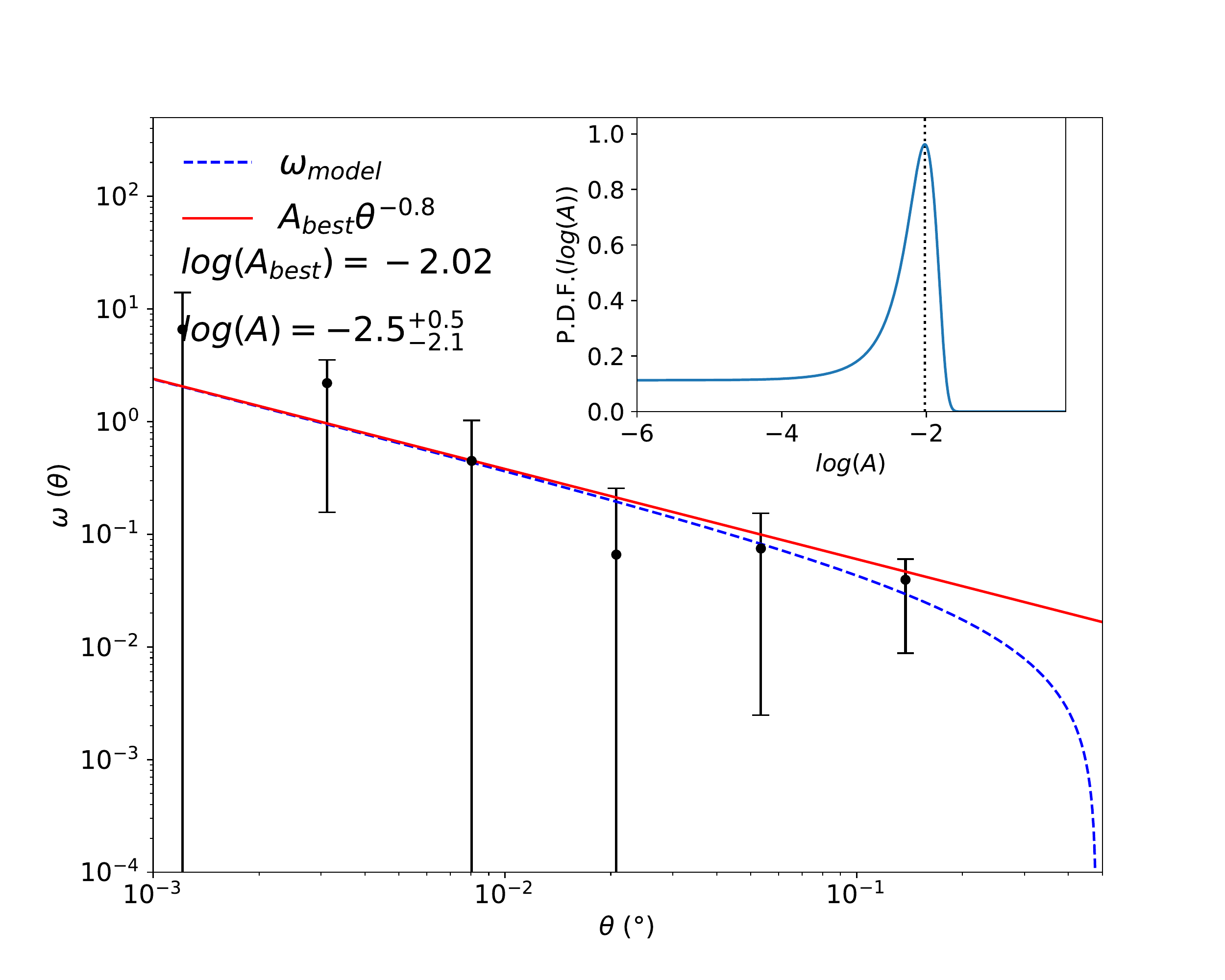}
\end{minipage}%
\begin{minipage}[b]{.5\textwidth}
\centering
\includegraphics[height=7.cm]{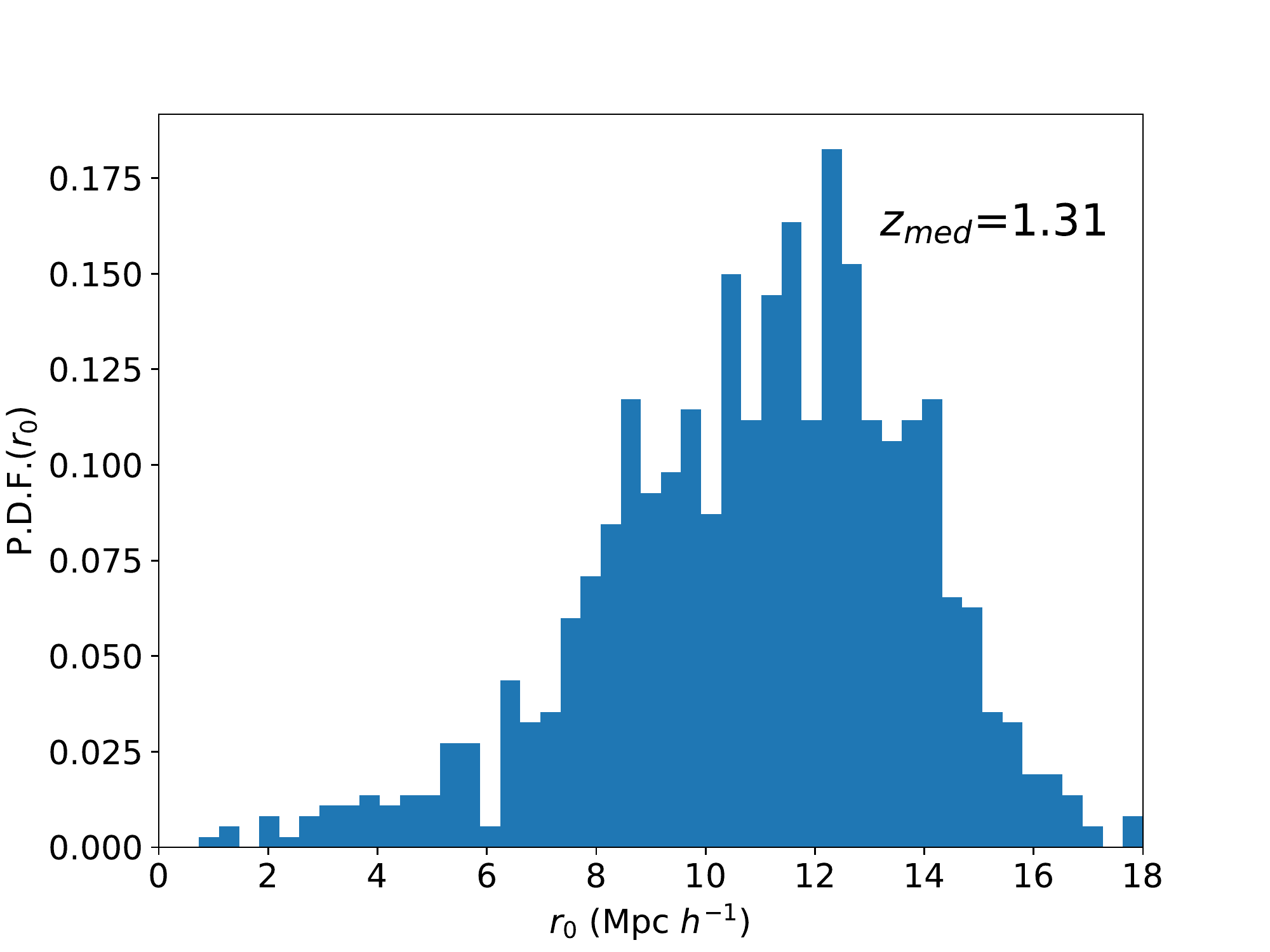}
\end{minipage}%
\end{center}
\caption{TPCF and fits for All MLAGN (top panel), MLAGN with $z<1$ (middle panel) and MLAGN with $z \ge 1$ (bottom panel). The left panel shows the TPCF, the red solid line represents the power law best fit to $\omega(\theta)$ and the dashed blue line is the same as the red line but with the integral constraint, $\sigma^2$ subtracted from it. The inset in the top right corner shows the probability density function for the value of $A$ to fit the TPCF, with the dashed line showing the best fit value. The right panel shows the corresponding P.D.F. for $r_0$. }
\label{fig:clusteringMLAGN}
\end{figure*}

As highlighted in Section~\ref{sec:introduction}, in the past decade it has become clear that although radio luminosity can provide a link to the accretion rate of the AGN, a more robust physical separation of AGN can be made using indicators of their accretion mode. 
We therefore measure how the clustering depends on the accretion efficiency of the AGN. To do this we use the HLAGN and MLAGN sub-samples as proxies for efficient and inefficient accreters, the results of which can be seen in the top panels of Figures~\ref{fig:clusteringHLAGN} for HLAGN and \ref{fig:clusteringMLAGN} for MLAGN. We find that the HLAGN have a clustering length of $r_0 = 9.8 ^{+ 1.3 }_{- 1.4 }$\,Mpc\,$h^{-1}$ ($b= 3.9 \pm 0.5$, $z_{med} = 1.35$). MLAGN, on the other hand, have a much lower median redshift, yet we determine a similar clustering length, $r_0 = 9.5 ^{+ 1.5 }_{- 1.8 }$\,Mpc\,$h^{-1}$ ($b=3.1 ^{+ 0.4 }_{- 0.5 }$, $z_{med} \sim 0.9$). 

Again to provide a direct comparison between the two sub-populations, we restrict the redshift range to $z<1$ (Figures \ref{fig:clusteringHLAGN} and \ref{fig:clusteringMLAGN}, middle panel).
In this case the HLAGN and MLAGN samples both have $z_{med} \sim 0.7$, and we find $r_{0}=5.8 ^{+ 1.4 }_{- 1.8 } $\,Mpc\,$h^{-1}$ ($b=1.8 ^{+ 0.4}_{- 0.5 }$) for HLAGN and $r_{0}=9.7^{+ 1.2 }_{- 1.3 }$\,Mpc\,$h^{-1}$ for MLAGN ($b=2.9 \pm 0.3$). As the redshift distributions are now comparable, this implies MLAGN reside in significantly more clustered environments than HLAGN. We therefore find strong evidence that the halo mass and the observed efficiency of the AGN are related. We also  measure the clustering for HLAGN and MLAGN at high redshifts ($z \ge 1$), to investigate the evolution of the two populations. In this case HLAGN have $r_0 = 6.5 ^{+ 1.5 }_{- 1.9 }$\,Mpc\,$h^{-1}$ ($b=3.2 ^{+ 0.7 }_{- 0.9 }$ , $z_{med}=1.85$), whereas MLAGN have $r_0 = 11.3 ^{+ 2.5 }_{- 3.0 }$\,Mpc\,$h^{-1}$ ($b=4.3 ^{+ 0.9}_{- 1.1 }$, $z_{med} \sim 1.3$). Although the median redshift of the HLAGN is much larger than that of the MLAGN and so their clustering is not directly comparable, we still find evidence that the MLAGN reside in more massive haloes at these high redshifts.

We note that the $S^3$ simulation does not separate AGN into HLAGN and MLAGN, but uses the radio morphology characterisation of FRIs and FRIIs. We therefore do not have a completely reliable estimate of the flux distribution of these individual AGN populations. This will affect which sources are recovered in the  random catalogue after being injected into the noisy map, but mainly affects those random sources with flux densities near the 5.5$\sigma$ flux limit. To investigate whether this would have an effect on our clustering results, we tested this on our whole redshift and $z<1$ sub-sample, using a more conservative, fixed flux limit of 22$\mu$Jy. With this higher flux limit, we also found MLAGN are appearing more clustered than HLAGN, suggesting the assumed source flux distribution in generating the random catalogue is not influencing our comparisons.

%%%%%%%%%%%% AGN FIGURES %%%%%%%%%%%%%%%%

%%%%%%%%%%%%%%%%%%%%

\section{Discussion}
\label{sec:discussions}

%%%%%%%%% COMPARISON FIGURES %%%%%%%%%%%
\begin{figure*}
\begin{center}
\begin{minipage}[b]{\textwidth}
\centering
\includegraphics[width=10.cm]{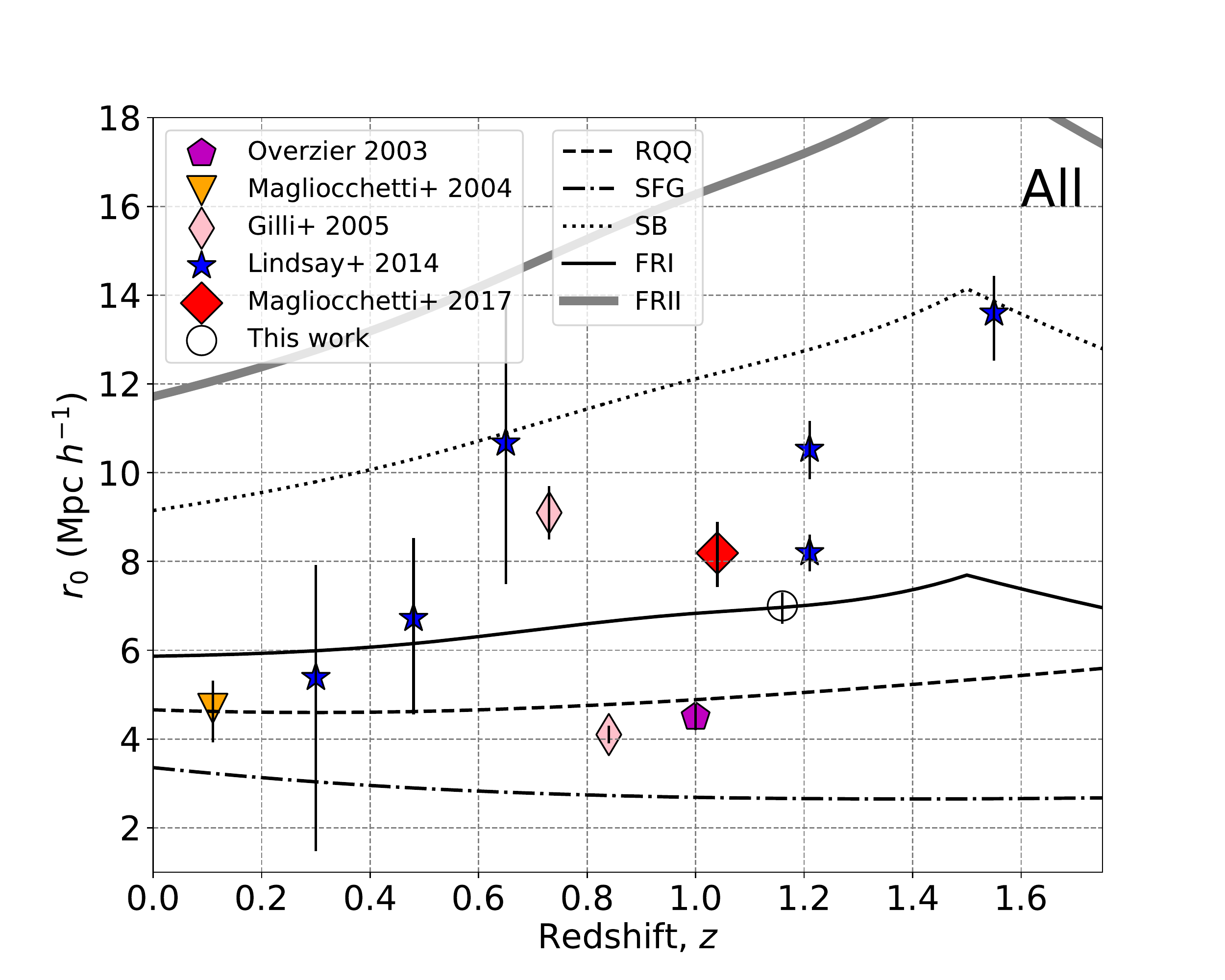}
\end{minipage}%
\newline 
\begin{minipage}[b]{.5\textwidth}
\centering 
\includegraphics[width=9.62cm]{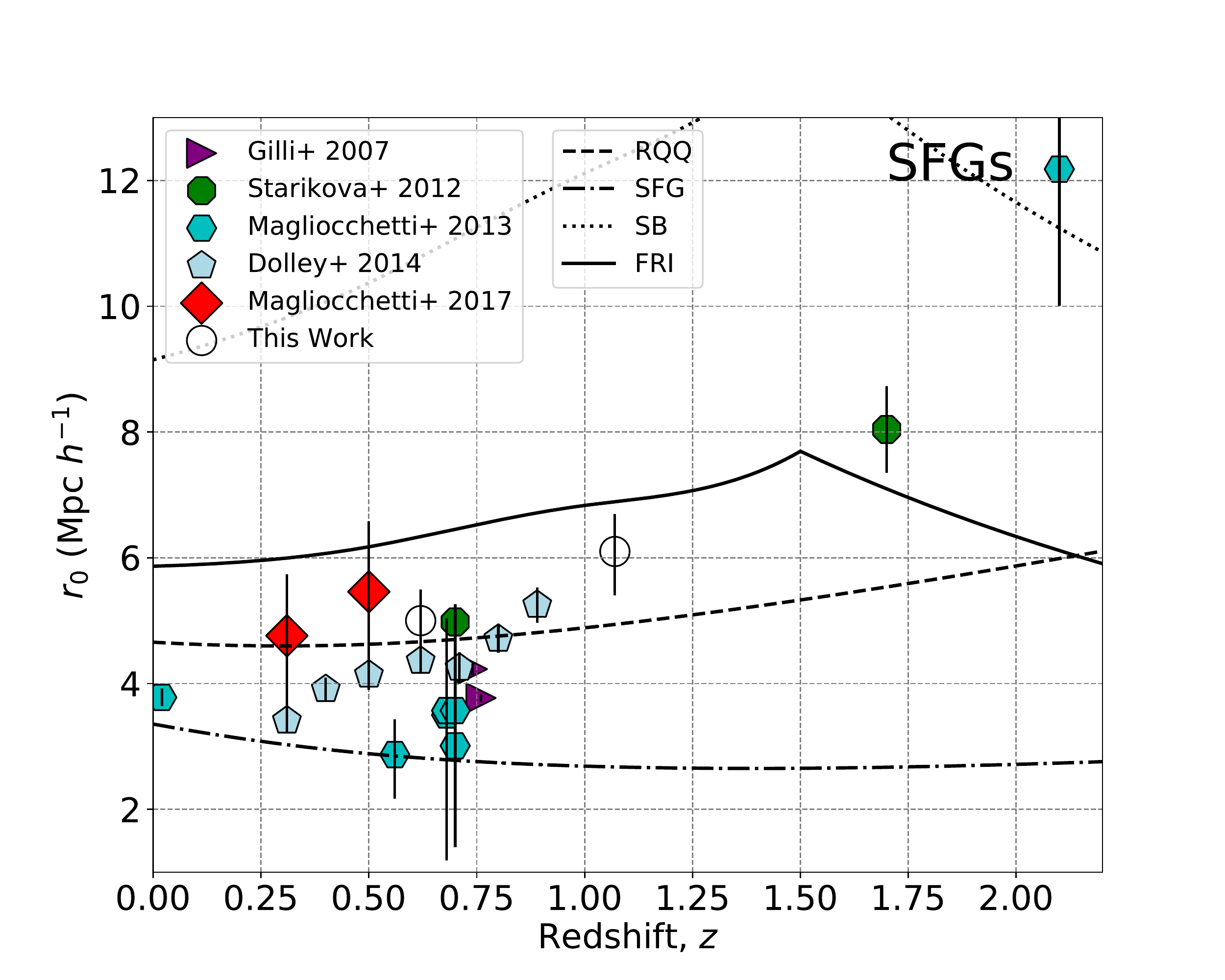}
\end{minipage}% 
\begin{minipage}[b]{.5\textwidth}
\centering 
\includegraphics[width=9.62cm]{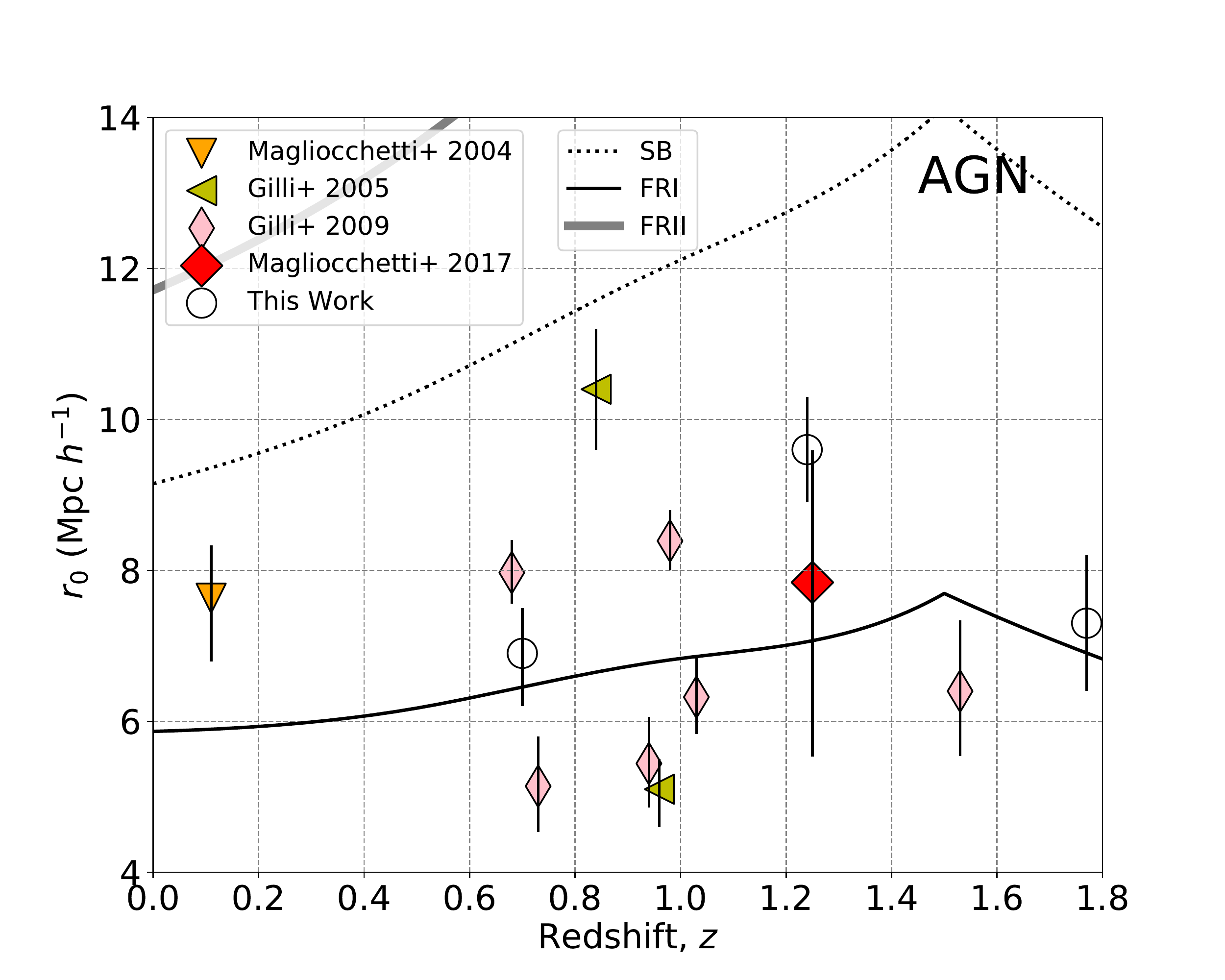}
\end{minipage}%
\end{center}
\caption{Comparison of our $r_0$ values (open circles) to previous studies for: All galaxies (top); SFGs (bottom left) and AGN (bottom right). These are for for all $z$, $z<1$ and $z \ge 1$ where appropriate. Previous work is from: \protect \cite{Overzier2003} (magenta pentagon); \protect \cite{Magliocchetti2004} (orange downwards triangles); \protect \cite{Gilli2005} (yellow left triangles - for the Chandra Deep Field North and South fields); \protect \cite{Gilli2007} (purple right triangles - for the GOODS-North and GOODS-South fields); \protect \cite{Gilli2009} (pink thin diamonds - for: All AGN; All AGN with a spike in galaxies at $z \sim 0.36$ removed; $z \sim 1$ AGN; $z<1$ AGN, $z<1$ AGN with a spike in galaxies at $z \sim 0.36$ removed; AGN with $z \ge 1$ ); \protect \cite{Starikova2012} (green octagon); \protect \cite{Magliocchetti2013} (cyan hexagons - showing points for IRAS galaxies; the COSMOS field; Extended Groth Strip; GOODS-South field); \protect \cite{Lindsay2014a} (blue stars); \protect \cite{Dolley2014} (pale blue pentagons - over a range of redshifts) and \protect \cite{Magliocchetti} (red diamonds). This is adapted from Figure 4 in \protect \cite{Magliocchetti}. For all studies shown, the redshift presented are the median redshift of the sample, however the samples will span a large range in redshifts, which is not shown. The lines show the evolution of $r_0$ as expected from \protect \cite{Wilman2008} for: Radio Quiet Quasars (RQQ, dashed); Star Forming Galaxies (SFG, dot-dashed); StarBursts (SB, dotted); Fanaroff-Riley Type 1 AGN (FRI, solid black) and Fanaroff-Riley Type 2 AGN (FRII, solid grey). The decline in $r_0$ for some of these lines (e.g. for SBs and FRIs) at $z>1.5$ is related to the constant bias imposed by \protect \cite{Wilman2008} at high redshift. }
\label{fig:r0distributioncomp}
\end{figure*}

\begin{figure*}
\begin{center}
\includegraphics[width=14cm]{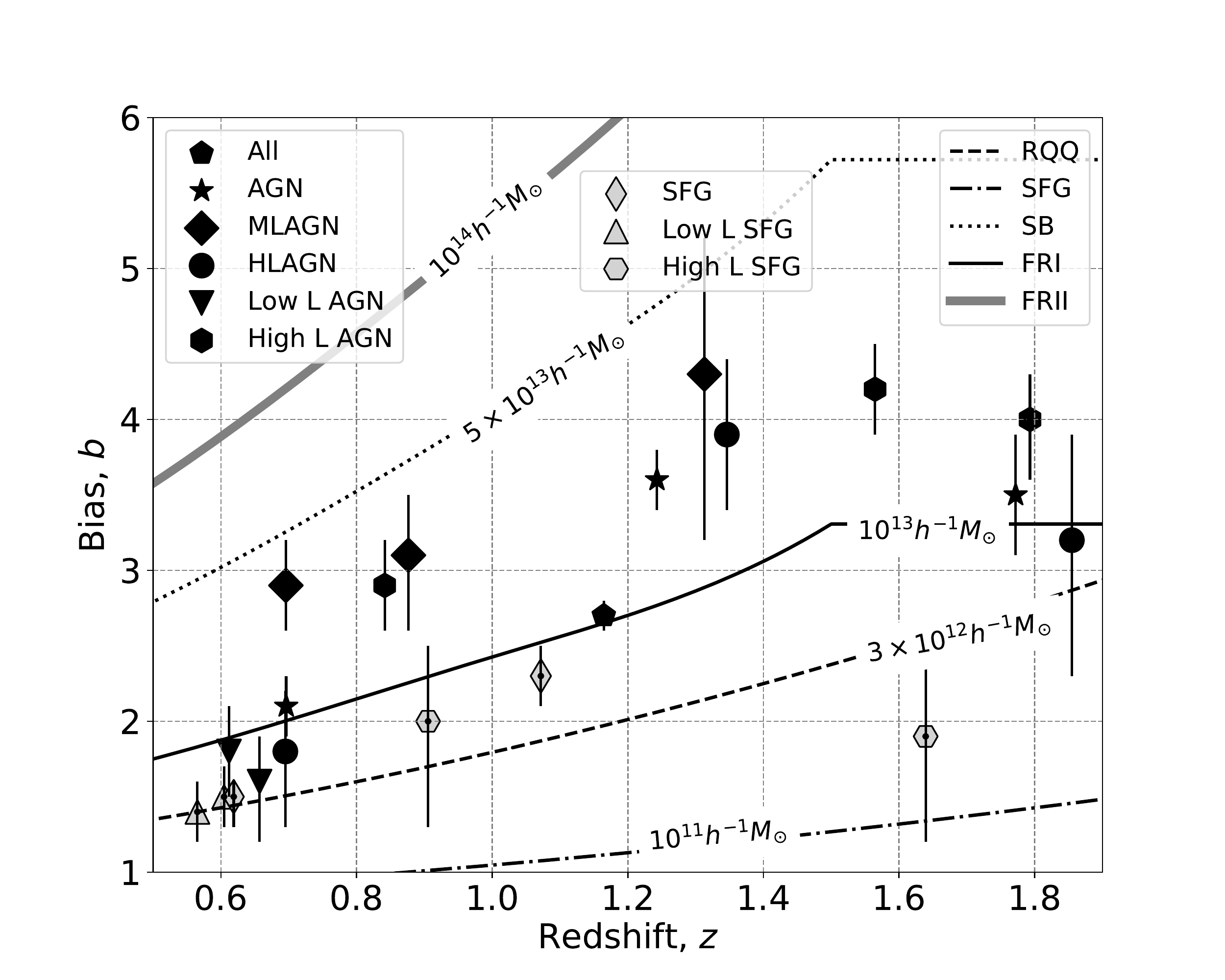}
\end{center}
\caption{The bias as a function of redshift for the different galaxy types investigated here. For all the galaxy sub-populations the results including sources of all redshifts as well as sources with $z<1$ and $z \ge1$ (except the SFGs and low luminosity AGN) are shown. The lines marked are the expected evolution of bias with redshift for different galaxy types from \protect \cite{Wilman2008}. These are for Radio Quiet Quasars (RQQ, dashed); Star Forming Galaxies (SFG, dot-dashed); StarBursts (SB, dotted); Fanaroff-Riley Type 1 AGN (FRI, solid black) and Fanaroff-Riley Type 2 AGN (FRII, solid grey). The flattening in the bias for some of the galaxy types from \protect \cite{Wilman2008} at $z>1.5$ is due to the constant bias imposed by \protect \cite{Wilman2008} at these high redshifts.}
\label{fig:bias}
\end{figure*}

%%%%%%%%%%%%%%%%%%%%%%%%%%%%%%%%

\subsection{Clustering of All Radio Sources}
\label{sec:discuss_clus_all}

Our measurements of the clustering length of all radio sources, along with the SFGs and AGN versus redshift are shown in Figure \ref{fig:r0distributioncomp}, along with some previous measurements from the literature. These include clustering measurements made using radio surveys as well as samples selected from other wavelengths.

Figure~\ref{fig:r0distributioncomp} (top) shows our results compared to those in \cite{Lindsay2014a} who used the radio survey data from the FIRST survey with a flux limit of 1\,mJy and \cite{Magliocchetti} who used a flux limit of 0.15\,mJy, both at 1.4 GHz. The flux limit used here, $S_{\rm 3GHz} \sim 13~\mu$Jy, is equivalent to $S_{\rm 1.4GHz}\sim 0.02$ mJy (assuming $\alpha = 0.7$). \cite{Lindsay2014a} found $r_0 = 8.20^{+0.41}_{-0.42}$\,Mpc\,$h^{-1}$ at $z_{med} \sim 1.2$, for all the radio  sources in their sample, similar to \cite{Magliocchetti}, who found $r_0=8.19^{+0.70}_{-0.77}$\,Mpc\,$h^{-1}$ at $z_{med} \sim 1.0$ ($\sim 11.7$\,Mpc).  Our value, $r_0= 7.0 ^{+ 0.3 }_{- 0.4 }$\,Mpc\,$h^{-1}$ at $z\sim 1.2$ is lower than these values. Given that the median redshift is similar, then we can attribute this lower clustering signal to the fact that the sample used here has a lower flux limit and thus contains, at each redshift, fainter sources. These will have lower luminosities and presumably inhabit lower mass galaxies (assuming radio luminosity traces stellar mass at some level, which in turn traces halo mass). These radio sources with low flux densities are also more likely to be SFGs at relatively low redshift, rather than the more strongly clustered high-redshift AGN. 

Indeed, using the $S^3$ simulation as a guide, the fraction of SFGs in the study of \cite{Lindsay2014a} is $\sim$8 per cent, and for \cite{Magliocchetti} is $\sim$31 per cent, compared to $\sim$60 per cent for the flux limit used in this work. 
Our measured clustering strength of all sources is also consistent with the average bias used in $S^3$ simulation. Our measured bias for the whole radio source population lies close to that for FRIs in the framework of the $S^3$ simulation, lying between the low bias of SFGs and high bias FRIIs (Figure~\ref{fig:bias}).

\subsection{Clustering by Type}
\label{sec:discuss_clus_type}
Our measurements of the clustering of AGN and SFGs show conclusively that AGN are more biased than SFGs (Figure~\ref{fig:bias}). This implies AGN reside in significantly more massive haloes than SFGs across all redshifts. The fact that the AGN reside in more massive haloes is not surprising given the extensive literature that shows that the host galaxies of powerful AGN are predominantly massive elliptical galaxies \citep[e.g.][]{Best1998,McLure1999,Jarvis2001kz,Seymour2007}, which we know from optical surveys are more strongly clustered than their lower mass (and bluer) counterparts \citep[e.g.][]{Croton2007,Zehavi2002,Zehavi2011,Hatfield2017}. However, it is only by splitting up the different populations of radio AGN, and measuring their clustering separately can we learn about how the mass of the halo influences the type of AGN that we observe at radio wavelengths. In the following, we discuss the clustering of SFGs and AGN separately. 

The $S^3$ simulations \citep[][]{Wilman2008}, use fixed halo masses for the different radio source populations. The bias evolution for the different types of radio source (with the different halo masses for each source type) are shown in Figure~\ref{fig:bias}, and are based on \cite{Mo1996}\footnote{We note that in Figures~\ref{fig:r0distributioncomp} and \ref{fig:bias} the predicted lines for some types of radio source from \cite{Wilman2008} appear flat in $b$ vs $z$ above $z>1.5$. This was imposed in the simulation and is discussed in Section~\ref{sec:discuss_clus_agn}.}. By comparing the bias evolution in these simulations to the bias evolution observed here, we are able to infer a halo mass for our different samples. It also allows us to compare to simulations that have been used to make cosmological predictions for observations of future telescopes.

\subsubsection{Clustering of SFGs}
\label{sec:discuss_clus_sfg}
Our results show that at the redshift ranges investigated here, the evolution of clustering for SFGs mirrors that expected for galaxies that reside in similar mass haloes across cosmic time i.e. it has a similar slope with redshift compared to the simulations in \cite{Wilman2008} for a halo with mass $M_{h} \sim 3-5 \times 10^{12}$\,$h^{-1}$\,M$_{\odot}$. 

Comparing to previous work (Figure~\ref{fig:r0distributioncomp}, bottom left), \cite{Magliocchetti} found $r_{0} =5.46^{+1.12}_{-2.10}$\,Mpc\,$h^{-1}$ (at $z_{med} \sim 0.5$) in close agreement with our value of $r_0 = 5.0^{+ 0.5 }_{- 0.6 }$\,Mpc\,$h^{-1}$ for our redshift-limited sample with $z<1$, with a median redshift of $z_{med} \sim 0.6$. This is also similar to the values of $r_{0}$ found by selecting SFGs at a range of wavelengths \citep[e.g.][]{Magliocchetti2013,Dolley2014,Starikova2012}, although we note that the $r_{0}$ values for the radio surveys do tend to be slightly higher than the studies at other wavelengths. This is most likely due to the fact that the other wavelength studies have a redshift distribution that is skewed to lower redshifts than the star-forming radio sources, and thus the clustering length for the radio sources is also skewed to slightly higher values even though the median redshift is similar. Indeed, as can be seen in Figure~\ref{fig:redshiftdistributionall}(b) the redshift distribution of the SFGs is still rising to the redshift limit of $z<1$.  This rise in source density towards $z\sim 1$ also means that the average star-formation rate of the sources in the SFG sample will also be skewed to higher SFRs than for a survey with a lower mean (rather than median) redshift distribution. Given the relationship between stellar mass and star-formation rate \citep[or the star-formation main sequence; ][]{Noeske2007,Daddi2007,Whitaker2012,Johnston2015}, and the strong correlation between stellar mass and halo mass at these redshifts \cite[e.g.][]{McCracken2015,Hatfield2016}, it is not surprising that the radio-selected SFGs have a higher clustering length.

The clustering measurements of our SFGs are also higher than predicted from the $S^3$ simulations. However the assumed bias in the $S^3$ simulation separates normal SFGs from Starbursts (SBs), whereas we do not make a separation on this basis. SBs are a lot less numerous than the normal SFGs and so their contribution to the bias is expected to be relatively small.
To test this, we split the SFG sample into high-luminosity and low-luminosity subsamples, with the split at SFR$\sim 100$\,M$_{\odot}$yr$^{-1}$. With this split, we find that the highly star-forming galaxies ($L_{\rm 3GHz} > 10^{23}$~WHz$^{-1}$), which are biased towards higher redshifts, have an approximately flat bias with large errors that correspond to halo masses in the range of $M_{h} \sim 10^{11} - 10^{13} $\,$h^{-1}$\,M$_{\odot}$ at the different redshift epochs. 
The low luminosity SFGs predominately lie at lower redshifts, and we therefore do not have a large enough sample to test the evolution in the bias for these sources. To investigate this, deeper and/or wider observations would be required. At the low redshifts investigated for these, they appear to inhabit haloes with  $M_{h} \sim 3 \times10^{12}$\,$h^{-1}$\,M$_{\odot}$.

\cite{Dolley2014} have used mid-IR data to investigate the evolution in the clustering of SFGs with redshift. They find a slow evolution of the clustering of SFGs with redshift, which is in agreement with the values presented here at $z \sim 0.6$ and $z \sim 1.1$ and also in \cite{Magliocchetti}. Extrapolating this trend in $r_0$ to higher redshifts, this also extends out to the measurement by \cite{Starikova2012} at $z \sim 1.7$ and \cite{Magliocchetti2013} at $z \sim 2.1$. 

The results by \cite{Dolley2014} also suggest that the halo mass can vary by 1-2 orders of magnitude, and can be significantly higher than assumed by \cite{Wilman2008}. Our results also seem to suggest higher halo masses than assumed for SFGs in their simulations. This may suggest that the halo masses assumed in \cite{Wilman2008}, which were forced to agree with the $z=0$ clustering measurements of {\em IRAS} galaxies, are too low for the general SFG population.  There will be a contribution from SBs that affects our results, especially when we consider the full SFG population and so we must consider our results as a combination of SBs and normal SFGs. Nevertheless, the halo masses used in \cite{Wilman2008}, as well as in various cosmology forecast papers based on the $S^3$ simulations \citep[e.g.][]{Ferramacho2014,Raccanelli2015}, may not be representative of the halo masses of SFGs as a whole.

\subsubsection{Clustering of AGN}
\label{sec:discuss_clus_agn}
Our results from Section~\ref{sec:clusAGN} have shown that the clustering length for AGNs increases from $z\sim 0.7$ to $z\sim 1.3$.
This is comparable to previous work (Figure \ref{fig:r0distributioncomp}, bottom right), by \cite{Magliocchetti} who investigated the clustering of radio AGN over a similar redshift range ($z_{med} \sim 1.2$) and found $r_{0}=7.84^{+1.75}_{-2.31}$\,Mpc\,$h^{-1}$. This is slightly lower, but consistent with that obtained here over all redshifts ($9.6 \pm 0.7$\,Mpc\,$h^{-1}$). At low redshift, our measured clustering length is in agreement with the work of \cite{Gilli2009}, who investigated the clustering of X-ray selected AGN in the COSMOS field and its evolution. When \cite{Gilli2009} first considered the clustering of all X-ray detected AGN in their field, they obtained a value of $r_0=8.39^{+0.41}_{-0.39}$\,Mpc\,$h^{-1}$ at $z_{med} \sim 1.0$, consistent with the evolution in $r_0$ found for the radio AGN in our sample. At high redshifts ($z \ge 1$) we measure a clustering length of $r_0 = 7.3\pm 0.9$\,Mpc\,$h^{-1}$ ($b=3.5 \pm 0.4$, $z_{med} \sim 1.8$). This suggests a flattening of the bias at high redshift.

It is worth mentioning that \cite{Gilli2009} found a spike in the population of AGN at $z \sim 0.36$ and so also removed this from their analysis and remeasured the clustering. This resulted in a lower value for the clustering length of $r_0=6.32^{+0.53}_{-0.49}$\,Mpc\,$h^{-1}$, at $z\sim1$. Our radio sample is also derived from the COSMOS field and we also see a peak in the redshift distribution at $z \sim 0.35$ (visible in Figure~\ref{fig:redshiftdistributionall}(c) and clearly visible in the low luminosity sample in Figure~\ref{fig:redshiftdistributionall}(h)). This suggests that such a structure in the field could be affecting our clustering measurements, with enhanced clustering at $z<1$ increasing our $r_0$ values. We do not attempt to remove this from our analysis, but emphasises the necessity to carry out similar analysis to this work over larger areas of sky in order to minimize the effect that cosmic variance in this relatively small field could have on our clustering measurements \citep[see e.g.][]{Heywood2013, Jarvis2017}. 
However, more importantly, \cite{Gilli2009} selected AGN in the X-ray, i.e. those AGN with appreciable accretion rates, thus it would be fairer to compare the clustering of the X-ray selected AGN to that of the HLAGN population.

The results for HLAGN and MLAGN at low redshift ($z<1$) show that MLAGN are significantly more clustered than HLAGN.We find that, at $z<1$, HLAGN have $r_0 = 5.8 ^{+ 1.4 }_{- 1.8 }$\,Mpc\,$h^{-1}$ whereas for MLAGN $r_0 = 9.7 ^{+ 1.2 }_{- 1.3 }$\,Mpc\,$h^{-1}$. This value for HLAGN is lower than found by \cite{Gilli2009} for X-ray selected AGN, however \cite{Gilli2009} have a slightly higher average redshift, $z \sim 1$. Assuming a shallow evolution in $r_{0}$ with $z$ for our HLAGN, suggests that the 
clustering measurement by \cite{Gilli2009} and our work are consistent.

We also find that at high redshifts ($z \ge 1$) the MLAGN show evidence for evolution in the clustering length with $r_0 = 11.3 ^{+ 2.5 }_{- 3.0 }$\,Mpc\,$h^{-1}$ at $z \sim 1.3$, consistent with a constant halo mass with redshift. On the other hand, for the HLAGN we find $r_0 = 6.5 ^{+ 1.5}_{- 1.9 }$\,Mpc\,$h^{-1}$ at $z = 1.85$. This is an increase in $r_0$ compared to our $z<1$ sub-sample, however is a decrease compared to the $r_0$ value for our whole HLAGN population at $z_{med}=1.35$. This decrease suggests a flattening in the bias of HLAGN towards high redshift compared to that for the whole HLAGN population.

Comparing the clustering of MLAGN to HLAGN at $z_{\rm med} \sim 1.3$ (see Figure \ref{fig:bias}), we continue to find tentative evidence that the MLAGN are more clustered than their HLAGN counterparts. However the errors for the MLAGN are large and the values are formally consistent with one another. At low redshifts LERGs are known, however, to inhabit a large range of environments which can be similar to those for HERGs \citep[see e.g. ][]{Best2004, Hardcastle2004, Gendre2013, Ineson2015}. It may therefore not be surprising that at high redshifts the difference between the two populations are less. 

Comparing our results to the assumed bias of different galaxy populations in the $S^3$ simulation (at $z<1.5$) we find (see Figure \ref{fig:bias}) that the evolution in the bias for all AGN is similar to that expected from \cite{Wilman2008}, with a constant halo mass of $\sim 1 - 2 \times 10^{13}$\,$h^{-1}$\,$M_{\odot}$. This is larger than that measured for SFGs, emphasising that AGN inhabit more biased and more massive haloes. At $z>1.5$ the measured bias of these AGN appears to flatten. 

The evolution in the bias follows a similar shape to e.g. the FRI line from $S^3$ however the flattening of these lines in $S^3$ is due to an imposed constant bias evolution at large redshifts \citep[for further details see][]{Wilman2008}. For a fixed halo mass, the actual bias at a given redshift should be higher than the lines from \cite{Wilman2008}. The fact that we observe a flattening in the bias for the AGN suggests that at high redshifts the typical halo mass hosting AGN inhabit declines. This suggests lower mass haloes can host AGN at high redshifts.

In their simulations, \cite{Wilman2008} assumed halo masses of $M_{h} =10^{13}$\,$h^{-1}$\,M$_{\odot}$ for FRI radio galaxies, corresponding to a bias at $z \sim 0.7$ of $b\sim 2$. Our measurements of the bias of MLAGN show a similar evolution to that for FRI source in $S^3$ however at a higher clustering amplitude, with a bias closer to $b \sim 3$ at $z\sim 0.7-0.8$, corresponding to a roughly constant mass of $M_{h} \sim 3 - 4\times 10^{13}$\,$h^{-1}$~M$_{\odot}$. 

At $z<1.5$, the bias values for HLAGN lie significantly lower than adopted in the $S^3$ simulation for the powerful FRII radio galaxies, where a halo mass of $M_{h} =10^{14}$\,$h^{-1}$\,M$_{\odot}$ was assumed. At the median redshift ($z_{med}=1.35$) of the whole HLAGN sample we find a bias of $b \sim 3.9$ and when we restrict the sample to $z<1$ we find $b \sim 1.8$, with a median redshift of $z_{med} = 0.7$. Thus the halo masses of the efficient HLAGN are closer to $M_{h} \sim 1 - 2 \times10^{13}$\,$h^{-1}$~M$_{\odot}$, an order of magnitude lower than assumed for the powerful AGN, FRIIs, in the $S^3$ simulation. 

The bias measurements for HLAGN at $z>1.5$ (in Figure \ref{fig:bias}) appear to show a flattening towards high $z$ (at $z \sim 1.8$). The flattening of the bias to high redshift again implies that the halo mass required to host a powerful HLAGN becomes lower towards $z\sim 2$. A similar decrease with redshift of the halo mass of AGN (although at higher redshift, $z>3$) has been recently found by \cite{Allevato2016} in their analysis of X-ray selected AGN in the COSMOS field. This could be related to the fact that the gas supply in lower mass haloes is plentiful at such high redshifts, even in the lower mass haloes, and therefore a powerful AGN can be fuelled efficiently. This would also fit in with measurements of the strong evolution in the co-moving space density of radio sources to $z\sim 2$ \citep[e.g.][]{DP90, Willott2001,Jarvis2001}, with the more abundant lower mass haloes being able to host powerful radio-loud AGN.

Our results show that MLAGN are significantly more clustered (Figure \ref{fig:bias}) at $z<1$ than HLAGN, and reside in much more massive haloes, reflecting the conclusions drawn in \cite{Hardcastle2004,  Tasse2008, RamosAlmeida2013, Gendre2013}, who suggest inefficient accreters (MLAGN) reside in denser environments.
The fuelling mechanism of MLAGN also leads us to expect that MLAGN reside in hotter surroundings \citep{Janssen2012}. In these hot and dense environments the thermal energy of the material hampers its ability to accrete onto the AGN. HLAGN, on the other hand are efficient at accreting, suggesting they are in cooler environments where the accreting material has lower thermal energy. \cite{Ineson2015} investigated the X-ray emission in the environments of a sample of $z\sim 0.1$ radio sources, and indeed found HERGs (efficient accreters) on average inhabiting lower temperature regions than LERGs (inefficient accreters), reinforcing the view that MLAGN reside in hotter environments than HLAGN.

However, we can also infer that there is not a one-to-one relation between the accretion mode or radio power. Furthermore, this work suggests the recipe used to relate the radio sources to their dark matter haloes in $S^3$ may be incorrect and requiring adjustment.

To further investigate whether the clustering is dependent on the radio luminosity, or on the accretion mode we split the AGN by their radio luminosity. We find that the low luminosity AGN  ($L_{3\rm GHz} < 10^{23}$~W~Hz$^{-1}$) have a clustering length and bias which corresponds to a halo mass of $M_{h} \sim 3\times 10^{12} - 10^{13}$\,$h^{-1}$\,M$_{\odot}$, at a redshift of $z\sim 0.6-0.7$. However, by isolating the high-luminosity radio AGN ($L_{3\rm GHz} > 10^{23}$\,WHz$^{-1}$), we find much higher values for the clustering length and bias, not only when considering all high-luminosity sources ($z_{med} = 1.56$), but also when restricting the analysis to $z<1$ and $z \ge 1$. We find that the observed evolution of bias is consistent with a constant halo mass of $M_{h} \sim 2-3\times 10^{13}$\,$h^{-1}$~M$_{\odot}$, a factor of 2 - 10 higher than found for the low-luminosity population.
This probably reflects the fact that higher radio luminosity AGN reside in more massive galaxies \citep[see e.g.][]{Jarvis2001kz,McLure2004}, which reside in more massive haloes. As such, they are more biased tracers of the dark matter distribution. 

\cite{Best2004}, \cite{Ineson2013} and \cite{Ineson2015} suggest there may be a relation between cluster richness and the radio luminosity.  We do not compare here how the luminosity of the HLAGN/MLAGN correlates with their clustering as larger samples of HLAGN and MLAGN are required to investigate this. However our comparison between the high and low luminosity AGN has shown that the luminosity is correlated with the clustering of sources. This may relate to the increased mass of the galaxies, and hence halo mass, needed to host higher luminosity radio galaxies.

It is important to note though that with all these measurements, we are highly affected by the redshift range used as well as the median redshift of the sample. This can have an effect on the values of $r_0$ and $b$, particularly for the high median redshifts investigated here. For smaller redshift ranges the effects of cosmic variance are more likely to affect our results as we are more likely to see effects of over/under-dense regions in our clustering measurements that are smoothed out in our modelled redshift distributions. At low redshifts ($z<1$) we are mainly reliant on spectroscopic redshifts, and so fine binning here is likely to be more reliable. However errors in the photometric redshifts at higher redshifts could affect our clustering measurements if finer redshift binning was used. The use of a large $z \ge 1$ bin should reduce errors that could arise from photometric redshift errors. Hence, deeper and wider surveys with deep multi-wavelength data are needed in the future to better understand the evolution of bias for different radio galaxy types. 

\section{Conclusions}
\label{sec:conclusions}
In this work we have investigated the clustering of radio sources in the COSMOS field, using the angular two-point correlation function, $\omega(\theta)$. We fit $\omega(\theta)$ as a power law ($\omega(\theta) = A \theta^{-0.8}$) and used the redshift distribution to determine the spatial correlation length, $r_0$ and bias, $b$. Using the wealth of ancillary data in this field, we measure clustering strength as a function of galaxy type (SFG or AGN), luminosity and AGN accretion efficiency. This allows us to understand how the haloes in which radio sources reside in affect their properties. We also considered how the measured clustering changes as a function of redshift using low ($z<1$) and high redshift ($z \ge 1$) sub-samples. 

We find that when constrained to $z<1$ Star Forming Galaxies (SFGs) have a clustering length $r_0=5.0 ^{+ 0.5 }_{- 0.6 }$\,Mpc\,$h^{-1}$ ($b=1.5 ^{+ 0.1 }_{- 0.2 }$, $z_{med} \sim 0.6$). AGN on the other hand have $r_0=6.9 ^{+ 0.6 }_{- 0.7 }$\,Mpc\,$h^{-1}$ ($b=2.1 \pm 0.2$, $z_{med} \sim 0.7$). Our results imply a halo mass of $M_{h} \sim 3-5\times 10^{12}$\,$h^{-1}$\,M$_{\odot}$ for SFGs and $M_{h} \sim 1 - 2 \times 10^{13}$\,$h^{-1}$~M$_{\odot}$ for AGN. Thus, as expected, we find SFGs reside in lower mass haloes than the general AGN population. Furthermore, we found that the clustering of SFGs appear to have little dependence on the Star formation Rate (SFR), at least over the redshift range considered ($z<1$).

We further investigated how the clustering of radiatively efficient AGN (HLAGN) and radiatively inefficient AGN (MLAGN) differ. For $z<1$, we find MLAGN have a  clustering length of $r_0 =9.7 ^{+ 1.2 }_{- 1.3}$\,Mpc\,$h^{-1}$ ($b= 2.9 \pm 0.3 $, $z_{med} \sim 0.7$), whereas HLAGN at a similar redshift exhibit weaker clustering, with $r_0= 5.8 ^{+ 1.4 }_{- 1.8 }$\,Mpc\,$h^{-1}$ ($b=1.8 ^{+ 0.4}_{- 0.5 }$, $z_{med} \sim 0.7$). 
MLAGN and HLAGN are thought to be analogues to L/HERGs (Low/High Excitation Radio Galaxies) which have physically different fuelling mechanisms, with HLAGN (HERGs) accreting more efficiently than MLAGN (LERGs). In more clustered (and hotter) environments, the material that would be accreted by the AGN has more kinetic energy, making it more difficult to accrete. AGN in these hot haloes (the MLAGN) would therefore be less efficient. HLAGN on the other hand in less clustered environments, and lower mass haloes, can accrete the cooler gas more easily. 

By considering the clustering for these different sources using the full sample, as well as those at low ($z<1$) and high redshifts ($z \ge 1$), we were also able to investigate the evolution of $r_0$ and $b$. The bias evolves with redshift for both the HLAGN and MLAGN sub-samples. The measured bias corresponds to a roughly constant dark matter halo mass of  $M_{h} \sim 3-4\times 10^{13}$\,$h^{-1}$~M$_{\odot}$ for MLAGN and $M_{h} \sim 1 - 2 \times 10^{13}$\,$h^{-1}$~M$_{\odot}$ for HLAGN when measured at $z<1.5$. At $z>1.5$, HLAGN showing a flattening in the bias at high redshift $(z>1.5)$, which suggests that the required halo mass of these sources may change at higher redshift, with lower halo masses being sufficient to be the host of a HLAGN, presumably due to the higher density of cold gas available for accretion. If the more numerous, less massive haloes are sufficient to host a powerful radio source at high redshift, then this may partly explain the strong evolution in the comoving space density of powerful radio galaxies \citep[e.g.][]{DP90,Willott2001,Jarvis2001}.
Our work also suggests that the halo masses assumed in the $S^3$ simulation may need reviewing and may be too low for FRI-type radio galaxies, and too high for the FRII-type sources, although we note that there is not a one-to-one relationship between the $S^3$ definition of FRI/FRII and the MLAGN/HLAGN classification.

The results presented here also have important consequences for using future wide-area and deep radio surveys for tracing the large-scale structure of the Universe, and using this to infer the level of e.g. non-Gaussianity using the multi-tracer technique \citep[see e.g.][]{Ferramacho2014,Raccanelli2015}. The fact that we find much lower bias for the HLAGN, which dominate the bright end of the radio source luminosity function would imply that the multi-tracer technique, which relies on a large difference in bias of distinct populations to overcome cosmic variance, could be less efficient than suggested in previous work. 

It is important to carry out similar analyses as performed in this paper over larger and deeper areas and obtaining more spectroscopic redshifts for galaxies, in order to fully constrain how bias scales with radio source luminosity for both the AGN and SFGs and how it evolves with redshift. Indeed, current and imminent surveys with LOFAR \citep{Shimwell2017}, ASKAP \citep{Norris2011} and MeerKAT \citep{Jarvis2017} should enable us to carry out in-depth studies of the relationship between radio AGN and SFGs and their dark matter haloes. In particular, using deep radio surveys that cover the best extragalactic deep fields \citep[e.g.][]{Mauduit2012,Oliver2012, Jarvis2013,Brandt2015}, will provide the opportunity to fully characterise the radio source population, as has been done with the VLA COSMOS survey used here, and measure the clustering properties.

\section*{Acknowledgements}

We would like to thank the referee for their helpful comments and suggestions. CLH would like to acknowledge the support given from the Science and Technology Facilities Council (STFC) for their support to the first author through an STFC studentship. This work was supported by the Oxford Centre for Astrophysical Surveys which is funded through generous support from the Hintze Family Charitable Foundation, the award of the STFC consolidated grant (ST/N000919/1). MJJ acknowledges support from the SA SKA. VS, ID, MN acknowledge funding by the European Union's Seventh Frame-work program under grant agreement 337595 (ERC Starting Grant, `CoSMass'). PWH wishes to acknowledge support provided through an STFC studentship. 

\bibliography{cosmostpcf_NEW}
\bibliographystyle{mnras}

\bsp	
\label{lastpage}

\end{document}